%% file: COMPLETE.TEX
\documentclass[11pt,oneside]{book}%
\usepackage[dvips]{graphics}%

\usepackage[centertags]{amsmath}
\usepackage{amsfonts}
\usepackage{amssymb}
\usepackage{amsthm}
\usepackage{newlfont}

\begin{document}%

\setlength{\unitlength}{1mm} \baselineskip .85cm
\input{chap01}
\newpage
\pagenumbering{arabic}

\include{chap1}
\include{chap2}
\include{chap3}
\include{chap4}
\include{chap5}
\include{chap6}
\include{chap7}

\include{chap8}

\include{chap04}
\include{chap05}
\include{chap06}

\end{document}

%% file: chap01.tex
\large \baselineskip .85cm
\begin{titlepage}
\title{\Huge\vspace{-3cm} {\bf ACCELERATING EXPANSION OF THE UNIVERSE} \vspace*{3cm} \\
\Large {\bf THESIS SUBMITTED FOR THE DEGREE OF\\
 DOCTOR OF PHILOSOPHY\\
 (SCIENCE)\\
    OF\\
    BENGAL ENGINEERING AND SCIENCE UNIVERSITY}\\\vspace{.1cm}
    }

\author{ {\bf BY} \vspace*{1cm} \\
\LARGE{\bf WRITAMBHARA CHAKRABORTY}  \vspace*{8cm}\\
\large {\bf DEPARTMENT OF MATHEMATICS} \\
\large {\bf BENGAL ENGINEERING AND SCIENCE UNIVERSITY}\\
\small  {\bf SHIBPUR, HOWRAH - 711 103} \\
\large {\bf INDIA}\\\\ {\bf APRIL, 2009}}
\date{}
\maketitle
\end{titlepage}
\pagenumbering{roman}
\newpage
\vspace*{2.5cm}
\begin{center}
 {\large {\bf\Large DECLARATION} }
\end{center}
\vspace*{1.5cm} This is to certify that the thesis entitled ``{\it
ACCELERATING EXPANSION OF THE UNIVERSE }" submitted by Writambhara
Chakraborty who got her name registered on 04.11.2006 for the
award of Ph.D. (Science) degree of Bengal Engineering and Science
University, is absolutely based upon her own work under the
supervision of Dr.Ujjal Debnath, Department of Mathematics, Bengal
Engineering and Science University, Howrah 711 103, and that
neither this thesis nor any part of its has been submitted for any
degree / diploma or any
other academic award anywhere before.\\

\vspace*{3cm}
\noindent Ujjal Debnath \\
(Supervisor)\\
Lecturer,                     \\
Department of Mathematics      \\
Bengal Engineering and Science University           \\
Shibpur   \\
Howrah - 711 103     \\
India                          \\


\newpage
\vspace*{1cm}
\begin{center}
 {\huge Acknowledgements}
\end{center}
\vspace*{1cm}

To start of I express my sincere gratitude to my supervisor Dr.
Ujjal Debnath for his help and support throughout my research
work. He has been a true friend, mentor and guide, teaching me all
the lessons I needed to learn during my entire research period,
sometimes scolding me and sometimes encouraging me to analyse and
solve the problems without being impatient. I thank him for all
the long but fruitful discussions we used to have regarding all my
seven papers with him. I am also grateful to Prof. Subenoy
Chakraborty for his valuable suggestions and advice during my
research period. He has also coauthored one of my papers. I would
like to thank the teachers and staff of the Department of
Mathematics, Bengal Engineering and Science University, for their
valuable resources and help. I would also like to thank the
members of Cosmology and Relativity Centre and the Department of
Mathematics, Jadavpur University for their cooperation. Thanks to
IUCAA, Pune, India, for their warm hospitality and facilities
during my visits there, as a lot of my work were carried out
there. I am grateful to Dr. Edmund J. Copeland of University of
Nottingham, UK, for his valuable suggestions and immense help
during my short period of visit there in 2008. I would like to
thank all the teachers and students of the Particle Theory Group
and the Astrophysics Group of University of
Nottingham for their cooperation during my stay over there.\\

I would like to thank my colleagues of New Alipore College,
Kolkata where I am presently employed as a lecturer. They have
been wonderful. Specially I would like to thank my Head of the
Department Prof. Dibes Bera and Principal Dr. Sujit Kumar Das,
who have helped me a lot by granting me leave whenever I needed.
I must mention Prof. B. B. Paira, Director of Heritage Institute
of Technology and Dr. Rasajit Kumar Bera, Head of the Department
of Mathematics, Heritage Institute of Technology, where I started
of my career as a lecturer and a research student. I am indebted
to them for encouraging me and helping me know the right path. I
should not forget my colleagues and friends of Heritage Institute
of Technology as well, who made that short phase of my life over
there a really cherishing one.\\

I would like to thank Ma and Babai for everything. It is they who
taught me all the lessons of life. It is they and their blessing
for which I am what I am right now. Thanks to my sister, my
parents-in-law and all the other members of my family for their help and support.\\

Thanks Anirban for being by my side through all the good and bad
times in life and constantly encouraging me not to lose hope. You
truly are the best friend and partner that I could have ever had.
And last but not the least, thanks to my little Rian, whose
arrival has truly been the happiest phase transition of my life.\\

\vspace*{2cm}
\noindent Writambhara Chakraborty\\
Research Scholar\\
Department of Mathematics      \\
Bengal Engineering and Science University           \\
Shibpur   \\
Howrah - 711 103     \\
India                          \\

\baselineskip .65cm
\tableofcontents

%% file: chap1.tex
\large \baselineskip .85cm

\chapter{General Introduction}
\setcounter{page}{2} \markright{\it CHAPTER~\ref{chap1}. General
Introduction}
\label{chap1}%

\section{Standard Cosmology and FRW Model}

The standard Cosmology assumes that at the beginning
(approximately 13 billion years ago) there must have been an
initial singularity from which the space time suddenly started
evolving. Since then the Universe has more or less gone through a
process of expansion and cooling from an extremely hot and dense
state to the present day cool state. In the first few seconds or
so there was a very fast expansion, known as Inflation
[http://cosmology.berkeley.edu], which is responsible for the
present homogeneous and isotropic Universe. Following this
inflationary phase, further expansion cooled down the Universe and
matter was produced in the process called baryogenesis. Various
light elements like deuterium, helium, lithium-7 were created in a
process called Big Bang Nucleosynthesis. The universe was still
very hot for the nuclei to bind electrons and form atoms,
therefore causing the Universe to be opaque to photons and other
electromagnetic radiation. Eventually the temperature drops enough
for free nuclei and electrons to combine into atoms in a process
called recombination. After the formation of atoms photons could
travel freely without being scattered. This caused the emission of
Cosmic Microwave Background Radiation (CMBR) which gives us the
information about the Universe at that time. Galaxies and stars
began to form after a few hundred million of years as a result of
gravitational collapse.\\

Modern Cosmology with the help of observational evidences has
within its reach billions of galaxies and all the heavenly bodies
spread all across the vast distances. Advanced observational
techniques have strengthened the particular branch of science,
sometimes supporting the conventional theories and sometimes
producing reverse results. As a consequence of these observational
advances cosmology has become more or less data driven, so that
all the theories need to be fitted with array of observations,
although there are still doubts and debates about the reliability
and interpretation of such data [Frieman, 1994].\\

The standard cosmological model which is very successful in
describing the evolution of the Universe, is based on homogeneous
and isotropic Friedmann-Robertson-Walker (FRW) spacetime.
Homogeneity and isotropy that we assume for this model is
supported by strong observational data [Smoot etal, 1992; Maddox
etal, 1990; Collins etal, 1992], CMBR measurements and galaxy
redshift surveys [Fisher etal, 1993; Geller etal, 1989]. This
global isotropy and homogeneity which is the foundation of the
standard cosmology is called Cosmological Principle. Cosmological
Principle leads to Hubble's Law, which says that the recession
velocity of galaxy is proportional to the distance from us, i.e.,
$V=HD$. The proportionality constant $H=\frac{\dot{a}}{a}$ is
known as Hubble constant and $a(t)$ is
the scale factor.\\

We now look at Einstein's model of the Universe. In 1932, Einstein
and de Sitter presented the Standard Cosmological Model of the
Universe, which has been the most favourite among the
cosmologists till 1980.\\

Initially Einstein assumed homogeneity and isotropy in his
cosmological problem. He chose a time coordinate $t$ such that the
line element of static space-time could be described by [Narlikar,
{\it An Introduction to Cosmology}],

\begin{equation}
ds^{2}=c^{2}dt^{2}-g_{\mu \nu}dx^{\mu}dx^{\nu}
\end{equation}

where $g_{\mu \nu}$ are functions of space coordinates
$x^{\mu}(\mu, \nu=1, 2, 3)$ only.

We can now construct the homogeneous and isotropic closed space
of three dimensions that Einstein wanted for his model of the
Universe. The equation of such a 3-surface of a four dimensional
hypersphere of radius $a$ is given in Cartesian coordinates
$x_{1}, x_{2}, x_{3}, x_{4}$ by

\begin{equation}
x_{1}^{2}+x_{2}^{2}+ x_{3}^{2}+ x_{4}^{2}=a^{2}
\end{equation}

Therefore the spatial line element on the surface is given by

\begin{equation}
d\sigma^{2}=(dx_{1})^{2}+(dx_{2})^{2}+(dx_{3})^{2}+(dx_{4})^{2}=a^{2}[d\chi^{2}+sin^{2}\chi(d\theta^{2}+sin^{2}\theta
d\phi^{2})]
\end{equation}

where $x_{1}=a ~sin\chi ~cos\theta,~x_{2}=a~sin\chi~ sin\theta
~cos\phi,~x_{3}=a~sin\chi ~sin\theta~ sin\phi,~x_{4}=a~cos\chi$
and the ranges of $\theta,~\phi$ and $\chi$ are given by
$0\leq\chi\leq\pi,~0\leq\theta\leq\pi,~0\leq\phi\leq2\pi$\\

Another way to express $d\sigma^{2}$ through coordinates
$r,~\theta,~\phi$ with $r=sin\chi$, (0$\leq r\leq 1$) is,

\begin{equation}
d\sigma^{2}=a^{2}\left[
\frac{dr^{2}}{1-r^{2}}+r^{2}(d\theta^{2}+sin^{2}\theta
d\phi^{2}))\right]
\end{equation}

The line element for the Einstein Universe is therefore given by

\begin{equation}
ds^{2}=c^{2}dt^{2}-d\sigma^{2}=c^{2}dt^{2}-a^{2}\left[
\frac{dr^{2}}{1-r^{2}}+r^{2}(d\theta^{2}+sin^{2}\theta
d\phi^{2}))\right]
\end{equation}

This line element is for + ve curvature only.\\

In general we have

\begin{equation}
ds^{2}=c^{2}dt^{2}-a^{2}\left[ \frac{dr^{2}}{1-k
r^{2}}+r^{2}(d\theta^{2}+sin^{2}\theta d\phi^{2})\right]
\end{equation}

where,  $k = 0 , +1, -1$  for  zero, +ve, -ve curvatures
respectively and are also  known  as  flat, closed, open  models
and  $a(t)$ is  known  as  the scale  factor  or  expansion
factor.\\

Thus for $c=1$, FRW line element reduces to,

\begin{equation}
ds^{2}=dt^{2}-a^{2}(t)\left(\frac{dr^{2}}{1-kr^{2}}+r^{2}d\theta^{2}+r^{2}sin^{2}\theta
d\phi^{2} \right)
\end{equation}

Now the energy-momentum tensor describing the material contents
of the Universe is given by

\begin{equation}
T_{\mu \nu}=(\rho c^{2}+p)u_{\mu}u_{\nu}-pg_{\mu \nu}
\end{equation}

where, $\rho=T_{00}$ is mean energy density of matter, $p=T_{ii}$
is the pressure, and $u_{\mu}= (c,0,0,0)$ is fluid four velocity.
Usually $\Omega_{i}=\frac{\rho_{i}}{\rho_{c}}$, where
$\rho_{c}=3H^{2}/8 \pi G$, is called the critical energy density.\\

Also the equation of motion describing the Universe, known as
Einstein field equations in general relativity are

\begin{equation}
G_{ik}=R_{ik}-\frac{1}{2}g_{ik}R=\frac{8 \pi G}{c^{4}}T_{ik}
\end{equation}

where, $G_{ik}$ = Einstein Tensor, $R_{ik}$ = Ricci Tensor, $R$ =
Ricci Scalar.\\

Thus for a static ($\dot{a}=0$), dust filled ($p=0$) and closed
($k=+1$) model of the universe, the field equations yield,
(choosing $8\pi G=c=1$)

\begin{equation}
\frac{3}{a^{2}}=\rho,~\frac{1}{a^{2}}=0
\end{equation}

Clearly no feasible solution is possible from these equations,
thus suggesting that no static homogeneous isotropic and dense
model of the Universe is possible under the regime of Einstein
equations stated above.\\

For this reason Einstein later modified his field equations as

\begin{equation}
G_{ik}=\frac{8 \pi G}{c^{4}}T_{ik}+\Lambda g_{ik}
\end{equation}

Thus introducing the famous $\Lambda$-term, known as Cosmological
Constant. With this, the picture changes to,
$\frac{3}{a^{2}}=\rho+\Lambda,~\frac{1}{a^{2}}=\Lambda$, so that
$a=\frac{1}{\sqrt{\Lambda}}=\sqrt{\frac{2}{\rho}}$.

This $\Lambda$ is universal constant like $G,~ c$, etc. To
estimate the value of $\Lambda$ the value of $\rho$ was used in
1917, which are given as follows,\\
$a\sim 10^{29}~ cm$, ~$\rho\sim 10^{-31} ~gm~ cm^{-3}$, ~$\Lambda\sim 10^{-58}~cm^{-2}$.\\
The $\Lambda$ - term introduces a force of repulsion between two
bodies that increases in proportion to the distances  between
them.\\

Einstein first proposed the cosmological constant as a
mathematical fix to the theory

\begin{equation}
\frac{\dot{a}^{2}}{a^{2}}+\frac{k}{a^{2}}=\frac{1}{3}\rho
\end{equation}

\begin{equation}
\frac{\ddot{a}}{a}=-\frac{1}{6}(\rho+3p)
\end{equation}

in general relativity. In its simplest form, general relativity
predicted that the universe must either expand or contract.
Einstein thought the universe was static, so he added this new
term to stop the expansion. Friedmann, a Russian mathematician,
realized that this was an unstable fix and proposed an expanding
universe model called {\bf Friedmann model} of the Universe.\\
For expanding Universe, $\dot{a}>0$. But since for normal matter
$\rho>0,~p\geq 0$, hence the second equation gives $\ddot{a}<0$.
So that, $\dot{a}$ is decreasing, i.e., expansion of Universe is
decelerated. This model is known as {\bf Standard Cosmological Model} (SCM).\\

Teams of prominent American and European Scientists using both
Hubble Space Telescope (HST) and Earth based Telescopes had
announced in 1998 the results of their many years observation and
measurement of the expansion of the Universe. Their collective
announcement was that the Universe is not just expanding, it is
in fact, expanding with ever increasing speed. This combined
discovery has been a total surprise for Cosmology! The SCM states
that Universe is decelerating but recent high redshift type Ia
Supernovae (explosion) observation suggests the Universe is
accelerating [Perlmutter et al, 1998, 1999; Riess et al, 1998;
Garnavich et al, 1998]. So there must be some matter field, either
neglected or unknown, which is responsible for accelerating
Universe.\\

For accelerating Universe, $\ddot{a}>0$, i.e., $\rho+3p<0$, i.e.,
$p<-\frac{\rho}{3}$. Hence, the matter has the property $-$ve
pressure. This type of matter is called Quintessence  matter
(Q-matter)  and the problem is called Quintessence  problem. The
missing energy in Quintessence problem  can  be  associated  to a
dynamical time dependent and  spatially  homogeneous /
inhomogeneous scalar field evolving  slowly  down  its potential
$V(\phi)$. These types of cosmological models are known as
quintessence models. In this models the scalar field can be seen
as a perfect fluid with a negative  pressure  given  by
$p=\gamma\rho$, ($-1\le\gamma\le$
1).\\

Introducing $\Lambda$ term,

\begin{equation}
\frac{\dot{a}^{2}}{a^{2}}+\frac{k}{a^{2}}=\frac{1}{3}\rho+\frac{1}{3}\Lambda=\frac{1}{3}(\rho+\Lambda)
\end{equation}

\begin{equation}
\frac{\ddot{a}}{a}=-\frac{1}{6}(\rho+3p)+\frac{1}{3}\Lambda=-\frac{1}{6}[(\rho+\Lambda)+3(p-\Lambda)]
\end{equation}

If $\Lambda$ dominates, $\ddot{a}>0$ , i.e, Universe will be
accelerating.\\

For normal fluid,  relation between $p$ and $\rho$ is given by,
$p=\gamma \rho$, ($0\leq\gamma\leq 1$) which is called  equation
of state.\\

For dust, $\gamma=0$; for radiation, $\gamma=\frac{1}{3}$.\\
Here $\Lambda$ satisfies an equation  of state $p=-\rho$, so
pressure is negative. Therefore for accelerating Universe we need
such type  of  fluid  which  generates  negative pressure. The
most puzzling questions that remain to be explained in cosmology
are the questions about the nature of these types of matter or the
mystery of ``missing mass", that is, the ``dark energy" and
``dark matter" problem.\\\\

\section{Dark Energy and Dark Matter}

In 1998, two teams studying distant Type Ia supernovae presented
independent results of observation that the expansion of the
Universe is accelerating [Perlmutter et al, 1998, 1999; Riess et
al, 1998; Garnavich et al, 1998]. For the last few decades
cosmologists had been trying to measure the deceleration of the
Universe caused by the gravitational attraction of matter,
characterized by the deceleration parameter $q=-\frac{a
\ddot{a}}{\dot{a}^{2}}$, as suggested by SCM. Therefore, discovery
of cosmic acceleration has been proved to be one of the most
challenging and important development in Cosmology, which
evidently indicates existence of some matter field, either unknown
or neglected so far, responsible for this accelerated
Universe [Bachall et al, 1999].\\

Observations suggest that $\Omega_{baryon} \simeq 0.04$ and
$\Omega_{total}=1.02 \pm 0.02$. That is ordinary matter or
baryons (atoms) described by the standard model of particle
Physics is only approximately $4\%$ of the total energy of the
Universe, another $23\%$ is some {\bf dark matter} and $73 \%$ is
{\bf dark energy}, yet to be discovered. Thus dark matter and dark
energy are considered to be the missing pieces in the cosmic
jigsaw puzzle [Sahni, 2004]\\
$\Omega_{total}-\Omega_{baryons} = ?$ \\

Although the nature of neither dark matter (DM) nor dark energy
(DE) is currently known, it is felt that both DM and DE are
non-baryonic in nature.  Non-baryonic dark matter does not emit,
absorb or scatter light and also it has negligible random motion
[Springel et al, Bennett, 2006]. Thereafter it is called cold dark
matter (CDM). Although, depending on the mass of the particle it
is usually differentiated between hot dark matter and cold dark
matter. Non-baryonic Hot Dark Matter (HDM) particles are assumed
to be relativistic while decoupling from the masses around and so
have a very large velocity dispersion (hence called hot), whereas,
CDM particles have a very small velocity dispersion and decouple
from the masses around when they are non-relativistic.
Non-baryonic dark matters do not seem to interact with light or
any known baryonic matter, therefore they are called weakly
interacting massive particles (WIMP). The standard cold dark
matter (SCDM) paradigm assumes $\Omega_{m}=1$, whereas, LCDM, a
modification of SCDM, which consists of CDM and a Lambda-term or
cosmological constant $\Omega_{m}=0.3$ and the hot dark matter
scenario is constrained due to structure formation calculations. A
variety of dark matter candidates are present in literature, among
which a WIMP called nutralino is one of the most favoured one.
This is a 100-1000 GeV particle and fermionic partner to the gauge
and Higgs bosons [Sahni, 2004]. Another candidate for CDM is an
ultra-light pseudo-Goldstone boson called an axion, which arises
in the solution of the strong CP problem of particle physics
[Masso, 2003]. Other candidates for CDM are string theory
motivated modulii fields [Brustein, 1998]; non-thermally produced
super-heavy particles having mass $\sim 10^{14}$ GeV and dubbed
Wimpzillas [Maartens, 2003 ]; axino (with a mass m $\sim$ 100keV
and a reheating temperature of 106K) and gravitino, superpartners
of the axion and graviton respectively [Roszkowski, 1999].
Although dark matter is assumed to be comprised of particles which
are pressureless cold dark matter with an equation of state
$\omega_{CDM} = 0$, nature of dark matter is yet to be discovered.\\

Existence of dark energy has been driven by the recently
discovered accelerated expansion of the Universe. Equation (1.13)
shows that acceleration of the Universe is to be expected when
pressure is sufficiently negative. Also expansion of the Universe
should decelerate if it is dominated by baryonic matter and CDM.
Two independent groups observed SNe Ia to measure the expected
deceleration of the Universe in 1998. Instead, both the groups
surprisingly discovered that the expansion of the Universe is
accelerating. This discovery hinted at a new negative pressure
contributing to the mass-energy of the Universe in equation
(1.13). Now, equation (1.13) that a universe consisting of only a
single component will accelerate if $\omega < -1/3$. Fluids
satisfying $\rho+3p \geq 0$ or $\omega \geq -1/3$ are said to
satisfy the {\bf strong energy condition} (SEC). We therefore find
that, in order to accelerate, dark energy must violate the SEC.
Another condition is the {\bf weak energy condition} (WEC) $\rho+
p \geq 0$ or $\omega \geq -1$. If WEC is not satisfied, Universe
can expand
faster than the exponential rate causing cosmic {\it Big Rip}.\\

Now the greatest challenge for cosmologists was to find an
explanation for this accelerated expansion of the Universe. One
modification was made by Einstein himself in1917 by introducing
the cosmological constant term, to act as a gravitational
repulsive term, to achieve a static Universe, as seen in the
previous section. Later it was dropped after Hubble's discovery
of the expansion of the Universe in 1929. In some models of the
dark energy, it is identified with this cosmological constant
$\Lambda$. However, particle physics looks at cosmological
constant as an energy density of the vacuum. Moreover, the energy
scale of $\Lambda$ should be much larger than that of the present
Hubble constant $H_{0}$, if it originates from the vacuum energy
density. This is the
"cosmological constant problem" [Weinberg, 1989].\\

Thus cosmological constant with equation of state $\omega=-1$ may
play the role to drive the recent cosmic acceleration. Another
possibility is that there exists a universal evolving scalar field
with equation of state $\omega<-1/3$, called quintessence field.
Also a few more models have been proposed recently in support of
dark energy analysis. We discuss a few of these models below.\\

\subsection{Cosmological Constant}

Cosmological constant is the simplest form of dark energy
($\omega=-1$). Also several cosmological observations suggest that
cosmological constant should be considered as the most natural
candidate for dark energy [Padmanabhan, 2006]. As discussed above
Einstein [1917] introduced cosmological constant $\Lambda$ in
order to make the field equations of GR compatible with Mach's
Principle and thus producing a static Einstein Universe. Later in
1922, Friedmann constructed a matter dominated expanding universe
without a cosmological constant [Sahni etal, 2000]. Friedmann
cosmological model was accepted as a standard cosmological model
after Hubble's discovery of expansion of Universe [Weinberg, 1989]
and thus Einstein dropped the idea of a static Universe and
henceforth positive $\Lambda$-term. Later Bondi [1960] and McCrea
[1971] realized that if the energy density of the cosmological
constant is comparable to the energy density of the present epoch,
the cosmological model takes a very reliable form [Padmanabhan,
2003]. But the importance of cosmological constant was first
noticed when Zeldovich [1968] showed that zero-point vacuum
fluctuations have Lorentz invariant of the form $p_{vac} =
-\rho_{vac}$, which is the equation of state for $\Lambda$,
therefore $T^{vac}_{\mu \nu}=\Lambda g_{\mu \nu}$, which shows
that the stress-energy of
vacuum is mathematically equivalent to $\Lambda$.\\

In 1970, the discovery of supersymmetry became a turning point
involving the cosmological constant problem, as the contributions
to vacuum energy from fermions and bosons cancel in a
supersymmetric theory. However supersymmetry is expected to be
broken at low temperatures at that of the present Universe. Thus
cosmological constant is expected to vanish in the early universe
and exist later when the temperature is sufficiently low. Which
poses a new problem regarding cosmological constant as a large
value of $\Lambda$ at an early time is useful from the viewpoint
of inflation, whereas a very small current value of $\Lambda$ is
in agreement with observations [Sahni etal, 2000]. A positive
$\Lambda$ term was still of interest as the inflationary models
constructed during 1970 and 1980's demanded matter, in the form of
vacuum polarization or minimally coupled scalar field, behaving as
a weakly time-dependent $\Lambda$-term. However, recently
constructing de-Sitter vacua in string theory or supergravity has
been very useful in solving cosmological
constant problem [Copeland etal, 2006].\\

Now introducing $\Lambda$ in Einstein field equations of GR (1.9),
we get,

\begin{equation}
G_{ik}+\Lambda g_{ik}=8 \pi G T_{ik}
\end{equation}

Thus the Friedmann equations become,

\begin{equation}
H^{2}+\frac{k}{a^{2}}=\frac{8 \pi G}{3}\rho+\frac{\Lambda}{3}
\end{equation}

\begin{equation}
\frac{\ddot{a}}{a}=-\frac{4 \pi G}{3}(\rho+3p)+\frac{\Lambda}{3}
\end{equation}

and the energy conservation law reads,

\begin{equation}
d(\rho a^{3})=-pda^{3}
\end{equation}

which on further reduction gives,

\begin{equation}
\dot{\rho}+3 \frac{\dot{a}}{a}(\rho+p)=0
\end{equation}

Now we define [Carroll etal, 1992],

\begin{equation}
\Omega_{\rho}=\frac{8 \pi
G}{3{H_{0}}^{2}}\rho_{0},~\Omega_{\Lambda}=\frac{\Lambda}{3{H_{0}}^{2}},~\Omega_{k}=-\frac{k}{{H_{0}}^{2}{a_{0}}^{2}}
\end{equation}

Thus equation (1.17) in combination with equation (1.21) gives,

\begin{equation}
\Omega_{\rho}+\Omega_{\Lambda}+\Omega_{k}=1
\end{equation}

and, we assume,

\begin{equation}
\Omega_{\rho}+\Omega_{\Lambda}=\Omega_{tot}
\end{equation}

so that, $\Omega_{tot}<1~(>1)$ implies a spatially open (closed)
Universe.\\

Now as we have seen before,

\begin{equation}
p_{vac} = -\rho_{vac}
\end{equation}

and thus the relation between $\Lambda$ and vacuum energy density
is,

\begin{equation}
\Lambda=8 \pi G \rho_{vac}
\end{equation}

Thus the deceleration parameter ($q=-\frac{a
\ddot{a}}{{\dot{a}}^{2}}$) becomes,

\begin{equation}
q=\frac{(1+3\omega)}{2}\Omega_{\rho}-\Omega_{\Lambda}
\end{equation}

so that, for the present universe consisting of pressureless dust
(dark matter) and $\Lambda$, the deceleration parameter takes the
form,

\begin{equation}
q_{0}=\frac{1}{2}\Omega_{\rho}-\Omega_{\Lambda}
\end{equation}

Recent cosmological observations also suggest the existence of a
non-zero cosmological constant. One of the most reliable
observation is that of high redshift Type Ia supernovae
[Perlmutter et al 1998, Riess et al 1998], which hints at a
recently accelerating Universe with a large fraction of the
cosmological density in the form of a cosmological constant. Also
models based on the cold dark matter model (CDM) with $\Omega_{m}
= 1$ does not meet up the values obtained from cosmological and
CMBR observations, whereas, a flat low density CDM+$\Lambda$
universe ($\Lambda$CDM) with $\Omega_{m} \sim 0.3$ and
$\Omega_{\Lambda} \sim 0.7$, such that $\Omega_{tot}\approx 1$
agrees remarkably well with a wide range of observational data
ranging from large and intermediate angle CMB anisotropies to
observations of galaxy clustering on large scales [Sahni etal,
2000]. This is evidence enough for a non-zero $\Lambda$-term.\\

Now, the vacuum energy density $\rho_{vac}$, proportional to
$\Lambda$, obeys [Cohn, 1998],\\
$\frac{{\rho}_{vac}^{observed}}{{\rho}_{vac}^{observed}} <
10^{-52}$\\

This disparity between the expected and the observed value of
$\Lambda$ poses the famous cosmological constant problem.\\

One solution to this problem of very large $\Lambda$-term
(predicted by field theory) and a small one (suggested by
observations) is to make $\Lambda$ time-dependent. An initial
large value of $\Lambda$ would explain inflation and galaxy
formation, while subsequent slow decay of $\Lambda(t)$ would
produce a small present value $\Lambda(t_{0})$ to be reconciled
with observations suggesting $\Omega_{\Lambda} \sim 0.7$ [Sahni
etal, 2000]. For this purpose a few phenomenological models have
been introduced, which Sahni has classified into three categories
[Sahni, 2000], viz, (1) {\it Kinematic models} where $\Lambda$ is
a function of cosmic time $t$ or the scale factor $a(t)$ in FRW
cosmology; (2) {\it Hydrodynamic models} where $\Lambda$ is
described by a barotropic fluid with some equation of state and
(3) {\it Field-theoretic models} where $\Lambda$ is assumed to be
a new physical classical field with some
phenomenological Lagrangian.\\

Now keeping $\Lambda$ to be time-dependent and moving the
cosmological term on the right hand side of equation (1.16)
[Overduin and Cooperstock, 1998], we have,

\begin{equation}
G_{ik}=8 \pi G \tilde{T}_{ik}
\end{equation}

where,

\begin{equation}
\tilde{T}_{ik}=T_{ik}-\frac{\Lambda}{8 \pi G}g_{ik}
\end{equation}

This implies that the effective energy-momentum tensor is
described by effective pressure $\tilde{p}=p-\frac{\Lambda}{8 \pi
G}$ and effective energy-density
$\tilde{\rho}=\rho+\frac{\Lambda}{8 \pi G}$. Thus the energy
conservation law (1.19) becomes,

\begin{equation}
d\left[(\rho+\frac{\Lambda}{8 \pi
G})a^{3}\right]=-(p-\frac{\Lambda}{8 \pi G})da^{3}
\end{equation}

along with equation (1.17) and (1.18).\\

A few phenomenological models of time-variant $\Lambda$-term are
given in table I followed by the one presented by
Overduin and Cooperstock.\\

{\bf Table I} \underline{}
\begin{center}
\begin{tabular}{lccccc}
\hline
$\Lambda$-decay law~~~~~~~~~~~~~~~~~~~~~~~~~~~~~~~Reference\\
\hline\\
$\Lambda\propto H^{2}$~~~~~~~~~~~~~~~~~~~~~~~~~Lima and Carvalho
(1994);
Wetterich (1995); Arbab (1997)\\
\\
$\Lambda\propto \frac{\ddot{a}}{a}$~~~~~~~~~~~~~~~~~~~~~~~~~~~Arbab (2003, 2004); Overduin and Cooperstock (1998)\\
\\
$\Lambda\propto \rho$~~~~~~~~~~~~~~~~~~~~~~~~~~~Viswakarma (2000)\\
\\
$\Lambda\propto t^{-2}$~~~~~~~~~~~~~~~~~~~~~~~~~Endo and
Fukui (1977); Canuto, Hsieh and Adams (1977);\\
~~~~~~~~~~~~~~~~~~~~~~~~~~~~~~~~~~~Bertolami (1986); Berman and
Som (1990); Beesham (1994);\\~~~~~~~~~~~~~~~~~~~~~~~~~~~~~~~~~~
Lopez and Nanopoulos (1996); Overduin and
Cooperstock (1998)\\
\\
$\Lambda\propto t^{-\alpha}$~~~~~~~~~~~~~~~~~~~~~~~~Kalligas,
Wesson and
Everitt (1992, 1995); Beesham (1993)\\ ($\alpha$ being a constant)\\
\\
$\Lambda\propto a^{-2}$~~~~~~~~~~~~~~~~~~~~~~~Ozer and
Taha (1986, 1987); Gott and Rees (1987);Kolb (1989)\\
~~~~~~~~~~~~~~~~~~~~~~~~~~~~~~~~~~Chen etal (1990); Abdel-Rahman
(1992);
Abdussattar etal (1997)\\
\\
$\Lambda\propto a^{-\alpha}$~~~~~~~~~~~~~~~~~~~~~~~Olson and
Jordan (1987); Pavon (1991); Maia
and Silva (1994)\\
$\alpha$ being a constant~~~~~~~~~~Silveira and Waga (1994, 1997);
Torres and
Waga (1996)\\
\\
$\Lambda\propto f(H)$~~~~~~~~~~~~~~~~~~~~~~Lima and Maia (1994); Lima and Trodden (1996)\\
\\

\hline\\

\end{tabular}
\end{center}

The $\Lambda$-decay laws that have been analyzed in the theory are
mostly based on scalar fields or derived from the modified version
of the Einstein action principle and they show the decay of the
cosmological term is consistent with the cosmological
observations. Many of these works do not exhibit the energy
transfer between the decaying $\Lambda$-term and other forms of
matter [Ratra and Peebles, 1988]. Some models overlook this
energy-exchange process assuming that equal amounts of matter and
antimatter are being produced (if the decay process does not
violate the baryon number)[Freese etal, 1987]. These models are
constrained by diffuse gamma-ray background observations
[Matyjasek, 1995]. Some models refer to this energy exchange
process as production of baryons or radiation. These models are
constrained by CMB anisotropies [Silveira and Waga, 1994, 1997]
and nucleosynthesis arguments [Freese etal, 1987]. Irrespective of
the sources, these models (many of which are independently
motivated), in general, solve the cosmological problems in a
simpler way and satisfy the cosmological observational
constraints [Overduin and Cooperstock, 1998].\\

Some authors have incorporated a variable gravitational constant
also while retaining the usual energy conservation law [Arbab,
1997, 1998, 2002]. A possible time-variable $G$ was suggested by
Dirac [1988] on the basis of his large number hypothesis and since
then many workers have extended GR with $G=G(t)$ to obtain a
satisfactory explanation for the the present day acceleration
[Abdel-Rahman, 1990; Abbussattar and Vishwakarma, 1997; Kalligas
etal, 1992]. Since $G$ couples geometry to matter, a
time-dependent $G$ as well as a time-dependent $\Lambda$ can
explain the evolution of the Universe, as the variation of $G$
cancels the variation of $\Lambda$, thus retaining the energy
conservation law. There also have been works considering $G$ and
$\Lambda$ to be coupled scalar fields. This approach is similar
to that of Brans-Dicke theory [Brans and Dicke, 1961]. This keeps
the Einstein field equations formally unchanged [Vishwakarma,
2008].\\

In FRW model the Einstein field equations with variable $G$ and
$\Lambda$

\begin{equation}
G_{\mu \nu}+\Lambda(t) g_{\mu \nu}=8 \pi G(t)T_{\mu \nu}
\end{equation}

take the forms [Abdel-Rahman, 1990]:

\begin{equation}
H^{2}+\frac{k}{a^{2}}=\frac{8 \pi G}{3}\rho+\frac{\Lambda}{3}
\end{equation}

\begin{equation}
2\frac{\ddot{a}}{a}+H^{2}+\frac{k}{a^{2}}=-8 \pi G p +\Lambda
\end{equation}

Elimination of $\ddot{a}$ and $k$ gives,

\begin{equation}
\dot{\rho}+3 H (\rho+p)+\frac{\dot{\Lambda}}{8 \pi
G}+\frac{\dot{G}}{G}\rho=0
\end{equation}

Assuming that the usual conservation law holds we can split this
equation as,

\begin{equation}
\dot{\Lambda}+8 \pi \dot{G} \rho=0
\end{equation}

and equation (1.20).\\

Equation (1.35) represents a coupling between vacuum and gravity
[Arbab, 2001]. This also shows that gravitation interaction
increases as $\Lambda$ decreases and thus causing the accelerated
expansion of the Universe, i.e., while $\Lambda$ decreases with
time, the gravitational constant is found to increase with time,
causing the Universe to have accelerated expansion in order to
overcome the increasing gravity [Arbab, 1999]. Arbab has commented
in his paper that this big gravitational force could have been the
reason to stop inflation and thereafter creation of matter in the
early Universe by forcing the small particles to form big ones.
Unlike the Dirac model, this model guarantees the present
isotropic Universe as the anisotropy decreases as $G$ increases
with time. In a closed Universe with variable $\Lambda$ and $G$,
Abdel-Rahman has shown that $a \propto t$, $G \propto t^{2}$ and
$\rho \propto t^{-4}$ in the radiation dominated era. Again
Kalligas etal have obtained a static solution with variable $G$
and $\Lambda$. Berman, Abdel-Rahman and Arbab have independently
obtained the solution $a \propto t$ in both matter and radiation
dominated era. Berman and Arbab have independently remarked that
$\Lambda \propto t^{-2}$ plays an important role in evolution of
the Universe. Varying $G$ theories have also been studied in the
context of induced gravity model where $G$ is generated by means
of a non-vanishing vacuum expectation value of a scalar field
[Zee, 1979; Smolin, 1979; Adler, 1980; Vishwakarma, 2008]. Another
theory uses a renormalization group flow near an infrared
attractive fixed point [Bonanno and Reuter, 2002; Vishwakarma,
2008]. Mostly the varying $G$ models are consistent with redshift
SNe Ia observations. Some of these models solve the horizon and
flatness problem also without any unnatural fine tuning of the
parameters [Bonanno and Reuter, 2002].\\

\subsection{Quintessence Scalar Field}

The fine tuning problem facing dark energy models with a constant
equation of state can be avoided if the equation of state is
assumed to be time dependent. An important class of models having
this property is scalar quintessence field proposed by Wetterich
[1988] and Ratra and Peebles [1988] which slowly rolls down its
potential such that the potential term dominates over the kinetic
term and thus generates sufficient negative pressure to drive the
acceleration. This Q-field couple minimally to gravity so that the
action for this field is given by

\begin{equation}
S=\int d^{4}x \sqrt{-g}\left(-\frac{1}{2}g^{\mu \nu} \partial
_{\mu} \phi \partial _{\nu} \phi -V(\phi)\right)
\end{equation}

where, $V(\phi)$ is the potential energy and the field $\phi$ is
assumed to be spatially homogeneous. Thus the energy-momentum
tensor is given by, [Copeland etal,2006],

\begin{equation}
T_{\mu \nu}=\partial _{\mu} \phi \partial _{\nu} \phi-g_{\mu
\nu}\left(\frac{1}{2}g^{\alpha \beta} \partial _{\alpha} \phi
\partial _{\beta} \phi+V(\phi) \right)
\end{equation}

For a scalar field $\phi$, with Lagrangian density $\cal
L=\frac{1}{2} \partial^{\mu} \phi \partial_{\mu} \phi-V(\phi)$ in
the background of flat FRW Universe, we have the pressure and
energy density are respectively,

\begin{equation}
p=-T^{\mu}_{\mu}=\frac{1}{2}{\dot{\phi}}^{2}-V(\phi)
\end{equation}

\begin{equation}
\rho=T^{0}_{0}=\frac{1}{2}{\dot{\phi}}^{2}+V(\phi)
\end{equation}

Hence the equation of state (EOS) parameter is,

\begin{equation}
\omega=\frac{p}{\rho}=\frac{\frac{1}{2}{\dot{\phi}}^{2}-V(\phi)}{\frac{1}{2}\dot{\phi}^{2}+V(\phi)}
\end{equation}

where an overdot means derivative with respect to cosmic time and
prime denotes differentiation w.r.t. $\phi$. Thus if
$\dot{\phi}^{2}<<V(\phi)$, that is, Q-field varies very slowly
with time, it behaves as a cosmological constant,
as $\omega\approx -1$, with $\rho_{VAC}\simeq V[\phi(t)]$.\\
Now the equation of motion for the quintessence field is given by
the Klein-Gordon equation,

\begin{equation}
\ddot{\phi}+3H\dot{\phi}+\frac{dV(\phi)}{d\phi}=0
\end{equation}

From equation (1.40) we see that $\omega$ can take any value
between $-1$ (rolling very slowly) and $+1$ (evolving very
rapidly) and varies with time [Frieman etal, 2008]. Also,
$\omega<- 1/3$ if ${\dot{\phi}}^{2} < V(\phi)$ and $\omega<- 1/2$
if
${\dot{\phi}}^{2} < \frac{2}{3}V(\phi)$.\\

Various quintessence potentials have been introduced in order to
explain recent cosmic acceleration. {\it Inverse Power Law
Potential}, given by $V(\phi)=V_{0} \phi^{-\alpha}$ is one of the
simplest of the lot [Ratra etal, 1988] . {\it Exponential
Potential} [Wetterich, 1995] given by, $V (\phi) = V_{0}~{\text
exp}(-\frac{ \sqrt{ 8 \pi}\alpha \phi}{M_{P}})$, where $M_{P}$ is
the reduced Planck mass, is one of the most favoured one among
cosmologists. But in this case
$\frac{\rho_{\phi}}{\rho_{total}}<0.02$, suggesting that
exponential potential cannot make the transition from subdominant
to dominant energy density component of the universe in late
times. Sahni and Wang proposed a model in 2000 as $V (\phi) =
V_{0}({\text cosh}~ \alpha \phi - 1)^{p}$. This model interpolates
from $V \propto ~{\text exp}(p \alpha \phi)$ to $V \propto (\alpha
\phi)^{2p}$, thereby preserving some of the properties of the
simpler exponential potential but allowing a different late time
behavior. This potential describes quintessence for $p \leq 1/2$
and pressureless CDM for $p = 1$. Thus the cosine hyperbolic
potential is able to describe both dark matter and dark energy
within a tracker framework [Sahni etal, 2000]. Sahni [2004] has
presented a few quintessence potentials proposed in literature in
a tabular form as below:\\

{\bf Table II} \underline{}

\begin{center}
\begin{tabular}{lcccccc}   \hline
Quintessence Potential~~~~~~~~~Reference\\
\hline\\
$V_{0}$ exp $(-\lambda \phi)$~~~~~~~~~~Ratra and Peebles (1988),
Wetterich (1988), Ferreira and Joyce (1998)\\
\\
$m^{2} {\phi}^{2}$, $\lambda \phi^{4}$~~~~~~~~~~~~~~~Frieman etal (1995)\\
\\
$V_{0}/\phi^{\alpha}$, $\alpha>0$~~~~~~~~~~~Ratra and Peebles (1988)\\
\\
$V_{0}$ exp $(\lambda {\phi}^{2})$~~~~~~~~~~~~Brax and Martin (1999, 2000)\\
\\
$V_{0}$(cosh $\alpha \phi - 1)^{p}$~~~~~~Sahni and Wang (2000)\\
\\
$V_{0}$ sinh $^{-\alpha}(\lambda \phi)$~~~~~~~~~Sahni and Starobinsky (2000), Ure$\tilde{n}$a-L$\acute{o}$pez and Matos (2000)\\
\\
$V_{0}(e^{\alpha \kappa \phi}+e^{\beta \kappa \phi})$~~~~~~~~~Barreiro, Copeland and Nunes (2000)\\
\\
$V_{0}(exp M_{P}/\phi-1)$~~~~~~Zlatev, Wang and Steinhardt (1999)\\
\\
$V_{0}[(\phi-B)^{\alpha}+A]e^{-\lambda \phi}$~Albrecht and Skordis (2000)\\

\hline\\

\end{tabular}
\end{center}

In order to obtain a feasible dark energy model, the energy
density of the scalar field should be subdominant during the
radiation and matter dominating eras, emerging only at late times
to give rise to the current observed acceleration of the universe
[Copeland etal, 2006]. Therefore, we introduce a barotropic fluid
in the background, with EOS given by,
$\omega_{m}=\frac{p_{m}}{\rho_{m}}$, where $p_{m}$ and $\rho_{m}$
are the pressure and energy density of the barotropic fluid
respectively.\\

A homogeneous and isotropic universe is characterized by the
Friedmann-Robertson-Walker (FRW) line element is given by equation
(1.7) with $c=1$. Thus, the over all stress-energy tensor of the
scalar field $\phi$ in presence of barotropic fluid the has the
form,

\begin{equation}
T_{\mu\nu}=(\rho+p)u_{\mu}u_{\nu}+pg_{\mu\nu},~~~~u_{\mu}u^{\mu}=-1
\end{equation}

where $\rho=\rho_{m}+\rho_{\phi}$ and $p=p_{m}+p_{\phi}$. Here
$\rho_{\phi}$ and $p_{\phi}$ are the energy density and pressure
of the Q-field given by equations (1.39) and (1.38) respectively.
If $\omega_{m}$ is assumed to be a constant, the fluid energy will
be given by $\rho_{m}= \rho_{0} a^{-3(1+\omega_{m})}$ and
$\omega_{\phi}$
dynamically changes in general.\\

The Friedmann equations together with the energy conservation of
the normal matter fluid and quintessence (Klein-Gordon equation)
are,

\begin{equation}
H^{2}+\frac{k}{a^{2}}=\frac{1}{3}(\rho_{m}+\rho_{\phi})
\end{equation}

\vspace{-5mm}

\begin{equation}
\dot{\rho}_{m}+3H\gamma_{m}\rho_{m}=0
\end{equation}

together with equation (1.41).\\

where $H\equiv \frac{\dot{a}}{a}$ denotes the Hubble factor.
Introducing $\Omega_{m}\equiv \frac{\rho_{m}}{\rho_{c}},~
\Omega_{\phi}\equiv
\frac{\rho_{\phi}}{\rho_{c}},~\Omega_{k}=-\frac{k}{(aH)^{2}}$ and
$\Omega=\Omega_{m}+\Omega_{\psi}$ with $\rho_{_{c}}\equiv 3H^{2}$
as the critical density, Ellis et al [1997] showed that the matter
violates the strong energy condition $\rho+3p\le 0$ and so the
deceleration parameter $q=-\frac{a\ddot{a}}{\dot{a}^{2}}<0$. Hence
as a consequence the universe accelerates its expansion.\\

In the investigation of cosmological scenarios we are also
interested about those solutions in which the energy density of
the scalar field mimics the background fluid energy density,
called {\it scaling solutions} [Copeland etal, 1998]. Exponential
potentials have been proved to give rise to scaling solutions and
so can play an important role in quintessence scenarios, allowing
the field energy density to mimic the background being
sub-dominant during radiation and matter dominating eras. In this
case, as long as the scaling solution is the attractor, then for
any generic initial conditions, the field would sooner or later
enter the scaling regime, thereby opening up a new line of attack
on the fine tuning problem of dark
energy [Copeland etal, 2006].\\

Going on with the analysis, very interesting question is whether
it is possible to construct a successful common scheme for the two
cosmological mechanisms involving rolling scalar fields i.e.,
quintessence and inflation. This perspective has the appealing
feature of providing a unified view of the past and recent history
of the universe, but can also remove some weak points of the two
mechanisms when considered separately. In general, scalar fields
tend to be of heavy (high energy). When renormalized, scalar
fields tend to become even heavier. This is acceptable for
inflation, because Universe was in a very high energy state at
that epoch, but it seems highly unnatural for the recent Universe
[Bennet, 2006]. Inflation could provide the initial conditions for
quintessence without any need to fix them by hand and quintessence
could hope to give some more hints in
constraining the inflation potential on observational grounds.\\

From theoretical point of view a lot of works [Caldwell et al,
1998; Ostriker et al, 1995; Peebles, 1984; Wang et al, 2000;
Perlmutter et al, 1999; Dodelson et al, 2000; Faraoni, 2000] have
been done to solve the quintessence problem. Scalar fields
[Peebles et al, 1988, 2002; Ratra et al, 1988; Ott, 2001; Hwang et
al, 2001; Ferreira et al, 1998] with a potential giving rise to a
negative pressure at the present epoch, a dissipative fluid with
an effective negative stress [Cimento et al, 2000] and more exotic
matter like a frustrated network of non-abelian cosmic strings or
domain wall [Bucher et al, 1999; Battye et al, 1999], scalar
fields with non-linear kinetic term, dubbed K-essence model
[Armendariz-Picon etal, 2001], are plausible candidates of
Q-matter. Also, there exist models of quintessence in which the
quintessence field is coupled to dark matter and/or baryons
[Amendola, 2000].\\

Scalar fields, although, being very popular in theory, they have
several shortcomings, as they need some suitable potential to
explain the accelerated expansion, they need to assume
cosmological constant to be zero [Padmanabhan, 2006]. Also, most
of the fields (Q-matter field, tracker field) work only for a
spatially flat ($k=0$) FRW model. Recently, Cimento et al [2000]
showed that a combination of dissipative effects such as a bulk
viscous stress and a quintessence scalar field gives an
accelerated expansion for an open universe ($k=-1$) as well. This
model also provides a solution for the `coincidence problem' as
the ratio of the density parameters corresponding to the normal
matter and the quintessence field asymptotically approaches a
constant value. Bertolami and Martins [2000] obtained an
accelerated expansion for the universe in a modified Brans-Dicke
(BD) theory by introducing a potential which is a function of the
Brans-Dicke scalar field itself. Banerjee and Pavon [2001] have
shown that it is possible to have an accelerated universe with BD
theory in Friedmann model without any matter.\\

\subsection{Chaplygin Gas}

In recent years a lot of other models, other than cosmological
constant and quintessence scalar fields, have also been proved to
provide plausible explanation for dark energy. One of the most
popular among these models is Chaplygin gas. Chaplygin Gas
unifies dark matter and dark energy under same EOS given by,

\begin{equation}
p=-A/\rho
\end{equation}

where $A$ is a positive constant.\\

Chaplygin [1904] introduced this EOS to calculate the lifting
force on a wing of an airplane in aerodynamics. Later Tsien [1939]
and Karman [1941] proposed the same model. Also Stanyukovich
[1960] showed that the same EOS can describe certain deformable
solids. Jackiw [2000] showed that this is the only gas to admit a
supersymmetric generalization. This invokes interest in string
theory as well since in a D-brane configuration in the D+2
Nambu-Goto action, the employment of the light-cone
parameterization leads to the action of a newtonian fluid with the
EOS (1.45), whose symmetries are the same as the
Poincar$\acute{e}$ group [Colistete Jr., 2002] and also is linked
with Born-Infeld model as both have the same D-brane ancestor
[Jackiw, 2000] [the parametrization invariant Nambu-Goto D-brane
action in a (D + 1, 1) spacetime leads, in the light-cone
parametrization, to the Galileo-invariant Chaplygin gas in a (D,
1) spacetime and to the Poincar$\acute{e}$-invariant Born-Infeld
action in a (D, 1) spacetime]. Thus there exists a mapping between
these two systems and their solutions. Thus the Born-Infeld
Lagrangian density

\begin{equation}
{\cal L_{BI}}=-\sqrt{A} \sqrt{1-g^{\mu \nu}\theta,_{\mu}
\theta,_{\nu}}
\end{equation}

gives rise to the EOS (1.45).\\

As seen before, the metric of a homogeneous and isotropic universe
in FRW model (without the $\Lambda$-term) is given by equation
(1.7). The Einstein field equations are (choosing $8\pi G=c=1$)
given by equations (1.12) and (1.13). The energy conservation
equation ($T_{\mu;\nu}^{\nu}=0$) is given
by equation (1.20).\\

The EOS (1.45) together with these equations give,

\begin{equation}
\rho=\sqrt{A+\frac{B}{a^{6}}}
\end{equation}

where $B$ is an arbitrary integration constant.\\

Thus for small values of $a$, $\rho\sim \frac{\sqrt{B}}{a^{3}}$
and $p\sim -\frac{A}{\sqrt{B}}a^{3}.$ which implies a dust like
matter for small values of $a$. Also for large values of $a$,
$\rho \sim \sqrt{A}$ and $p\sim -\sqrt{A}$, which implies an empty
Universe with cosmological constant $A$, i.e., the
$\Lambda$CDM model.\\

Kamenshchik etal [2001] showed that Chaplygin gas cosmology can
interpolate between different phases of the Universe, starting
from dust dominated phase to a de-Sitter Universe passing through
an intermediate phase which is a mixture of a cosmological
constant and a stiff matter (given by the EOS $p=\rho$) and thus
can explain the evolution of the Universe from a decelerated phase
to a stage of cosmic acceleration for a flat or open Universe,
i.e., for $k=0$ or $k=-1$. For a closed Universe with $k=1$ they
obtained a static Einstein Universe with $B=\frac{2}{3
\sqrt{3}A}$.\\

Now to find a homogeneous scalar field $\phi(t)$ and a
self-interacting potential $V(\phi)$ corresponding to the
Chaplygin gas cosmology, we consider the Lagrangian of the scalar
field as,

\begin{equation}
{\cal L_{\phi}}=\frac{1}{2}\dot{\phi}^{2}-V(\phi)
\end{equation}

The analogous energy density $\rho_{\phi}$ and pressure
$p_{\phi}$ for the scalar field then read,

\begin{equation}
\rho_{\phi}=\frac{1}{2}\dot{\phi}^{2}+V(\phi)=\rho=\sqrt{A+\frac{B}{a^{6}}}
\end{equation}

and

\begin{equation}
p_{\phi}=\frac{1}{2}\dot{\phi}^{2}-V(\phi)=-A/\rho=-\frac{A}{\sqrt{A+\frac{B}{a^{6}}}}
\end{equation}

Hence for flat universe (i.e., $k=0$) we have

\begin{equation}
\dot{\phi}^{2}=\frac{B}{a^{6}\sqrt{A+\frac{B}{a^{6}}}}
\end{equation}

and

\begin{equation}
V(\phi)=\frac{1}{2} \sqrt{A}\left({\text Cosh}~ 3\phi+\frac{1}{
{\text Cosh}~ 3\phi} \right)
\end{equation}

Fabris etal [2001] showed the density perturbations to this model,
but, their unperturbed Newtonian equations cannot reproduce the
energy density solution of the Chaplygin gas cosmology given by
equation (1.47) due to choice of lightcone gauge, whereas, Bilic
etal [2002] extending this model to large per-turbations by
formulating a Zeldovich-like approximation showed that
inhomogeneous Chaplygin
gas can combine the roles of dark energy and dark matter.\\

Later Bento etal [2002] generalized the EOS (1.45) to,

\begin{equation}
p=-A/\rho^{\alpha}
\end{equation}

with $0\le\alpha\leq 1$, $A>0$ and obtained generalized Chaplygin
gas. It can be seen that for $\alpha=1$, the above EOS reduces to
the pure Chaplygin gas with EOS (1.45).\\

This is obtained from the generalized Born-Infeld Lagrangian
density given by,

\begin{equation}
{\cal L_{GBI}}=-A^{\frac{1}{1+\alpha}}\left[1-(g^{\mu \nu}
\theta,_{\mu} \theta,_{\nu})^{\frac{1+\alpha}{2
\alpha}}\right]^{\frac{\alpha}{1+\alpha}}
\end{equation}

This EOS together with the Einstein equations (1.12) and (1.13)
and the conservation law (1.20), gives

\begin{equation}
\rho=\left(A+\frac{B}{a^{3(1+\alpha)}}\right)^{\frac{1}{1+\alpha}}
\end{equation}

where $B$ is an arbitrary positive integration constant.\\

Again for small values of $a$, $\rho\sim
\frac{B^{\frac{1}{1+\alpha}}}{a^{3}}$ and $p\sim
-\frac{A}{B^{\alpha/(1+\alpha))}}a^{3 \alpha}.$ which implies a
dust like matter for small values of $a$. Also for large values
of $a$, $\rho \sim A^{1/(1+\alpha)}$ and $p \sim
-A^{1/(1+\alpha)}$, which implies an empty Universe with
cosmological constant $A^{1/(1+\alpha)}$, i.e., the
$\Lambda$CDM model.\\

Bento etal [2002] showed that Generalized Chaplygin gas (GCG)
cosmology can also explain the evolution of the Universe from dust
dominated phase to a de-Sitter Universe passing through an
intermediate phase which is a mixture of a cosmological
constant and a soft EOS (given by the EOS $p= \alpha\rho$).\\

Now we look at the field theoretic approach of this model.
Considering the Lagrangian of the scalar field $\phi$ with
potential $V(\phi)$, the corresponding energy density and pressure
will be given by,

\begin{equation}
\rho_{\phi}=\frac{1}{2}\dot{\phi}^{2}+V(\phi)=\rho=\left[A+\frac{B}{a^{3(1+\alpha)}}
\right]^{\frac{1}{1+\alpha}}
\end{equation}

and

\begin{equation}
p_{\phi}=\frac{1}{2}\dot{\phi}^{2}-V(\phi)=-A
\left[A+\frac{B}{a^{3(1+\alpha)}}
\right]^{-\frac{\alpha}{1+\alpha}}
\end{equation}

Thus for flat Universe ($k=0$), the scalar field and the
potential will be given by,

\begin{equation}
\phi=\frac{2}{\sqrt{3}(1+\alpha)}~{\text
Sinh}^{-1}\left\{\sqrt{\frac{B}{A}}\frac{1}{a^{\frac{3}{2}(1+\alpha)}}
\right\}
\end{equation}

and

\begin{equation}
V(\phi)=\frac{1}{2}A^{\frac{1}{1+\alpha}}{\text
Cosh}^{\frac{2}{1+\alpha}}
\left\{\frac{\sqrt{3}(1+\alpha)}{2}\phi\right\}+ \frac{1}{2}
A^{\frac{1}{1+\alpha}}{\text Cosh}^{-\frac{2\alpha}{1+\alpha}}
\left\{\frac{\sqrt{3}(1+\alpha)}{2}\phi\right\}
\end{equation}

which reduces to that corresponding to pure Chaplygin gas model
if $\alpha=1$.\\

Introducing inhomogeneities, Bento etal [2002] have shown that the
model evolves consistently with the observations (specially, CMBR
peaks, such as Archeops for the location of the first peak and
BOOMERANG for the location of the third peak, supernova and
high-redshift observations [Bento etal, 2003]) and that the
density contrast in this model is less than the CDM model and even
gives a better approximation of the $\Lambda$CDM  model compared
to the pure Chaplygin gas model. Later Makler etal [2003] showed
that GCG is consistent with SNIa data for any value of $0\leq
\alpha\le 1$, although for $\alpha \sim 0.4$, the case is most
favoured. They also examined that the presence of Cosmological
constant should rule out the pure Chaplygin gas as per SNAP data,
whereas, presence of pure Chaplygin gas should rule out the
possibility of cosmological constant in the
Universe.\\

Later Benaoum [2002] further modified this model and proposed to
modified Chaplygin gas (MCG) with EOS,

\begin{equation}
p=A\rho-\frac{B}{\rho^{\alpha}} ~~~~\text{with}~~~~ 0\le \alpha
\le 1
\end{equation}

With this EOS the energy density takes a much generalized form,

\begin{equation}
\rho=\left[\frac{B}{1+A}+\frac{C}{a^{3(1+A)(1+\alpha)}}
\right]^{\frac{1}{1+\alpha}}
\end{equation}

where $C$ is an arbitrary integration constant.\\
Substituting $A=0$ and $B=A$ we get back the results of GCG.\\

Debnath etal [2004] have shown that in this model for small value
of scale factor Universe will decelerate and for large values of
scale factor Universe will accelerate and the transition occurs
when
$a=\left[\frac{C(1+A)(1+3A)}{2B}\right]^{\frac{1}{3(1+\alpha)(1+A)}}$.
They have also examined that this model can describe the evolution
of the Universe from radiation era ($A=\frac{1}{3}$ and $\rho$ is
very large) to $\Lambda$CDM ($\rho$ is very small constant) model
and thus can explain the evolution of the Universe
to a larger extent than the pure Chaplygin Gas or GCG models.\\

Considering a scalar field $\phi$ with self-interacting potential
$V(\phi)$, the corresponding energy density and pressure will be,

\begin{equation}
\rho_{\phi}=\frac{1}{2}\dot{\phi}^{2}+V(\phi)=\rho=\left[\frac{B}{1+A}+\frac{C}{a^{3(1+A)(1+\alpha)}}
\right]^{\frac{1}{1+\alpha}}
\end{equation}

and

\begin{equation}
p_{\phi}=\frac{1}{2}\dot{\phi}^{2}-V(\phi)=A\rho-\frac{B}{\rho^{\alpha}}
=A \left[\frac{B}{1+A}+\frac{C}{a^{3(1+A)(1+\alpha)}}
\right]^{\frac{1}{1+\alpha}}-B
\left[\frac{B}{1+A}+\frac{C}{a^{3(1+A)(1+\alpha)}}
\right]^{-\frac{\alpha}{1+\alpha}}
\end{equation}
\\
Hence for flat Universe, the scalar field and the potential will
be given by,

\begin{equation}
\phi=\frac{2}{\sqrt{3}(1+\alpha)(1+A)}~Sinh^{-1}\left\{\sqrt{\frac{C(1+A)}{B}}\frac{1}{a^{\frac{3}{2}(1+\alpha)(1+A)}}
\right\}
\end{equation}

and

\begin{eqnarray*}
V(\phi)=\frac{1}{2}(1-A)\left(\frac{B}{1+A}\right)^{\frac{1}{1+\alpha}}Cosh^{\frac{2}{1+\alpha}}
\left\{\frac{\sqrt{3}\sqrt{1+A}(1+\alpha)}{2}\phi\right\}
\end{eqnarray*}

\begin{equation}
+\frac{1}{2}B\left(\frac{B}{1+A}\right)^{-\frac{\alpha}{1+\alpha}}Cosh^{-\frac{2\alpha}{1+\alpha}}
\left\{\frac{\sqrt{3}\sqrt{1+A}(1+\alpha)}{2}\phi\right\}
\end{equation}

For small values of the scale factor $a(t)$ and large values of
$\rho$, Debnath etal [2004] obtained two different qualitative
nature of the potential for $A=1$ and $A\neq 1$. For $A=1$,
$V(\phi)\rightarrow 0$ as $\rho \rightarrow \infty$ and for $A\neq
1$, $V(\phi)\rightarrow \infty$ as $\rho \rightarrow \infty$. For
large values of the scale factor $V(\phi)\rightarrow
\left(\frac{B}{1+A}\right)^{\frac{1}{1+\alpha}}$ for both the cases.\\

Chimento and Lazkoz [2005] studied the large-scale perturbations
in this model using a Zeldovich-like approximation  and showed
that this model evolve from a phase dominated by non-relativistic
matter to an asymptotically  de Sitter phase and that the
intermediate regime is described by the combination of dust and a
cosmological constant. They also showed that the inhomogeneities
introduced in this model evolve consistently with the observations
and the density contrast is less than the CDM model and are more
similar to $\Lambda$CDM or GCG model. Dao-jun Liu and and Xin-zhou
Li [2005] investigated this model using the location of the peak
of CMBR spectrum and obtained the range for a non-zero
$A$ to be, $-0.35\lesssim A \lesssim 0.025$.\\

Recently, Jianbo Lu etal [2008] have shown that according to
Akaike Information Criterion (AIC) of model selection, recent
observational data support the MCG model as well as other popular
models.\\

\subsection{Tachyonic Field}

Recently rolling tachyon condensate has been studied as a source
of dark energy. This is a scalar field of non-standard form
motivated by string theory as the negative-mass mode of the open
string perturbative spectrum [Calacagni etal, 2006]. For strings
attached to Dirichlet D-branes such tachyonic modes reflect
instability [Das etal, 2005]. The basic idea is that the usual
open string vacuum is unstable but there exist a stable vacuum
with zero energy density [Gibbons, 2002]. The unstable vacuum
corresponds to rolling tachyon. Sen showed that this tachyonic
state is associated with the condensation of electric flux tubes
of closed strings described by Dirac-Born-Infeld action. For
strings attached to Dirichlet  D-branes such tachyonic modes
reflect D-brane instability [Das etal, 2005]. Tachyonic field has
a potential which rolls down from an unstable maximum to a stable
minimum with a stable vacuum expectation value. This is known as
tachyon condensate [Das etal, 2005]. Sen [2002] has shown that the
energy momentum tensor for rolling tachyon solution in D-branes in
bosonic string theory is described by a pressureless gas with
non-zero energy density, which is stored in open string field,
although there are no open string degrees of freedom around the
minimum of tachyonic potential. Thus it represents dust, which can
be considered as a candidate for CDM. Also the energy-momentum
tensor of tachyon condensate can be split into two parts, one with
$\omega=0$ and the other with $\omega=-1$. This has led a lot of
authors to construct cosmological models with tachyonic field as a
candidate for dark energy, as dark matter and dark energy thus can
be described by a single scalar field. Hence in cosmology rolling
of tachyon is analogous to the expansion of the Universe [Gibbons,
2002]. Also Sen [2002] showed that the supersymmetry breaking by
tachyon matter can be adjusted since the total energy of the
tachyon matter is adjustable and is
determined by the initial position and velocity of tachyon.\\

Now let us move to the dynamics of the tachyon condensate. The
Lagrangian density of tachyon condensate is given by the
Born-Infeld action

\begin{equation}
\cal L_{tach}=-V(T) \sqrt{1+g^{\mu \nu}\partial_{\mu} T
\partial_{\nu} T}\sqrt{-\text{det}(g_{\mu \nu})}=-V(T)\sqrt{-\text{det}(G_{\mu \nu})}
\end{equation}

where $T$ is the tachyonic field, $V(T)$ is the tachyon potential
having a local maximum at the origin and a global minimum at $T
=\infty$ [Gibbons, 2003] where
the potential vanishes.\\

The tachyon metric is thus given by,

\begin{equation}
G_{\mu \nu}=g_{\mu \nu}+\partial_{\mu} T
\partial_{\nu} T
\end{equation}

(Another tachyon model has been proposed with Lagrangian
$V(T)\sqrt{g^{\mu \nu}\partial_{\mu} T \partial_{\mu} T-1}$,
which has been proved to be more effective as to explore more
physical situations than quintessence [Chimento, 2003;
Srivastava, 2005].\\

Thus the stress tensor of the tachyonic field is given in the
form of a perfect fluid by,

\begin{equation}
T^{\mu}_{\nu}=(\rho+p)u^{\mu} u_{\nu}-p \delta^{\mu}_{\nu}
\end{equation}

where, $u_{\mu}=\frac{\partial_{\mu} T}{\sqrt{\partial^{\nu}T
\partial_{\nu}T}}$, hence

\begin{equation}
\rho=\frac{V(T)}{\sqrt{1-\partial^{\nu}T
\partial_{\nu}T}}
\end{equation}

and

\begin{equation}
p=-V(T)\sqrt{1-\partial^{\nu}T
\partial_{\nu}T}
\end{equation}

which for a homogeneous and time dependent tachyonic field reduce
to

\begin{equation}
\rho=\frac{V(T)}{\sqrt{1-\dot{T}^{2}}}
\end{equation}

and

\begin{equation}
p=-V(T){\sqrt{1-\dot{T}^{2}}}
\end{equation}

Hence,

\begin{equation}
p=-\frac{V^{2}(T)}{\rho}
\end{equation}

and the EOS parameter reads,

\begin{equation}
\omega=\frac{p}{\rho}=-(1-\dot{T}^{2})
\end{equation}

Thus $-1\leq \omega \leq 0$. Also if $V(T)$ is constant, equation
(1.73) reduces to the EOS of pure Chaplygin gas.\\

As for the strong energy condition
$\rho+3p=\frac{V(T)}{\sqrt{1-\dot{T}^{2}}}(3 \dot{T}^{2}-2)>0$,
i.e. SEC fails if $\mid \dot{T}\mid < \sqrt{\frac{2}{3}}$. As
seen from the above equations $\mid \dot{T}\mid <1$.\\

The equation of motion is

\begin{equation}
\left(g^{\mu \nu}- \frac{\partial^{\mu}T
\partial\partial^{\nu}T}{1+(\partial T)^{2}}\right) \partial_{\mu}
\partial_{\nu} T=-\frac{V'}{V}\left( 1+(\partial T)^{2}\right)
\end{equation}

Considering the gravitational field generated by tachyon
condensate and assuming that the cosmological constant term
vanishes in the tachyon ground state the action becomes [Gibbons,
2003],

\begin{equation}
S=\int d^{4}x \left[ \frac{R}{16 \pi G}\sqrt{- \text{det} (g_{\mu
\nu })} -V(T) \sqrt{-\text{det}(G_{\mu \nu})}\right]
\end{equation}

Using this the Raychaudhuri and Friedmann equations give,

\begin{equation}
\frac{\ddot{a}}{a}=\frac{8 \pi
G}{3}\left[\frac{V(T)}{\sqrt{1-\dot{T}^{2}}}-\frac{3}{2}\frac{V(T)
\dot{T}^{2} }{\sqrt{1-\dot{T}^{2}}} \right]
\end{equation}

and

\begin{equation}
\left(\frac{\dot{a}}{a}\right)^{2}=-\frac{k}{a^{2}}+\frac{8 \pi G
}{3}\frac{V(T)}{\sqrt{1-\dot{T}^{2}}}
\end{equation}

The equation of motion reads,

\begin{equation}
\ddot{T}=-(1-\dot{T}^{2})\left[\frac{V'(T)}{V(T)}+3\dot{T}\frac{\dot{a}}{a}\right]
\end{equation}

Also the conservation equation becomes,

\begin{equation}
\dot{\rho}+ 3 H \rho \dot{T}^{2}=0
\end{equation}

These equations show that tachyon field rolls down hill with an
accelerated motion and the universe expands [Gibbons, 2002].
Raychaudhuri equation shows that initially for small $T$, i. e.,
when $\mid T \mid <\sqrt{\frac{2}{3}}$, $\ddot{a}>0$, i.e., the
Universe accelerates and starts decelerating once $\mid T \mid
>\sqrt{\frac{2}{3}}$. For flat space-time, i. e., for $k=0$, $a(t)$
approaches a constant value, for $k = -1$, $a \rightarrow t$ and
for $k = 1$ re-collapse will take place [Gibbons, 2002]. \\

Tachyonic field can be treated as dark energy or dark matter
depending on the form of the potential associated with it. Tachyon
potential is usually assumed to be exponentially decaying or
inversely quadratic. However, Copeland etal [2005] and Calacagni
etal [2006] have carried out the analysis for a wide range of
potentials, as given below, although more or less all these models
face the problem of fine tuning or are constrained by
observational data.\\

1) $V=V_{0} T^{-n}$: For $n<0$, the model shows instability; for
$0<n<2$, there is s stable late time attractor solution and
$V_{0}$ does not need to be fine tuned as super-Planckian problem
does not affect this model; for $n=2$, one gets a power law
solution of the form $a=t^{m}$ and $V_{0}$ needs to be fine tuned
in order to get present day acceleration, and hence is not a good
candidate for dark energy; for $n>2$, the model has a dust
attractor.\\

2) $V=V_{0} e^{1/ \alpha T}, ~\alpha>0$: This model gives an
asymptotic de-Sitter solution with $V_{0}$ representing the
effective cosmological constant.\\

3) $V=V_{0} e^{\alpha^{2} T^{2}},~ \alpha>0$: This model gives an
oscillating field around the origin with $V_{0}$ being the
effective cosmological constant.\\

4) $V=V_{0} e^{- \alpha T}, ~\alpha>0$: This model has a stable
dust attractor after a period of acceleration. Also this is
large-field approximation of $V=V_{0}/\text{cosh}(\alpha T)$.

5) $V=V_{0} e^{-\alpha^{2} T^{2}},~ \alpha>0$: This model is
similar to model 4.\\

Bagla etal [2003] studied the effects of homogeneous tachyon
matter in the background of non-relativistic matter and radiation,
choosing the inverse square potential and the exponential
potential for the tachyonic field and showed that for both these
models the density parameter for matter and the tachyons are
comparable even in the matter dominated phase. For the exponential
potential, we get a phase where $a\propto t^\frac{2}{3}$ as
$t\rightarrow \infty$ preceded by an accelerated expansion and
thus eliminating the event horizon present in $\Lambda$CDM model.
They also carried a supernova Ia data analysis and showed that
both the potentials present models where the Universe accelerates
at low redshifts and are also consistent
with requirements of structure formation.\\

To study the tachyon driven cosmology Sen [2003] has considered
the effective action, taking into account the cosmological
constant term, to be,

\begin{equation}
S=-\frac{1}{16 \pi G}\int d^{4}x \left[-\sqrt{- \text{det} (g)}~
R+V(T) \sqrt{-\text{det} (g_{\mu \nu}+\partial_{\mu} T
\partial_{\nu} T)}+\Lambda \sqrt{-\text{det} (g)} \right]
\end{equation}

with $V(T)=V_{0}/\text{cosh} (T/\sqrt{2})$ and $T=T(x^{0})$.\\

Implementing the FRW line element the Einstein's field equations
and the equations of motion of $T$ are,

\begin{equation}
\frac{\ddot{a}}{a}=\frac{8 \pi G}{3}\left[\Lambda
+\frac{V(T)}{\sqrt{1-\dot{T}^{2}}}-\frac{3}{2}\frac{V(T)
\dot{T}^{2} }{\sqrt{1-\dot{T}^{2}}} \right]
\end{equation}

\begin{equation}
\left(\frac{\dot{a}}{a}\right)^{2}=-\frac{k}{a^{2}}+\frac{8 \pi G
}{3}\left[ \Lambda +\frac{V(T)}{\sqrt{1-\dot{T}^{2}}}\right]
\end{equation}

and

\begin{equation}
\ddot{T}=-(1-\dot{T}^{2})\left[\frac{V'(T)}{V(T)}+3\dot{T}\frac{\dot{a}}{a}\right]
\end{equation}

with initial conditions

\begin{equation}
\dot{T}=0,~~~~ \dot{a}=0, ~~~~T=T_{0}~~~~ \text {at}~~~~ x^{0}=0
\end{equation}

Sen has used time reversal symmetry and concluded that if
$\Lambda$ is comparatively small rather negligible the universe
begins with a big bang and ends in a big crunch. Whereas, in
presence of a bulk cosmological constant Universe expands without
any singularity for some special range of initial conditions on
the tachyon.\\

Although a lot of shortcomings have been pointed out by many
authors [Linde etal, 2002; Shiu etal, 2003] regarding the fine
tuning of the model, Gibbons [2003] has stressed on the
possibility that the tachyon was important in a pre-inflationary
{\it Open-String Era} preceding our present {\it Closed String Era}.\\

\subsection{Inhomogeneous EOS}

The disturbing features of $\Lambda$CDM model discussed above
motivate the search for alternatives for standard $\Lambda$CDM
model, thus causing imhomogeity to be introduced in the EOS so as
to account for the present day acceleration. A lot of
imhomogeneous models are being studied recently for this purpose.\\

Garfinkle [2006] has shown that an inhomogeneous but spherically
symmetric cosmological model can account for the cosmic
acceleration without any dark energy. His model fits the
supernova data like the standard FRW model with $\Lambda$.\\

Capozzeillo etal [2006] have investigated the effects of viscosity
terms depending on the Hubble parameter and its derivatives in the
dark energy EOS. For this purpose they considered two EOS,
given by,\\
$$ p=-\rho-A \rho^{\alpha}-B H^{2 \beta}$$\\
and\\
$$f(p,~\rho,~H)=0$$\\
where, $A,~B,~\alpha,~\beta$ are constant and $f$ is a function
of its arguments.\\

They present the likelihood analysis to show that both models fit
the data given by SNeIa and radio galaxies and predict values of
the deceleration parameter and the age of the Universe almost
correctly. Nojiri etal [2005] considered a Hubble parameter
dependent EOS to construct a late time Universe with $\omega=-1$
crossing. Stefancic [2005] also considered a class of these
models to investigate the singularities of the Universe.\\

Brevik etal [2004] used Cardy-Verlinde formula in a FRW Universe
filled with dark energy and obtained the same resualts as modified
gravity, which is a gravitational alternative for dark energy.
Elizalde etal [2005] have also considered decaying vacuum
cosmology and holographic dark energy models motivated by vacuum
fluctuations and AdS/CFTlike holographic considerations
respectively and have shown that there is no need to introduce
exotic matter explicitly, as these models violate the basic
energy conditions.\\

\section{Brans-Dicke Cosmology}

Brans-Dicke (BD) theory has been proved to be very effective
regarding the recent study of cosmic acceleration. This theory has
very effectively solved the problems of inflation and the early
and the late time behaviour of the Universe. The starting point of
Brans-Dicke Theory [Brans et al, 1961] is Mach's Principle, that
the phenomenon of inertia ought to arise from acceleration w.r.t.
the general mass distribution of the Universe. Therefore
gravitational acceleration should be used to measure the absolute
scale of the elementary particle masses, as they are not
constants, rather, represents the particles' interaction with some
cosmic field [Weinberg, Gravitation and Cosmology]. Thus the
Gravitational Constant $G$ is related to the mass distribution in
an expanding Universe by the relation $\frac{G M}{R c^{2}}\sim 1$,
where $R$ is the radius of the Universe and $M$ is the mass of the
Universe, or rather,

\begin{equation}
G^{-1}\sim \sum _{i}(m_{i}/r_{i} c^{2})
\end{equation}

where the sum is over all the matter that contribute to the
inertial reaction, since both nearby and distant matter should
contribute to the inertial reaction. Now if $G$ is to vary it
should be a function of some scalar field variable. Thus if $\phi$
represents the scalar field coupled to the mass density of the
Universe, $G$ should be related to $\phi$ in some manner. Since a
wave equation for $\phi$ with a scalar matter density as source
gives an equation same as equation (1.86), a suitable relation
between $G$ and $\phi$ could be given by $\phi \simeq
\frac{1}{G}$. Brans and Dicke proposed a theory in which the
correct field equations for gravitation are obtained by replacing
$G$ with $\frac{1}{\phi}$. Thus Brans-Dicke Theory is a
generalization of the theory of general relativity. Here
gravitation effects are described by a scalar field in Riemannian
manifold, thus expressing the gravitational effects as both
geometrical and due to scalar interactions. They generalize the
usual variational principle of general relativity to obtain
equations of motion of matter and non-gravitational fields using
Einstein field equations, by

\begin{equation}
\delta\int[\phi R+\frac{16 \pi}{c^{4}}L-\omega(\phi,_{i}
\phi^{ij}/\phi)]=0
\end{equation}

where $R$ is the scalar curvature and $L$ is the Lagrangian
density of matter including all non-gravitational fields (and not
of $\phi$) and $\omega$ is a dimensionless constant.\\

Now the conservation laws give,

\begin{equation}
T^{ij}_{;j}=0
\end{equation}

where $T^{ij}$ is the energy momentum tensor of matter (excluding
$\phi$).\\

The wave equation for $\phi$ is given by (varying $\phi$ and
$\phi,_{i}$ in equation (1.87))

\begin{equation}
2 \omega \phi^{-1} \square \phi -
(\omega/\phi^{2})\phi^{,i}\phi,_{i}+R=0
\end{equation}

where the generally covariant D'Alembertian $\square$ is defined
to be the covariant divergence of $\phi^{,i}$,

\begin{equation}
\square\phi=\phi^{,i}_{;i}=(-g)^{-\frac{1}{2}}[(-g)^{-\frac{1}{2}}
\phi^{,i} ],_{i}
\end{equation}

Varying the components of the metric tensor and the first
derivatives in equation (1.87) the field equations for the metric
field are obtained as,

\begin{equation}
R_{ij}-\frac{1}{2}g_{ij} R=(8 \pi
\phi^{-1}/c^{4})T_{ij}+(\omega/\phi^{2})(\phi,_{i}
\phi,_{j}-\frac{1}{2}g_{ij} \phi,_{k}
\phi^{,k})+\phi^{-1}(\phi,_{i;j}-g_{ij}\square\phi)
\end{equation}

which while contracted gives,

\begin{equation}
-R=(8 \pi \phi^{-1}/c^{4})T-(\omega/\phi^{2})\phi,_{k}\phi^{,k}-3
\phi^{-1}\square\phi
\end{equation}

Combining this equation with equation (1.89), the wave equation
for $\phi$ is given by,

\begin{equation}
\square\phi=\frac{8 \pi}{(3+2\omega)c^{4}}T
\end{equation}

with $ds^{2}=g_{ij}dx^{i}dx^{j}$, $g_{00}<0$.\\

Now for a perfect fluid the energy-momentum tensor given by
equation (1.42). Thus we have,

\begin{equation}
T=3p-\rho
\end{equation}

where, $\rho$ and $p$ are respectively energy density
and pressure of the fluid.\\

Now to apply the Brans-Dicke theory to cosmology, we consider the
Universe to be homogeneous and isotropic.\\

The Robertson-Walker form of the metric is given by equation
(1.7).\\

We write the gravitational field equations (1.91) as,

\begin{equation}
R_{ij}=-\frac{8\pi}{\phi ~c^{4}}\left[T_{ij}-\left(
\frac{1+\omega}{3+2\omega}
\right)g_{ij~}T^{\mu}_{\mu}\right]-\frac{\omega}{\phi^{2}}\phi_{;i}\phi_{;j}-\frac{1}{\phi}\phi_{;i;j}
\end{equation}

The time-time component of equation (1.95) gives,

\begin{equation}
3\frac{\ddot{a}}{a}=-\frac{8 \pi}{(3+2\omega)\phi
c^{4}}\{(2+\omega)\rho+3(1+\omega)p\}-\omega\frac{\dot{\phi}^{2}}{\phi^{2}}-\frac{\ddot{\phi}}{\phi}
\end{equation}

and the space-space component is,

\begin{equation}
-\frac{\ddot{a}}{a}-2\frac{\dot{a}^{2}}{a^{2}}-2\frac{k}{a^{2}}=-\frac{8\pi}{(3+2\omega)\phi}\{(1+\omega)\rho-\omega
p\}+\frac{\dot{\phi}}{\phi}\frac{\dot{a}}{a}
\end{equation}

and the time-space gives,

\begin{equation}
0=0
\end{equation}

Also the field equation for $\phi$, i.e., equation (1.93) gives,

\begin{equation}
\frac{d}{dt}(\dot{\phi}a^{3})=\frac{8\pi}{(3+2\omega)c^{4}}(\rho-3p)a^{3}
\end{equation}

and the conservation law is given by equation (1.20).\\

Equation (1.99), on simplification gives,

\begin{equation}
\ddot{\phi}+3\frac{\dot{a}}{a}
\dot{\phi}=(\rho-3p)\frac{8\pi}{(3+2\omega)c^{4}}
\end{equation}

Using equations (1.96), (1.97) and (1.100), we get,

\begin{equation}
3\frac{\dot{a}^{2}+k}{a^{2}}=\frac{8\pi}{c^{4}}\frac{\rho}{\phi}
-3\frac{\dot{\phi}}{\phi}\frac{\dot{a}}{a}+\frac{\omega}{2}\frac{\dot{\phi}^{2}}{\phi^{2}}
\end{equation}

and

\begin{equation}
2\frac{\ddot{a}}{a}+\frac{\dot{a}^{2}+k}{a^{2}}=-\frac{8\pi}{c^{4}}\frac{p}{\phi}
-\frac{\omega}{2}\frac{\dot{\phi}^{2}}{\phi^{2}}-2\frac{\dot{\phi}}{\phi}\frac{\dot{a}}{a}-\frac{\ddot{\phi}}{\phi}
\end{equation}

Equations (1.99), (1.100), (1.101), (1.102) are the fundamental
equations of Brans-Dicke Cosmology. In these equations the
Brans-Dicke parameter `$\omega$' is kept
as a constant.\\

Solar system experiments impose a limit on the value of $\omega$,
i.e., $\mid \omega\mid \geq 500$, although $\omega$ is seen to
have a low negative value in order to solve the cosmic
acceleration and coincidence problem. Also a constant negative
$\omega$ fails to give a consistent radiation model which explains
the primordial nucleosysnthesis. Banerjee and Pavon [2001] have
shown that this can be solved by a varying $\omega$ theory where
$\omega$ is considered to be a function of the scalar field $\phi$
[Nordtvedt, 1970; Bergmann, 1968; Wagoner, 1970]. Therefore, the
action integral for this general class of scalar-tensor
gravitational theory is,

\begin{equation}
A=\int \left[16 \pi L+\phi R+ \frac{\omega(\phi)}{\phi}\phi_{;\mu}
\phi^{;\mu}\right]\sqrt{-g}~d^{4}x
\end{equation}

Thus the field equations for the tensor and scalar fields become,

\begin{equation}
\square \phi=\frac{8
\pi}{(3+2\omega)c^{4}}T-\frac{\omega'}{3+2\omega}\phi,_{\mu}
\phi^{,\mu}
\end{equation}

together with equation (1.91), where, $\omega'=\frac{d\omega}{d\phi}$.\\

Thus these equations can be combined to get,

\begin{equation}
R_{ij}=-\frac{8\pi}{\phi c^{4}}\left[
T_{ij}-\frac{1+\omega}{3+2\omega}g_{ij}T^{\mu}_{\mu}\right]-\frac{\omega}{\phi^{2}}\phi_{;i}
\phi_{;j}-\frac{1}{\phi}\phi_{;i}
\phi_{;j}+\frac{\omega'}{2\phi(3+2\omega)}\phi_{;i}
\phi^{;i}g_{ij}
\end{equation}

This modifies equation (1.100) by,

\begin{equation}
\ddot{\phi}+3\frac{\dot{a}}{a}
\dot{\phi}=(\rho-3p)\frac{8\pi}{(3+2\omega)c^{4}}-\frac{\dot{\omega}
\dot{\phi}}{3+2\omega}
\end{equation}

Now the effect of $\omega'\neq 0$ will be in $g_{\mu \nu}$ only in
the non-linear order in the mass source strength. Hence,
gravitational fields in linear mass are identical with the results
of Brans-Dicke theory with $\omega=$ constant. Banerjee and Pavon
[2001] have shown that this varying $\omega$-theory can
potentially solve the quintessence problem and give rise to a
non-decelerating radiation model also. On the other hand,
Bertolami and Martins [2000] obtained an accelerated expansion of
the Universe in a further modified form of Brans-Dicke Theory by
introducing a potential which is a function of the scalar field.
This self-interacting Brans-Dicke Theory is described by the
action,

\begin{equation}
S=\int d^{4} x \sqrt{-g}\left[\phi R- \frac{\omega(\phi)}{\phi}
{\phi}^{,\alpha} {\phi,}_{\alpha}-V(\phi)+ 16 \pi{\cal
L}_{m}\right]
\end{equation}

where, $V(\phi)$ is the self-interacting potential for the
Brans-Dicke scalar field $\phi$. Thus the field equations are
obtained as,

\begin{equation}
G_{\mu \nu}=\frac{\omega}{\phi^{2}}\left[ \phi,_{\mu}
\phi,_{\nu}-\frac{1}{2} g_{\mu \nu}\phi,_{\alpha}
\phi^{,\alpha}\right]+\frac{1}{\phi}\left[ \phi,_{\mu;\nu}-g_{\mu
\nu} \square \phi\right]-\frac{V(\phi)}{2 \phi}g_{\mu
\nu}+\frac{T_{\mu \nu}}{\phi}\frac{8 \pi}{c^{4}}
\end{equation}

\begin{equation}
\square \phi=\frac{8
\pi}{(3+2\omega)c^{4}}T-\frac{1}{3+2\omega}\left[2
V(\phi)-\phi\frac{d V(\phi)}{d \phi}\right]
\end{equation}

These field equations under Friedmann-Robertson-Walker geometry
modifies equations (1.100), (1.101) and (1.102) as,

\begin{equation}
\ddot{\phi}+3\frac{\dot{a}}{a}
\dot{\phi}=(\rho-3p)\frac{8\pi}{(3+2\omega)c^{4}}+\frac{1}{3+2\omega}\left[2
V(\phi)-\phi\frac{d V(\phi)}{d \phi}\right]
\end{equation}

\begin{equation}
3\frac{\dot{a}^{2}+k}{a^{2}}=\frac{8\pi}{c^{4}}\frac{\rho}{\phi}
-3\frac{\dot{\phi}}{\phi}\frac{\dot{a}}{a}+\frac{\omega}{2}\frac{\dot{\phi}^{2}}{\phi^{2}}
+\frac{V}{2\phi}
\end{equation}

and

\begin{equation}
2\frac{\ddot{a}}{a}+\frac{\dot{a}^{2}+k}{a^{2}}=-\frac{8\pi}{c^{4}}\frac{p}{\phi}
-\frac{\omega}{2}\frac{\dot{\phi}^{2}}{\phi^{2}}
-2\frac{\dot{\phi}}{\phi}\frac{\dot{a}}{a}-\frac{\ddot{\phi}}{\phi}+\frac{V}{2\phi}
\end{equation}

Bertolami and Martins have obtained the solution for accelerated
expansion with a quadratic self-coupling potential ($V(\phi) \sim
\phi^{2}$) and a negative coupling constant $\omega$, although
they have not considered the positive energy conditions for the
matter and scalar field. Amendola [1999] has shown that coupled
quintessence models are conformally equivalent to Brans-Dicke
Lagrangians with power-law potential given by, $V(\phi) \sim
\phi^{n}$. Sen etal [2001] have studied the late time
acceleration in the context of Brans Dicke (BD) theory with
potential $V(\phi)= \lambda \phi^{4}-\mu^{2} (t) \phi^{2}$, and
showed that a fluid with dissipative pressure can drive this late
time acceleration for a simple power law expansion of the
universe, whereas, a perfect fluid cannot support this
acceleration.  Later Chiba [2003] extended the gravity theories
and obtained a BD theory with potential for BD scalar field, but
found that this is not compatible with solar system experiments
if the field is very light.\\

\section{Statefinder Diagnostics}

Over the years a lot of models have proved to be viable candidates
of Dark Energy, thus leading to the problem of discriminating
between these models. For this purpose Sahni etal [2003] proposed
a new geometric diagnosis (dimensionless) to characterize the
properties of dark energy in a model independent manner. They
introduced a pair of parameters called {\it statefinder
parameters} depending on the scale factor and its derivatives,
defined by,

\begin{equation}
r=\frac{\dddot{a}}{a H^{3}},~~~~~s=\frac{r-1}{3(q-1/2)}
\end{equation}

where $q=-\frac{a \ddot{a}}{\dot{a}^{2}}$ is the deceleration
parameter. The parameter $r$ forms the next step in the hierarchy
of geometrical cosmological parameters after $H$ and $q$. In fact
trajectories in the $\{s,r\}$ plane corresponding to different
cosmological models demonstrate qualitatively different behaviour,
for example $\Lambda$CDM model correspond to the fixed point $s=0,~r=1$..\\

For spatially flat space-time ($k=0$), considering the Universe to
be consisted of non-relativistic matter $\Omega_{m}$, i.e., CDM
and baryons, and dark energy $\Omega_{x}=1-\Omega_{m}$, the
statefinder pair $\{r,s\}$ takes the form, [Sahni etal, 2003]

\begin{equation}
r=1+\frac{9}{2}\Omega_{x} \omega(1+\omega)-\frac{3}{2}
\Omega_{x}\frac{\dot{\omega}}{H}
\end{equation}

and

\begin{equation}
s=1+\omega-\frac{1}{3}\frac{\dot{\omega}}{\omega H}
\end{equation}

Thus for $\Lambda$CDM model with a non-zero $\Lambda$ ($\omega=-1$), $r=1$ and $s=0$.\\

If $\omega$ is constant these parameters reduce to

\begin{equation}
r=1+\frac{9}{2}\Omega_{x} \omega(1+\omega),~~~~~s=1+\omega
\end{equation}

Degeneracy occurs when $\omega=-1/3$ or $\omega=-2/3$, as
$r\rightarrow 1$  at earlier stage and $r\rightarrow 0$ at later
age, with $r\simeq 0.3$ for present age ($\Omega_{x}\simeq 0.7$)
at the present time.\\

For a quintessence scalar field these parameters take the forms

\begin{equation}
r=1+\frac{12 \pi G \dot{\phi}^{2}}{H^{2}}+\frac{8 \pi G
\dot{V}}{H^{3}}
\end{equation}

and

\begin{equation}
s=\frac{2(\dot{\phi}^{2}+2\dot{V}/3 H)}{\dot{\phi}^{2}-2V}
\end{equation}

For pure ($p=-B/\rho, (B>0)$) and generalized
($p=-B/\rho^{\alpha}, 0\le \alpha \le 1$) Chaplygin Gas, we get
respectively, [Gorini etal, 2002]

\begin{equation}
r=1-\frac{9}{2} s (1+s)
\end{equation}

and

\begin{equation}
r=1-\frac{9}{2} s (\alpha+s)/\alpha
\end{equation}

For modified Chaplygin gas ($p=A\rho-\frac{B}{\rho^{\alpha}}$ with
$0\le \alpha \le 1$), these parameters have rather an implicit
form given by, [Debnath etal, 2004]

\begin{equation}
18(r-1)s^{2}+18\alpha s(r-1)+4\alpha
(r-1)^{2}=9sA(1+\alpha)(2r+9s-2)
\end{equation}

In general, for one fluid model, these $\{r,s\}$ can be written as

\begin{equation}
r=1+\frac{9}{2}\left(1+\frac{p}{\rho}\right)\frac{\partial
p}{\partial\rho}
\end{equation}

and

\begin{equation}
s=\left(1+\frac{\rho}{p}\right)\frac{\partial p}{\partial\rho}
\end{equation}

Gorini etal have shown that for pure Chaplygin gas $s$ varies in
the interval $[-1, ~0]$ and $r$ first increases from r = 1 to its
maximum value and then decreases to the $\Lambda$CDM fixed point
$s = 0, ~ r = 1$. For generalized Chaplygin gas, the model becomes
identical with the standard $\Lambda$CDM model for small values of
$\alpha$ from statefinder viewpoint. Debnath etal have shown that
in case of modified Chaplygin gas the Universe can be described
from radiation era to $\Lambda$CDM with statefinder
diagnosis.\\

Alam etal [2003] have shown that $s$ is positive for quintessence
models, but negative for the Chaplygin gas models, whereas, $r$ is
$< 1$  or $> 1$ for quintessence or Chaplygin gas.\\

Later Shao and Gui [2007] carried out the statefinder diagnosis on
tachyonic field and showed that the tachyon model can be
distinguished from the other dark energy models by statefinder
diagnosis as the evolving trajectories of the attractor solutions
lie in the total region although they pass through the LCDM fixed
point.\\

Using SNAP data Alam etal have also demonstrated that the
Statefinder can distinguish a cosmological constant ($\omega =
-1$) from quintessence models with $\omega > -0.9$ and Chaplygin
gas models with $\kappa \leq 15$
($\kappa=\frac{\Omega_{m}}{1-\Omega_{m}}$) at the $3 \sigma$ level
if the value of $\Omega_{m}$ is known. Even if the value of
$\Omega_{m}$ is known to approximately $20\%$ accuracy statefinder
diagnosis rule out quintessence with $\omega > -0.85$ and the
Chaplygin gas with $\kappa\leq 7$ again at $3\sigma$ level. They
have shown that the statefinder diagnosis can differentiate
between various dark energy models at moderately high redshifts of
$z \lesssim 10$.

%% file: chap2.tex
\large \baselineskip .85cm
\chapter{Cosmological Dynamics of MCG in presence of Barotropic Fluid }
\label{chap2}\markright{\it CHAPTER~\ref{chap2}. Cosmological
Dynamics of MCG in presence of Barotropic Fluid}

\section{Prelude}

Chaplygin Gas Cosmology has been studied by a lot of authors to
get a plausible model of Dark Energy [Sahni etal, 2000; Peebles
etal, 2003; Padmanabhan, 2003]. Pure Chaplygin Gas [Kamenshchik
etal, 2001] with EOS $p=-B/\rho, (B>0)$ behaves as pressureless
fluid for small values of the scale factor and as a cosmological
constant for large values of the scale factor which tends to
accelerate the expansion. Subsequently the above equation was
generalized (GCG) to the form $p=-B/\rho^{\alpha}, 0\le \alpha
\le 1$ [Gorini etal, 2003; Alam etal, 2003; Bento etal, 2002] and
recently it was modified to the form $p=A\rho-B/\rho^{\alpha},
(A>0)$ [Benaoum, 2002; Debnath etal, 2004], which is known as
Modified Chaplygin Gas (MCG). Further Gorini etal considered a two
fluid model consisting of Chaplygin Gas and a dust component and
showed that for some particular values of the parameters this
model can solve the cosmic coincidence problem.\\

Over the time the statefinder diagnostics  have become very
popular to discriminate between various dark energy models. In
fact trajectories in the $\{r,s\}$ plane corresponding to
different cosmological models demonstrate qualitatively different
behaviour. Debnath etal [2004] carried out the statefinder
diagnostics for MCG and showed that this model shows a radiation
era ($A=1/3$) at one extreme and a $\Lambda$CDM model at the other
extreme. Gorini etal [2003] showed that their two fluid model is
indistinguishable from $\Lambda$CDM model for some values of the
parameters from statefinder point of view and thus is quite
different from the pure Chaplygin Gas.\\

In this chapter, we have generalized the model proposed by Gorini
etal [2003]. We have considered a two fluid model consisting of
modified Chaplygin gas and barotropic fluid. We have analysed
this model to study the cosmological evolution of the Universe.
Also we have carried out the statefinder analysis to describe the
different phases of the evolution and the significance of this
model in comparison with the one fluid model of MCG.\\

\section{Field Equations and Solutions}

The metric of a homogeneous and isotropic universe in FRW model
is given by equation (1.7). The Einstein field equations are
(1.12) and (1.13) and the energy conservation equation is (1.20).
For modified Chaplygin gas, the energy density is given by
equation (1.61).\\

Here we consider two fluid cosmological model which besides a
modified Chaplygin's component, with EOS (1.60) contains also a
barotropic fluid component with equation of state
$p_{_{1}}=\gamma\rho_{_{1}}$. Normally for accelerating universe
$\gamma$ satisfies $-1\le\gamma\le 1$. But observations state that
$\gamma$ satisfies $-1.6\le\gamma\le 1$ i.e., $\gamma<-1$
corresponds to phantom model. For these two component fluids
$\rho$ and $p$ should be replaced by $\rho+\rho_{_{1}}$ and
$p+p_{_{1}}$ respectively. Here we have assumed the two fluid are
separately conserved. For Chaplygin gas, the density has the
expression given in equation (1.62) and for another fluid, the
conservation equation gives the expression for density as
\begin{equation}
\rho_{_{1}}=\frac{d}{a^{3(1+\gamma)}}
\end{equation}

where $d$ is an integration constant.\\

\section{Field Theoretical Approach}

We can describe this two fluid cosmological model from the field
theoretical point of view by introducing a scalar field $\phi$ and
a self-interacting potential $V(\phi)$ with the effective
Lagrangian is given by (1.48). The analogous energy density
$\rho_{\phi}$ and pressure $p_{\phi}$ corresponding scalar field
$\phi$ having a self-interacting potential $V(\phi)$ are the
following:

\begin{equation}
\rho_{\phi}=\frac{1}{2}~\dot{\phi}^{2}+V(\phi)=\rho+\rho_{_{1}}=\left[\frac{B}{1+A}+\frac{C}{a^{3(1+A)(1+\alpha)}}
\right]^{\frac{1}{1+\alpha}}+\frac{d}{a^{3(1+\gamma)}}
\end{equation}
and
\begin{eqnarray*}
p_{\phi}=\frac{1}{2}~\dot{\phi}^{2}-V(\phi)=p+p_{_{1}}=A\rho-\frac{B}{\rho^{\alpha}}+\gamma\rho_{_{1}}
=A \left[\frac{B}{1+A}+\frac{C}{a^{3(1+A)(1+\alpha)}}
\right]^{\frac{1}{1+\alpha}}
\end{eqnarray*}
\begin{equation}
\hspace{3in} -B \left[\frac{B}{1+A}+\frac{C}{a^{3(1+A)(1+\alpha)}}
\right]^{-\frac{\alpha}{1+\alpha}}+\frac{\gamma~d}{a^{3(1+\gamma)}}
\end{equation}
\\

For flat Universe ($k=0$) and by the choice $\gamma=A$, we have
the expression for $\phi$ and $V(\phi)$:

\begin{equation}
\phi=-\frac{1}{\sqrt{3(1+A)}~(1+\alpha)}\int
\left[\frac{d+c\left(C+\frac{Bz}{1+A}
\right)^{-\frac{\alpha}{1+\alpha}}} {d+\left(C+\frac{Bz}{1+A}
\right)^{\frac{1}{1+\alpha}}} \right]^{\frac{1}{2}}~\frac{dz}{z}
\end{equation}
and
\begin{equation}
V(\phi)=A \left[\frac{B}{1+A}+\frac{C}{z}
\right]^{\frac{1}{1+\alpha}}-B \left[\frac{B}{1+A}+\frac{C}{z}
\right]^{-\frac{\alpha}{1+\alpha}}+\frac{(1-A)~d}{z^{\frac{1}{1+A}}}
\end{equation}

where $z=a^{3(1+A)(1+\alpha)}$.\\

\begin{figure}
\includegraphics{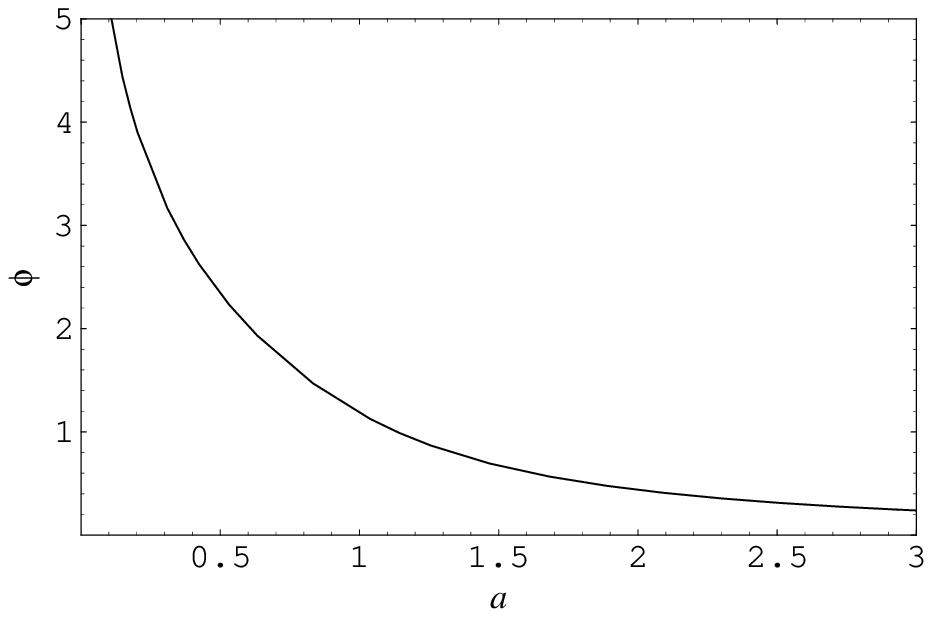}\\
Fig 2.1: Here variation of $\phi$ has been plotted against $a$ for
$A(=\gamma)=1/3$ and $\alpha= 1$ (values of other constants:
$B=1, C=1, d=1$)\\

\includegraphics{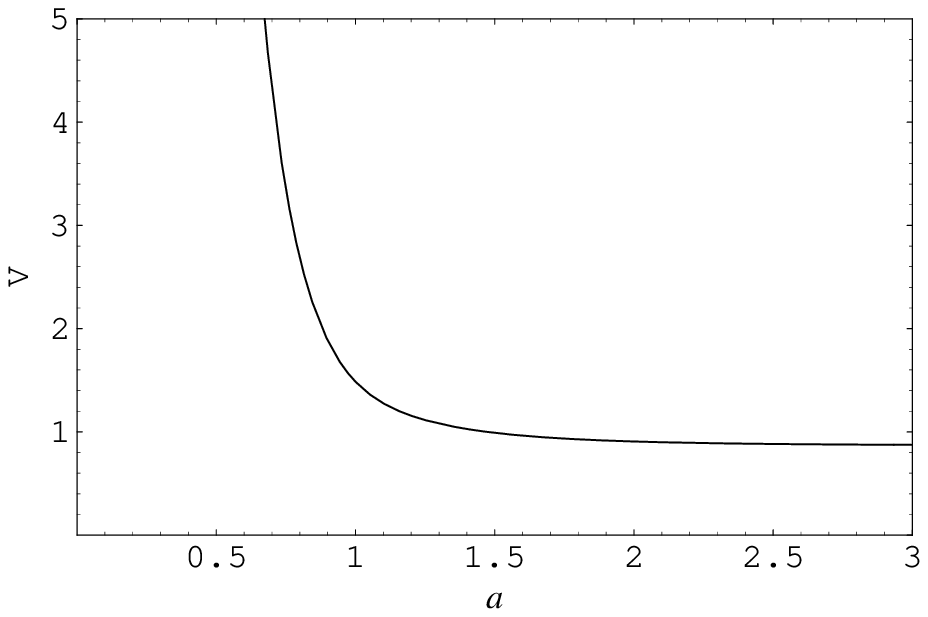}\\
Fig 2.2: Here variation of $V$ has been plotted against $a$ for
$A(=\gamma)=1/3$ and $\alpha= 1$ (values of other constants:
$B=1, C=1, d=1$)\\

\includegraphics{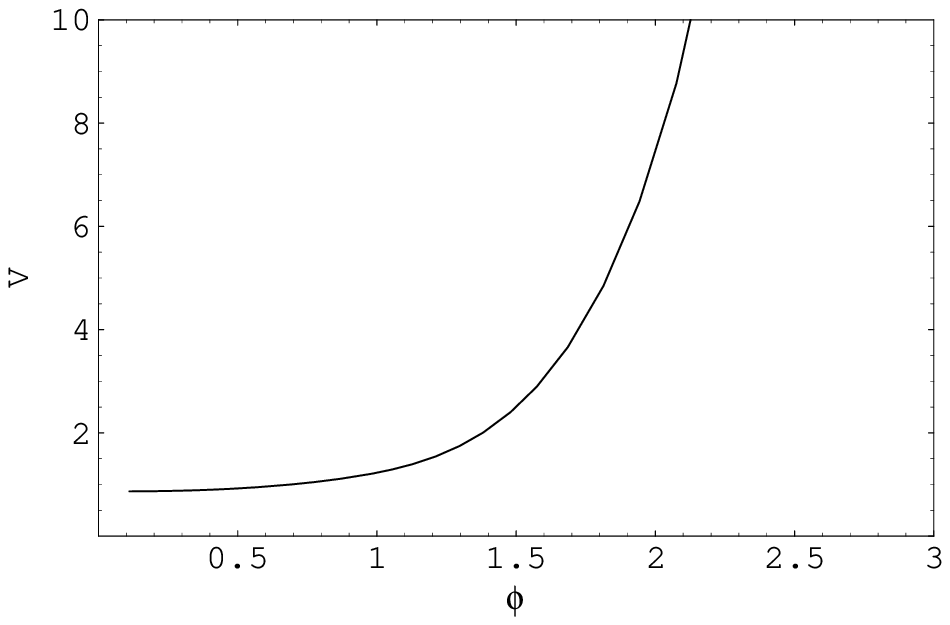}\\
Fig 2.3: Here variation of $V$ has been plotted against $\phi$ for
$A(=\gamma)=1/3$ and $\alpha= 1$ (values of other constants:
$B=1, C=1, d=1$)\\

\end{figure}

The graphical representation of $\phi$ against $a$ and $V(\phi)$
against $a$ and $\phi$ respectively have been shown in figures 2.1
- 2.3 for $A=1/3$ and $\alpha=1$. From figure 2.1 we have seen
that scalar field $\phi$ decreases when scale factor $a(t)$
increases for $A=1/3$. In figure 2.2, we see that potential
function $V(\phi)$ sharply decreases from extremely large value to
a fixed value for $A=1/3$. The potential function $V(\phi)$
increases to infinitely large value when scale factor $a(t)$
increases for $A=1/3$. So the
figures show how $V(\phi)$ varies with $\phi$ and $a(t)$.\\

\section{Statefinder Diagnosis}

We have already studied the significance of statefinder diagnosis
of the models. Let us now analyse our model using statefinder
parameters. The statefinder diagnostic pair has the form given by
equation (1.113). For one fluid model, these $\{r,s\}$ can be
given by equations (1.122) and (1.123).\\

For the two component fluids, these equations take the following
form:

\begin{equation}
r=1+\frac{9}{2(\rho+\rho_{_{1}})}\left[\frac{\partial
p}{\partial\rho}(\rho+p)+\frac{\partial
p_{_{1}}}{\partial\rho_{_{1}}}(\rho_{_{1}}+p_{_{1}})\right]
\end{equation}

and

\begin{equation}
s=\frac{1}{(p+p_{_{1}})}\left[\frac{\partial
p}{\partial\rho}(\rho+p)+\frac{\partial
p_{_{1}}}{\partial\rho_{_{1}}}(\rho_{_{1}}+p_{_{1}})\right]
\end{equation}

The deceleration parameter $q$ has the form:
\begin{equation}
q=-\frac{\ddot{a}}{aH^{2}}=\frac{1}{2}+\frac{3}{2}\left(\frac{p+p_{_{1}}}{\rho+\rho_{_{1}}}\right)
\end{equation}

For modified gas and barotropic equation states, we can set:

\begin{equation}
x=\frac{p}{\rho}=A-\frac{B}{\rho^{\alpha+1}}
\end{equation}
and
\begin{equation}
y=\frac{\rho_{_{1}}}{\rho}=\frac{\frac{d}{a^{3(1+\gamma)}}}{\left[\frac{B}{1+A}+\frac{C}{a^{3(1+A)(1+\alpha)}}
\right]^{\frac{1}{1+\alpha}}}
\end{equation}

\begin{figure}

\includegraphics{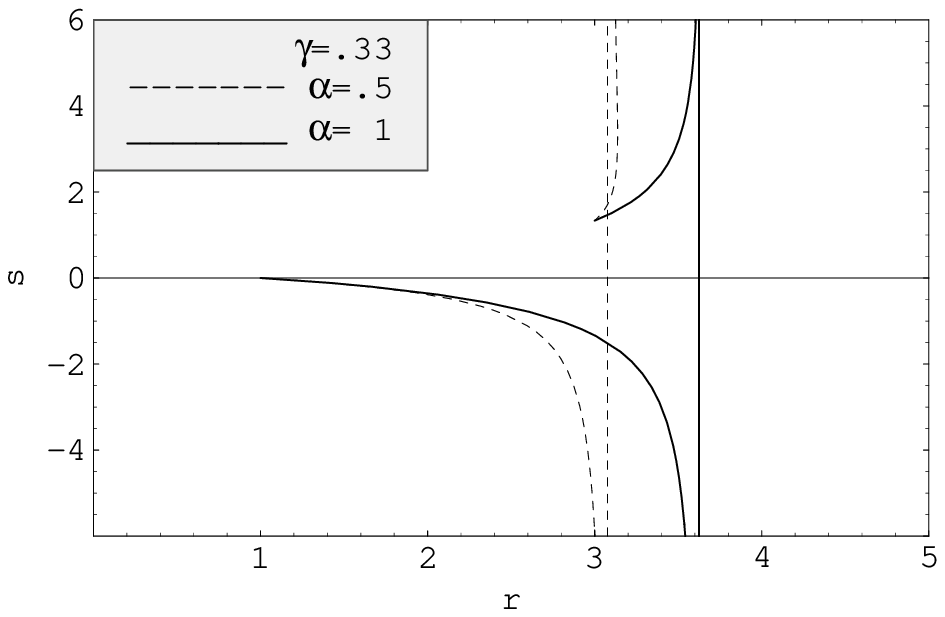}\\
Fig 2.4: Here variation of $s$ has been plotted against $r$ for
$\gamma=1/3$, $\alpha$ (= ~0.5,~1)and $A=1/3$.\\

\includegraphics{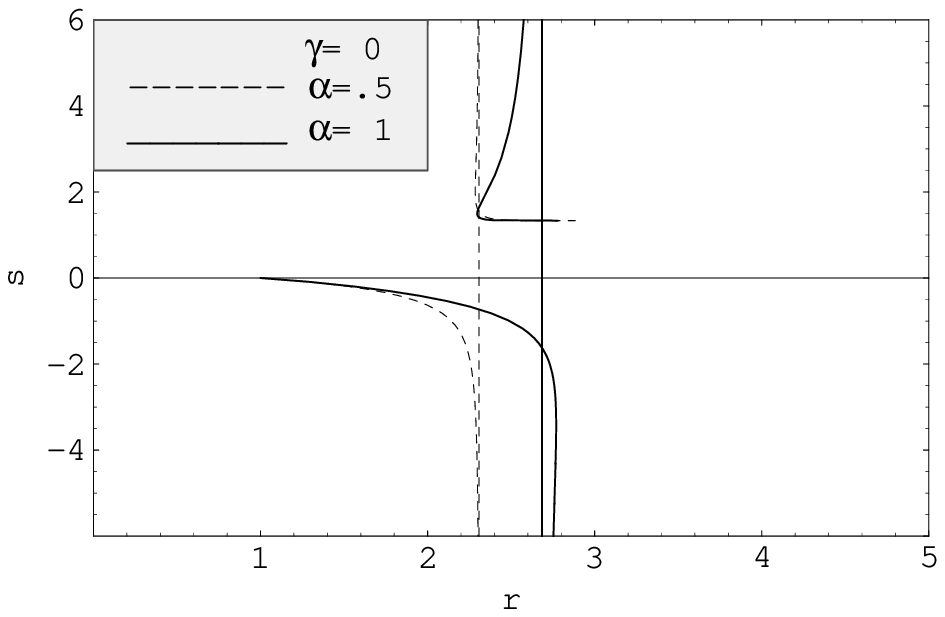}\\
Fig 2.5: Here variation of $s$ has been plotted against $r$ for
$\gamma=0$, $\alpha$ (= ~0.5,~1) and $A=1/3$.\\

\includegraphics{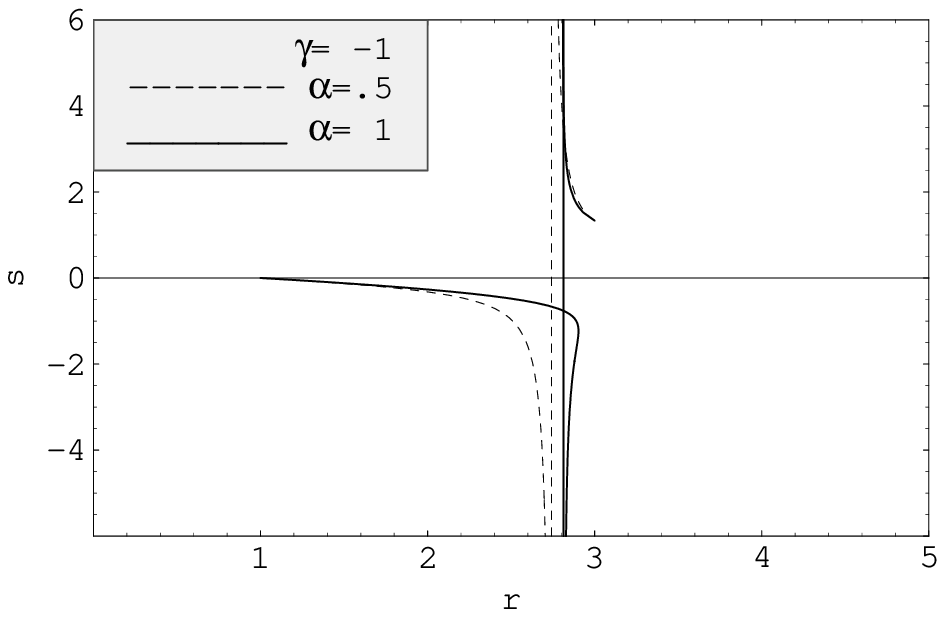}\\
Fig 2.6: Here variation of $s$ has been plotted against $r$ for
$\gamma=-1$, $\alpha$ (= ~0.5,~1) and $A=1/3$.\\

\end{figure}

\begin{figure}

\includegraphics{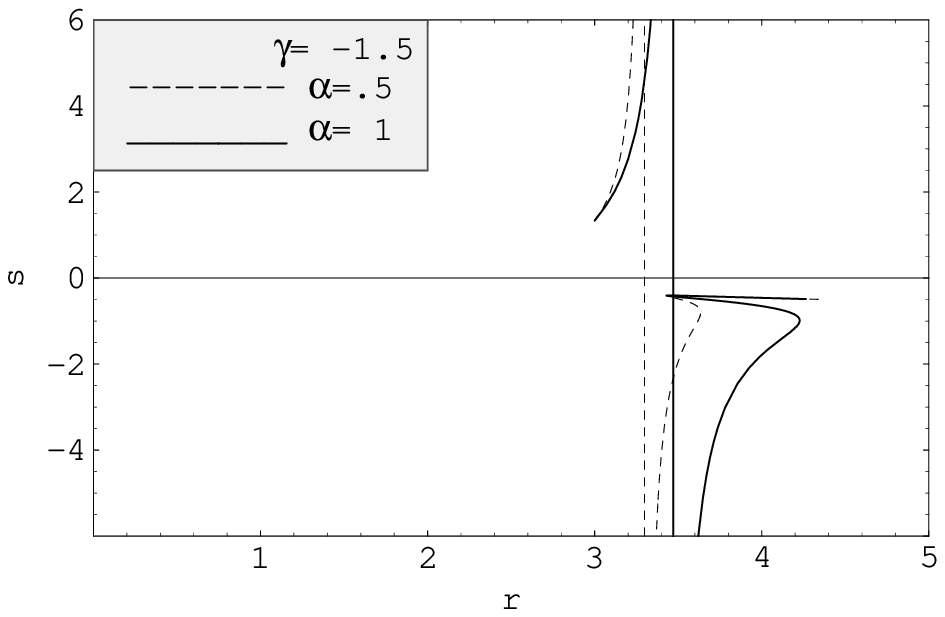}\\

\vspace{10mm}Fig 2.7: Here variation of $s$ has been plotted
against $r$ for
$\gamma=-1.5$, $\alpha$ (= ~0.5,~1)and $A=1/3$.\\

\vspace{1in}

\includegraphics{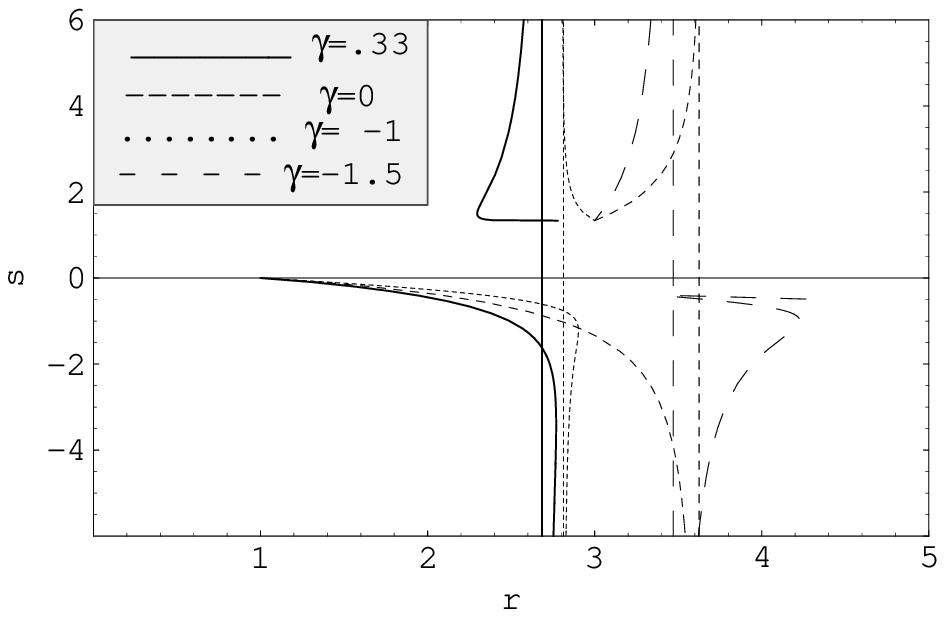}\\

\vspace{10mm}Fig 2.8: This figure shows the variation of $s$
against $r$ for different values of $\gamma=1/3,~0,~-1,~-1.5$ and
for
$\alpha$ (=~1), $A=1/3$.\\

\end{figure}

Thus equations (2.6) and (2.7) can be written as

\begin{equation}
r=1+\frac{9s}{2}\left(\frac{x+\gamma y}{1+y}\right)
\end{equation}

and

\begin{equation}
s=\frac{(1+x)\{A(1+\alpha)-\alpha x \}+\gamma(1+\gamma)y}{x+\gamma
y}
\end{equation}

with

\begin{equation}
y=\left[\frac{d^{(1+\alpha)(1+A)}B^{\gamma-A}(1+\gamma)^{1+\gamma}
} {C^{1+\gamma}(1+A)^{1+\gamma}(A-x)^{\gamma-A} }
\right]^{\frac{1}{(1+\alpha)(1+A)}}
\end{equation}

From the equations (2.11) and (2.12) we can not write the
relationship between $r$ and $s$ in closed form. Thus the relation
between the parameters $r$ and $s$ in $\{r,s\}$ plane for
different choices of other parameters are plotted in figures 2.4 -
2.8. The figures 2.4 - 2.7 shows the variation of $s$ against $r$
for different values of $\gamma=1/3,~0,~-1,~-1.5$ respectively and
for $\alpha$ (= ~0.5,~1), $A=1/3$. Fig 2.8 shows the variation of
$s$ against $r$ for different values of $\gamma=1/3,~0,~-1,~-1.5$
and for $\alpha$ (=~1), $A=1/3$. Thus the figures 2.4 - 2.6
represent the evolution of the universe starting from the
radiation era to the $\Lambda$CDM model for $\gamma=1/3,~0,~-1$
and the figure 2.7 represents the evolution of the universe
starting from the radiation era to the quiessence model for
$\gamma=-1.5$. Thus $\gamma$ plays an active role for the various
stages of the evolution of the universe. If we choose the
arbitrary constant $d$ is equal to zero, we recover the model of
Modified Chaplygin gas [Debnath etal, 2004]. If $A$ and the
barotropic index $\gamma$ are chosen to be zero, we get
back to the results of the works of Gorini etal [2003].\\

\section{Discussion}

In this chapter, we have analysed a model consisting of modified
Chaplygin gas and barotropic fluid. As We have shown that the
mixture of these two fluid models is valid from (i) the radiation
era to $\Lambda$CDM for $-1\le\gamma\le 1$  and (ii) the radiation
era to quiessence model for $\gamma<-1$. We have carried out the
statefinder diagnosis for this model and presented the result
graphically. The graphical representation show the validation of
this model for different phases of the evolution of the Universe.
During the various stages of the evolution we see that $\gamma$
plays a very important role. It depends on $\gamma$ whether the
barotropic fluid will behave as dark matter or dark energy. For
$\gamma=0$ this fluid is dust, for $\gamma=\frac{1}{3}$, it
represents radiation and for $\gamma<0$ it implies negative
pressure, thus determining the nature of the barotropic fluid,
whereas MCG unifies dark matter and dark energy under the same
umbrella. Also equation (2.10) shows that for $A=\gamma$ at the
initial stage with a parameter $\kappa=\frac{d}{C^{1/(1+\alpha)}}$
of order one, the initial energies of MCG and barotropic fluid are
of same order of magnitude, which may provide a solution to the
cosmic coincidence problem.\\

%% file: chap3.tex
\large \baselineskip .85cm
\chapter{Effect of Dynamical Cosmological Constant in presence of MCG} \label{chap3}\markright{\it
CHAPTER~\ref{chap3}. Effect of Dynamical Cosmological Constant in
presence of MCG}

\section{Prelude}

There are two parameters, the cosmological constant $\Lambda$ and
the gravitational constant $G$, present in Einstein's field
equations. The Newtonian constant of gravitation $G$ plays the
role of a coupling constant between geometry and matter in the
Einstein's field equations. In an evolving Universe, it appears
natural to look at this ``constant'' as a function of time.
Numerous suggestions based on different arguments have been
proposed in the past few decades in which $G$ varies with time
[Wesson, 1978, 1980]. Dirac [1979] proposed a theory with variable
$G$ motivated by the occurrence of large numbers discovered by
Weyl, Eddington and Dirac himself.\\

It is widely believed that the value of $\Lambda$ was large during
the early stages of evolution and strongly influenced its
expansion, whereas its present value is incredibly small
[Weinberg, 1989; Carroll etal, 1992]. We have already discussed
in the introduction that several authors [Freese etal, 1987; Ozer
and Taha, 1987; Gasperini, 1987, 1998; Chen and Wu, 1990] have
advocated a variable $\Lambda$ in the framework of Einstein's
theory to account for this fact. $\Lambda$ as a function of time
has also been considered in various variable $G$ theories in
different contexts [Banerjee etal, 1985; Bertolami, 1986;
Abdussattar and Vishwakarma, 1997; Kalligas etal, 1992]. For
these variations, the energy-momentum tensor of matter leaves
the form of the Einstein's field equations unchanged.\\

In attempt to modify the General Theory of Relativity, Al-Rawaf
and Taha [1996] related the cosmological constant to the Ricci
Scalar $\cal R$. This is written as a built-in-cosmological
constant, i.e., $\Lambda\propto\cal R$. Since the Ricci Scalar
contains a term of the form $\frac{\ddot{a}}{a}$, one adopts this
variation for $\Lambda$. We parameterized this as
$\Lambda\propto\frac{\ddot{a}}{a}$ [Arbab, 2003, 2004].
Similarly, we have chosen another two forms for $\Lambda :
\Lambda\propto\rho$ and $\Lambda\propto\frac{\dot{a}^{2}}{a^{2}}$
[Carvalho, 1992]; where $\rho$ is the
energy density. \\

In this chapter we have considered the Universe to be filled with
Modified Gas and the Cosmological Constant $\Lambda$ to be
time-dependent with or without the Gravitational Constant $G$ to
be time-dependent. We have considered various phenomenological
models for $\Lambda$ , viz., $\Lambda\propto\rho,
\Lambda\propto\frac{\dot{a}^{2}}{a^{2}}$  and
$\Lambda\propto\frac{\ddot{a}}{a}$. Also we have shown the
natures of $G$ and $\Lambda$ over the total age of the Universe
and analysed our models in the viewpoint of satefinder
diagnostics.\\

\section{Einstein Field Equations with Dynamic Cosmological Constant}

We consider the spherically symmetric FRW metric (1.7). The
Einstein field equations for a spatially flat Universe (i.e.,
taking $k=0$) with a time-dependent cosmological constant
$\Lambda(t)$ are given by (choosing $c=1$),

\begin{equation}
3\frac{\dot{a}^{2}}{a^{2}}=8\pi G \rho+\Lambda(t)
\end{equation}
and
\begin{equation}
2\frac{\ddot{a}}{a}+\frac{\dot{a}^{2}}{a^{2}}=-8\pi G p+\Lambda(t)
\end{equation}

where $\rho$ and $p$ are the energy density and isotropic pressure
respectively.\\

Let us choose MCG with EOS given by equation (1.60). Here, we
consider the phenomenological models for $\Lambda(t)$ of the forms
$\Lambda\propto\rho$, $\Lambda\propto\frac{\dot{a}^{2}}{a^{2}}$
and $\Lambda\propto\frac{\ddot{a}}{a}$.\\

First we will consider $G$ to be constant and try to find out the
solutions for density $\rho$ and the scale factor $a(t)$ and hence
study the cosmological models in terms of the statefinder
parameters $r$, $s$. Secondly we will consider $G$ to be variable
as well and study the various phases of the Universe represented
by the models.\\

\section{Models keeping $G$ constant and $\Lambda$ variable}

Taking $G$ to be constant and $\Lambda$ to be time dependent, the
energy conservation equation is,
\begin{equation} \dot{\rho}+3\frac{\dot{a}}{a}(\rho+p)=-\frac{\dot{\Lambda}}{8\pi G}
\end{equation}

\subsection{Model with $\Lambda\propto\rho$}

Here we consider \begin{equation} \Lambda=\beta_{1}~\rho
\end{equation}
where $\beta_{1}$ is a constant.\\

Equation (3.4) together with equations (1.61) and (3.3) yield the
solution for $\rho$ to be,
\begin{equation}\rho=\left(\frac{B}{1+A}+\frac{C}{a^{\frac{24\pi G(1+A)(1+\alpha)}{8\pi G+\beta_{1}}}}\right)^{\frac{1}{1+\alpha}}\end{equation}
where $C$ is an arbitrary constant.\\

Substituting equation (3.4) and (3.5) in equation (3.1), we get
the solution for the scale factor $a(t)$ as,
\begin{equation}
a^{f_{1}f_{2}}\sqrt{8 \pi G+\beta_{1}}~~_{2}F_{1}
[f_{2},f_{2},1+f_{2},-\frac{a^{f_{1} B}}{C
(1+A)}]=4\sqrt{3}(1+A)G\pi C^{f_{2}}~t
\end{equation}
where  $f_{1}=\frac{24(1+A)(1+\alpha)\pi G }{8 \pi G +\beta_{1}}$
and  $f_{2}=\frac{1}{2(1+\alpha)}$. Hence, for small values of
$a(t)$, we have,  $\rho\simeq\left(\frac{C}{a^{\frac{24 \pi
G(1+A)(1+\alpha}{8 \pi
G+\beta_{1}}}}\right)^{\frac{1}{1+\alpha}}$ which is very large
and the EOS (1.61) reduces to $p\simeq A \rho$. Again for large
values of $a(t)$, we get
$\rho\simeq\left(\frac{B}{1+A}\right)^\frac{1}{1+\alpha}$ and
$p\simeq-\left(\frac{B}{1+A}\right)^\frac{1}{1+\alpha}$ , i.e.,
$p\simeq-\rho$ which coincides with the result obtained for MCG
with $\beta_{1}=0$ [Gorini etal, 2003; Alam etal, 2003; Bento etal, 2002].\\

\begin{figure}

\includegraphics{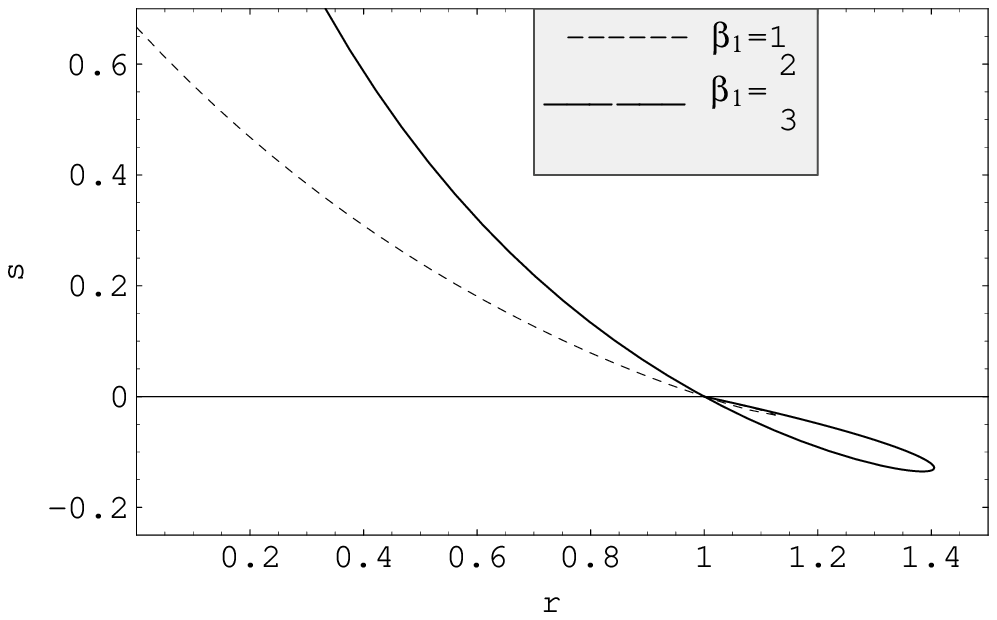}\\
Fig 3.1: This figure shows the variation of $s$ against $r$ for
different values of $\beta_{1}=1,~\frac{2}{3}$ respectively and
for $\alpha$= 1, $A=1/3$ and $8\pi G$=~1.\\

\end{figure}

Using equations (3.1) and (3.3) in statefinder equations (1.113)
we get,
\begin{equation}
r=1+\frac{36 \pi G(1+y)[8 \pi G
\{A(1+\alpha)-y\alpha\}-\beta_{1}]}{(8 \pi G+\beta_{1})^{2}}
\end{equation}

and

\begin{equation}
s=\frac{8\pi G (1+y)[8\pi G \{A(1+\alpha)-y\alpha\}-\beta_{1}]
}{(8\pi G +\beta_{1})(8\pi G y-\beta_{1})}
\end{equation}

where $y=\frac{p}{\rho}$ which can be further reduced to a single
relation between $r$ and $s$. Now
$q=-\frac{\ddot{a}}{aH^{2}}=\frac{8\pi G (1+3y)-2\beta_{1}}{2(8\pi
G+\beta_{1})}$. Therefore for acceleration
$q<0~\Rightarrow~y<\frac{\beta_{1}}{12\pi G}-\frac{1}{3}$. Also
for the present epoch
$q=-\frac{1}{2}~\Rightarrow~y=\frac{1}{3}(\frac{3\beta_{1}}{8\pi
G}-2)$. If we assume that the present Universe is dust filled, we
have $y=0$, i.e., $\beta_{1}=\frac{16\pi G}{3}$. Taking $8\pi G=1$
we get the best fit value to be $\beta_{1}=\frac{2}{3}$, which
gives $r=1$ ( choosing $A=\frac{1}{3},~\alpha=1$ ) for the present
Universe. That means the dark energy responsible for the the
present acceleration is nothing but $\Lambda$. Also $A=1,~
\alpha=1$ and $\beta_{1}=\frac{2}{3}$ give $r=2.16$ for the
present time. For this case $\beta_{1}>3$ gives non-feasible
solutions in the sense that the present values of $y$, i.e.,
$\frac{p}{\rho}$ becomes too large. For $\beta_{1}=1$, we get the
present value of $y$ to be $\frac{1}{3}$, but again $r=1$. In
either of the above cases we get accelerating expansion of the
Universe. These can be represented diagrammatically in the {$r,s$}
plane. This is shown in figure 3.1 (taking $A=\frac{1}{3},
\alpha=1, \beta_{1}=1, \frac{2}{3}, 8\pi G=1$ and $A=1, \alpha=1,
\beta_{1}=1,\frac{2}{3}, 8\pi G=1$). Figure 3.1 represents the
evolution of the Universe starting from radiation era to
$\Lambda$CDM model.
Here we get a discontinuity at $\beta_{1}=-8\pi G$. \\

Again for this model
\begin{equation}
\Lambda=\beta_{1}\left(\frac{B}{1+A}+\frac{C}{a^{\frac{24\pi
G(1+A)(1+\alpha)}{8\pi G+\beta_{1}}}}\right)^{\frac{1}{1+\alpha}}
\end{equation}
Variation of $\Lambda(t)$ against $a(t)$ is shown in figure 3.2
for different choices of $\beta_{1}$, which represents that
regardless the values of $\beta_{2}$, $\Lambda(t)$, i.e., the
effect of the cosmological constant decreases with time.\\

\begin{figure}

\includegraphics{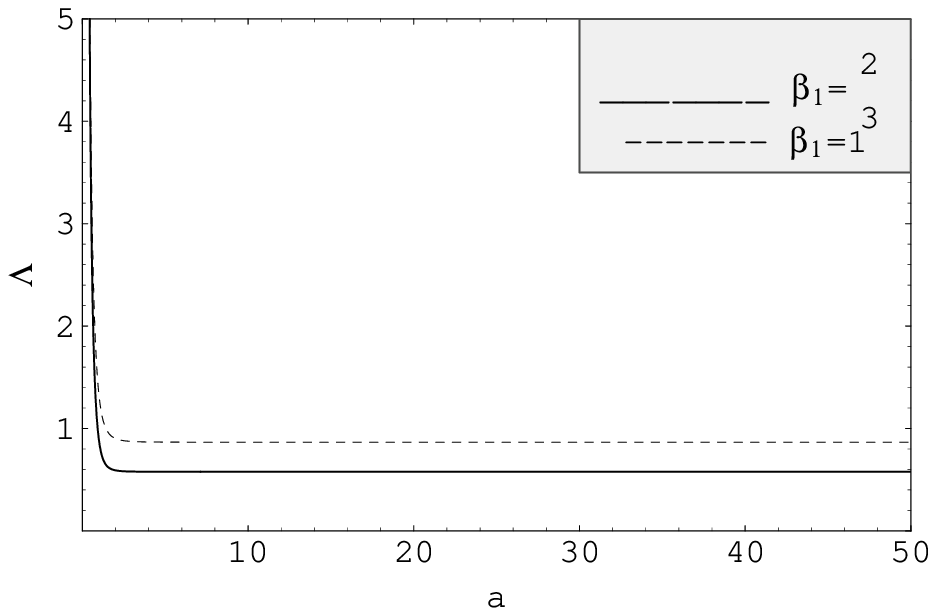}\\
Fig 3.2: Here the variation of $\Lambda$ has een plotted against
$a(t)$ for different values of $\beta_{1}=~1,~\frac{2}{3}$
respectively and for $\alpha$ = ~1, $A=1/3$, $8\pi G$ =~1, $B$
=~1, $C$ =~1.\\

\end{figure}

\subsection{Model with $\Lambda\propto H^{2}$}

Choosing \begin{equation} \Lambda(t)=\beta_{2}H^{2} \end{equation}
where $\beta_{2}$ is a constant and proceeding as above, we
obtain the solutions for $\rho,~ a(t), ~\Lambda$ as,

\begin{equation}
\rho=\left(\frac{B}{1+A}+\frac{C}{a^{(3-\beta_{2})(1+A)(1+\alpha)}}\right)^{\frac{1}{1+\alpha}}
\end{equation}

\begin{equation}
a^{f_{1}f_{2}}~_{2}F_{1} [f_{2},f_{2},1+f_{2},-\frac{a^{f_{1}
B}}{C (1+A)}]=\sqrt{2 \pi G }\sqrt{3-\beta_{2}}~(1+A) C^{f_{2}}~t
\end{equation}

where  $f_{1}=(3-\beta_{2})(1+A)(1+\alpha)$ and
$f_{2}=\frac{1}{2(1+\alpha)}$

\begin{equation}
\Lambda=\frac{8 \pi G
\beta_{2}}{3-\beta_{2}}\left(\frac{B}{1+A}+\frac{C}{a^{(3-\beta_{2})(1+A)(1+\alpha)}}\right)^{\frac{1}{1+\alpha}}
\end{equation}

Here for $\beta_{2} < 3$ we can check the consistency of the
result by showing $p\simeq A\rho$ at small values of $a(t)$ and
$p=-\rho$ for large values of $a(t)$. But if we take $\beta_{2} >
3$ we get opposite results which contradict our previous notions
of the nature of the EOS (1.61). Again for $\beta_{2} = 3$, we get
only $\Lambda$CDM point ,i.e., we get a discontinuity. Therefore,
we restrict our choice for $\beta_{2}$
in this case to be $\beta_{2} < 3 $.\\

\begin{figure}

\includegraphics{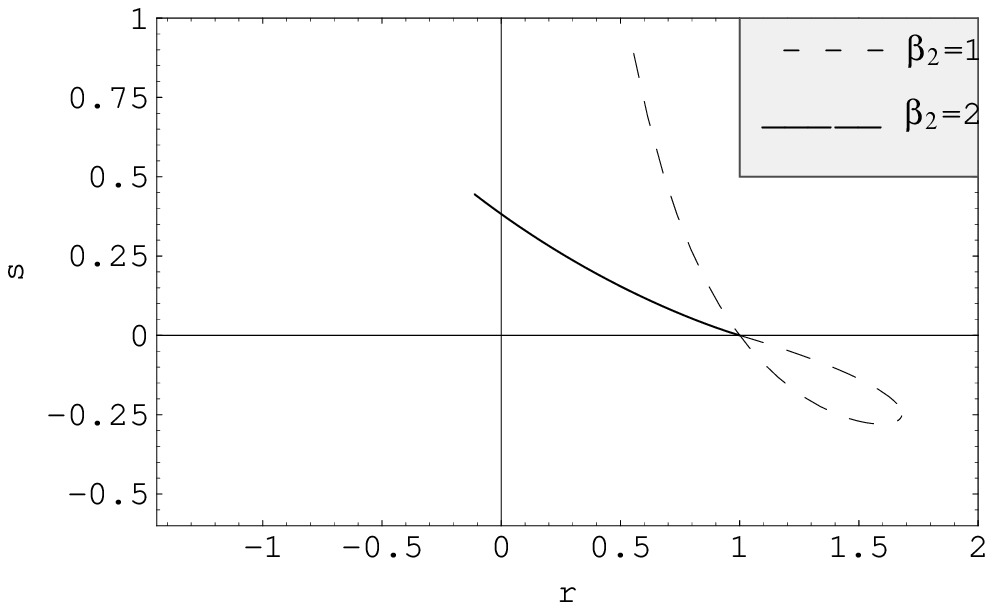}\\

\vspace{10mm}Fig 3.3: The variation of $s$ has been plotted
against $r$ for different values of $\beta_{2}=1,~2$ and for
$\alpha$ = ~1,
$A=1/3$, $8\pi G=1$.\\

\vspace{1in}

\includegraphics{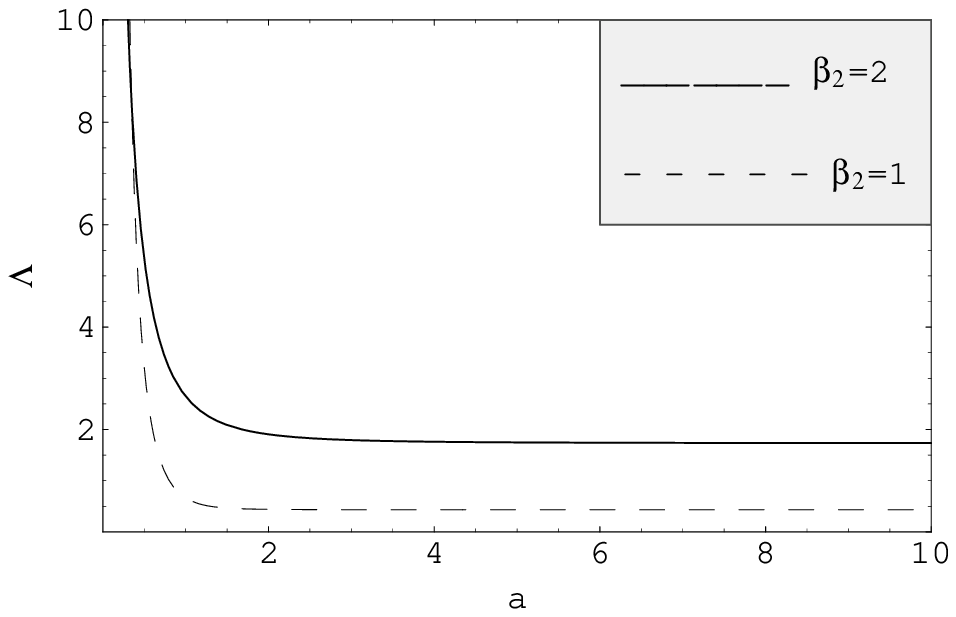}\\

\vspace{10mm}Fig 3.4: The variation of $\Lambda$ has been plotted
against $a(t)$ for different values of $\beta_{2}=1,~2$ and for
$\alpha$ = ~1, $A=1/3$, $8\pi G=1$.\\

\end{figure}

Computing the state-finder parameters given by equation (1.113),
we get the equations for $r$ and $s$ to be,

\begin{equation}
r=1+\frac{(3-\beta_{2})(1+y)[\{A(1+\alpha)-y\alpha\}(3-\beta_{2})-\beta_{2}]}{2}
\end{equation}

and

\begin{equation}
s=\frac{(3-\beta_{2})(1+y)][\{A(1+\alpha)-y\alpha\}(3-\beta_{2})-\beta_{2}]}{3\{(3-\beta_{2})y-\beta_{2}\}}
\end{equation}

(where $y=\frac{p}{\rho}$), which can still be resolved into a
single relation and can be plotted in the ${r,s}$ plane. Here
$q=\frac{1}{2}[(3-\beta_{2})y-(\beta_{2}-1)]$. Hence the Universe
will accelerate if
$q<0~\Rightarrow~y<\frac{\beta_{2}-1}{3-\beta_{2}}$. Again for the
present Universe
$q=-\frac{1}{2}~\Rightarrow~y=\frac{\beta_{2}-2}{3-\beta_{2}}$.
Assuming the present Universe to be dust dominated, i.e., $y=0$ we
get the best fit value for $\beta_{2}$ to be $2$. Taking
$A=\frac{1}{3},\alpha=1,8\pi G=1,\beta_{2}=2$ and $y=0$ (i.e.,
dust dominated present Universe) we get the present value to be
$r=1/3$, also the same values with $\beta_{2}=1$ gives the present
values to be $y=-\frac{1}{2},~r=\frac{5}{3}$ .This is shown in
figure 3.3 ($A=\frac{1}{3}, \alpha=1,\beta_{2}=1$ and $2, 8 \pi
G=1$), which explains the evolution of the Universe from radiation
era to $\Lambda$CDM model. Again  variation of $\Lambda$ against
time is shown in figure 3.4, where we can see
that $\Lambda$ decreases with time for whatever the value of $\beta_{2}$ be.\\

\subsection{Model with $\Lambda\propto \frac{\ddot{a}}{a}$}

Taking

\begin{equation}
\Lambda=\beta_{3}\frac{\ddot{a}}{a}
\end{equation}

(where $\beta_{3}$ is a constant), and proceeding as above we get
a relation for $\rho$ as,

\begin{equation}
\rho^{(\frac{2}{1+A}-\beta_{3})}\left(1+A-\frac{B}{\rho^{\alpha+1}}\right)^{(\frac{2}{(1+A)(1+\alpha)}-\beta_{3})}=\frac{C}{a^{2(3-\beta_{3})}}
\end{equation}

Unlike the previous two cases here we get a far more restricted
solution. Here the only choice of $\beta_{3}$ for which we get
the feasible solution satisfying  $p\simeq A\rho$ for small
values of $a(t)$ and $p\simeq -\rho$ for large values of $a(t)$ is

\begin{equation}
\beta_{3}<\frac{2}{(1+A)(1+\alpha)}~~~~~~\text
{or}~~~~~~ \beta_{3}>3
\end{equation}

Again since $q=-\frac{\ddot{a}}{a H^{2}}=-\frac{\Lambda}{\beta_{3}
H^{2}}=\frac{4\pi G(\rho+3p)}{(3-\beta_{3})H^{2}}=\frac{4\pi G
(\rho+3p)}{(3-\beta_{3})H^{2}}$, $\beta_{3}>3$ implies $q<0$
without even violating the energy-condition $\rho+3p\geq 0$.
Although $\beta_{3}<\frac{2}{(1+A)(1+\alpha)}$ causes the
acceleration of the Universe violating the energy-condition.
Taking $q=-\frac{1}{2}$ for the present epoch, we obtain
$y=\frac{(\beta_{3}-4)}{(6-\beta_{3})}$. Hence the present epoch
is dust filled if $\beta_{3}=4$ and thus giving the present value
of $r$ to be $-\frac{1}{7}$ for $A=\frac{1}{3},~\alpha=1,~8\pi
G=1$. On using relation (3.18), $\rho$ and therefore $a,~\Lambda$
cannot be expressed in an open form. We can rather derive a
solution for $\Lambda$ in terms of $p,~ \rho$ as,

\begin{equation}
\Lambda=\frac{4 \pi G \beta_{3}}{\beta_{3}-3}(\rho+3p)
\end{equation}

Using equations (1.113) we get the statefinder parameters as,

\begin{equation}
r=1-\frac{(1+y)(\beta_{3}-3)[\beta_{3}+(\beta_{3}+6)x]}{(\beta_{3}+\beta_{3}x-2)(\beta_{3}+\beta_{3}y-2)}
\end{equation}

and

\begin{equation}
s=\frac{2(1+y)(\beta_{3}-3)[\beta_{3}+(\beta_{3}+6)x]}{[\beta_{3}+(\beta_{3}+6)y][\beta_{3}+\beta_{3}x-2]}
\end{equation}

where $y=\frac{p}{\rho}$ and $x=\frac{\partial p}{\partial
\rho}$, i.e., $x=A(1+\alpha)-y\alpha$ [from equation (1.61)].\\

\begin{figure}

\includegraphics{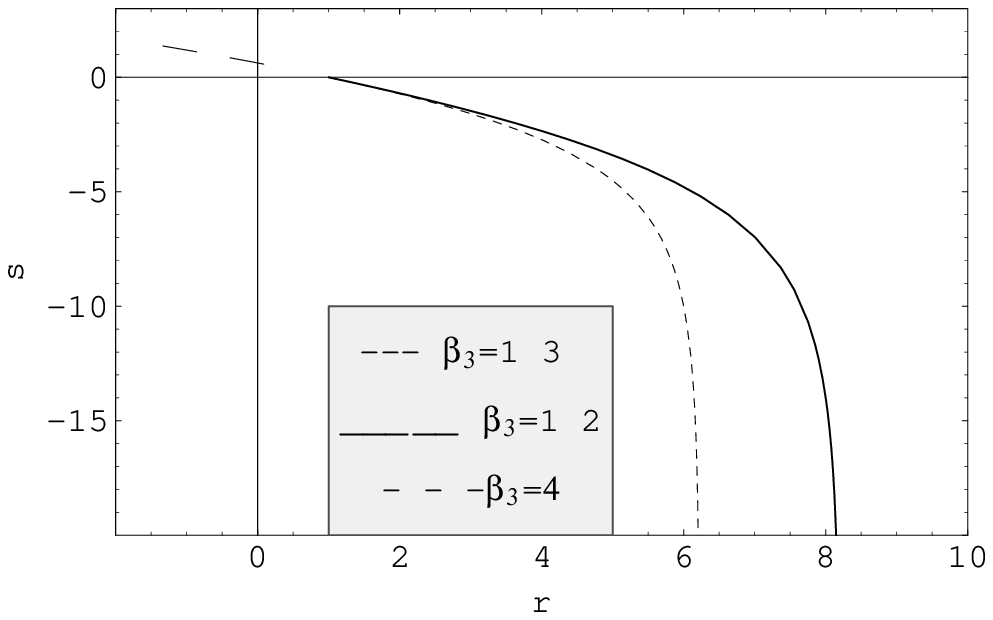}\\
\vspace{1mm} Fig 3.5: The variation of $s$ has been plotted
against $r$ for different values of $A=\frac{1}{3}, \alpha=1,
8\pi G=1, \beta_{3}=\frac{1}{2},4$ and
$\frac{1}{3}$\\
\vspace{5mm}

\end{figure}

Eliminating $y$ between the equations (3.20) and (3.21), we get a
single relation of $r$ and $s$, which can be represented
diagrammatically in the ${r,s}$ plane (figure 3.5). Here we have
taken $A=\frac{1}{3}, \alpha=1, 8\pi G=1,
\beta_{3}=\frac{1}{2},4$ and $\frac{1}{3}$, combining two cases.
Taking $\beta_{3}=\frac{1}{2},\frac{1}{3}$ we can explain the
evolution of the Universe starting from
$\frac{p}{\rho}=-\frac{1}{3}$ to $\Lambda$CDM model and
$\beta_{3}=4$ explains the evolution of the Universe starting
from radiation era to $y=-\frac{1}{3}$, as seen from the
expression for $q$. Considering the present epoch to be
dust-dominated, the present value of $r$ is given for
$\beta_{3}=4$ to be $-\frac{1}{7}$. As follows, the former two
cases cannot give the present value of $r$, as $y=0>-\frac{1}{3}$
for the present epoch. Here we have an infinite discontinuity at
$\frac{p}{\rho}=-\frac{1}{3}$, i.e., when $\rho+3p=0$. Also since
we do not get a closed from of $\rho$ here,
it is difficult to plot $\Lambda$ against the scale factor $a(t)$ .\\

\section{Models with $G$ and $\Lambda$ both variable}

Now we consider $G$ as well as $\Lambda$ to be variable. With
this the conservation law reads,
\begin{equation}
\dot{\rho}+3H(\rho+p)=0
\end{equation}
and
\begin{equation} \dot{\Lambda}+ 8 \pi \dot{G} \rho=0
\end{equation}
Now we study the various phases of the Universe represented by
these models.\\

Equation (3.22) together with equation (1.60) yield the solution
for $\rho$ as,

\begin{equation}
\rho=\left(\frac{B}{1+A}+\frac{C}{a^{3(1+A)(1+\alpha)}}\right)^{\frac{1}{1+\alpha}}
\end{equation}

where $C$ is an arbitrary constant. This result is consistent
with the results already obtained [Gorini etal, 2003].\\

\subsection{Model with $\Lambda\propto \rho$}

Here we consider

\begin{equation}
\Lambda= \gamma_{1}~ \rho
\end{equation}

where $\gamma_{1}$ is a constant.\\

Equation (3.22), (3.23) and (3.25) give,

\begin{equation}
G=C_{1}-\frac{\gamma_{1}}{8\pi}\log\rho
\end{equation}

where $C_{1}$ is a constant and $\rho$ is given by
equation (3.24).\\

Using equations (1.113), (3.1), (3.23) and (3.26), we get,

\begin{eqnarray}
\begin{array} {ccc}
G=C_{1}+\frac{\gamma_{1}(1+\alpha)}{8\pi}\log(\frac{B}{A-y})\\\\\
r=1+\frac{9(1+y)[8\pi G
\{A(1+\alpha)-y\alpha\}-\gamma_{1}(1+y)]}{2(8\pi
G+\gamma_{1})}\\\\\
s=\frac{(1+y)[8\pi G
\{A(1+\alpha)-y\alpha\}-\gamma_{1}(1+y)]}{(8\pi G y-\gamma_{1})}\\\\
\end{array}
\end{eqnarray}

where $y=\frac{p}{\rho}$.\\

\begin{figure}

\includegraphics{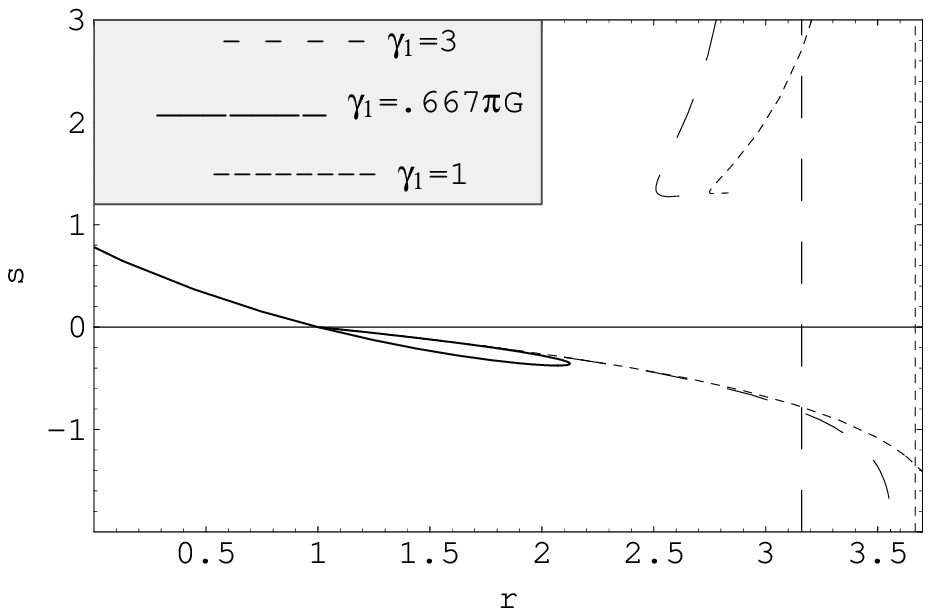}\\
\vspace{1mm} Fig 3.6: The variation of $s$ has been plotted
against $r$ for different values of $\gamma_{1}=1, 3$ and $3.5$
and $A=\frac{1}{3}, \alpha=1, B=1, C_{1}=1$ \vspace{4mm}

\end{figure}

Equation (3.27) cannot be resolved to get a single relation
between $r$ and $s$, rather we obtain a parametric relation
between the same with $y=\frac{p}{\rho}$ as the parameter. This
can be represented diagrammatically in the ${r,s}$ plane, which is
shown in figure 3.6 taking $\gamma_{1}=1, 3$ and $3.5$ and
$A=\frac{1}{3}, \alpha=1, B=1, C_{1}=1$. Now $q=\frac{4\pi G
(1+3y)-\gamma_{1}}{8\pi G+\gamma_{1}}$. Taking into account that
$q=-\frac{1}{2}$ for the present epoch, we get
$y=\frac{\gamma_{1}}{8\pi G}-\frac{2}{3}$. Therefore, for the
present dust-dominated era $y=0$ and $\gamma_{1}=\frac{16 \pi
G}{3}$. hence for the This models represents the Universe starting
from the radiation era to $\Lambda$CDM model. Again figure 3.7
represents the variation of $\Lambda$ against the scale factor
$a(t)$ with $\gamma_{1}=1, 3, 3.5$ and figure 3.8 represents the
variation $G$ against the scale factor $a(t)$. These figures show
that for this particular phenomenological model of $\Lambda$, $G$
starting from very low initial value increases largely and becomes
constant after a certain period of time, whereas $\Lambda$
starting from a very large decreases largely to
reach a very low value and becomes constant.\\

\begin{figure}
\includegraphics{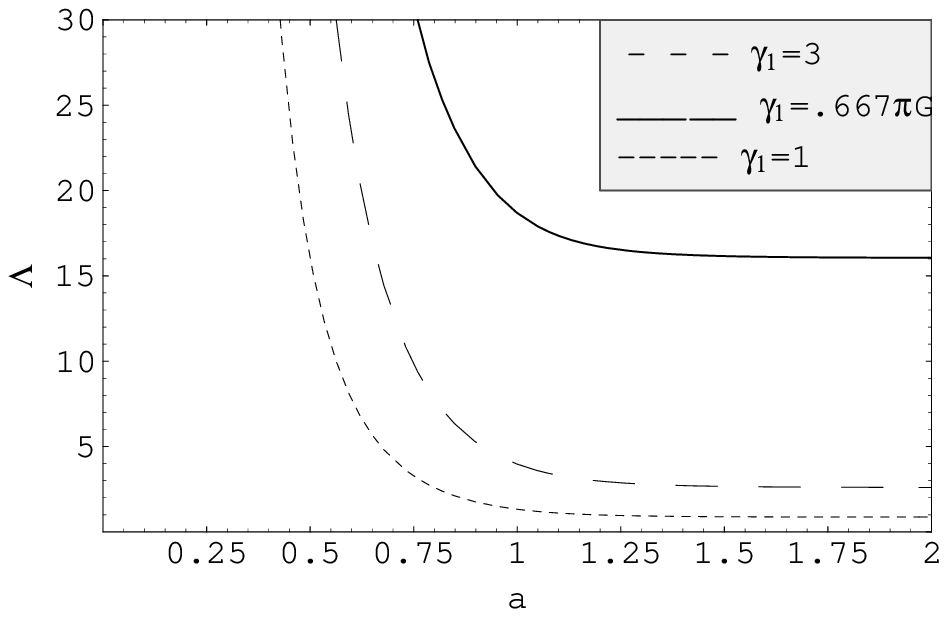}\\

\vspace{10mm} Fig 3.7: The variation of $\Lambda$ is plotted
against $a(t)$ for different values of $\gamma_{1}=1, 3, 3.5$ and
for $\alpha$ = ~1, $A=1/3$, $C_{1}=1$.\\

\vspace{1in}

\includegraphics{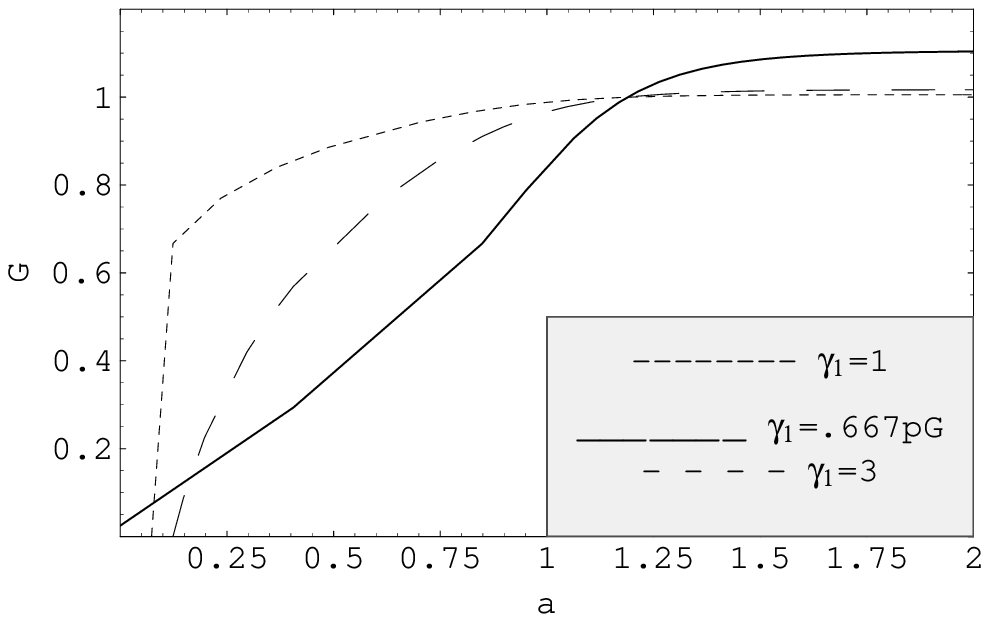}\\

\vspace{10mm} Fig 3.8: The variation of $G$ is plotted against
$a(t)$ for different values of $\gamma_{1}=1, 3, 3.5$ and for
$\alpha$ = ~1, $A~=~1/3$, $C_{1}~=~1, B~=~1,~C~=~1$.\\

\end{figure}

\subsection{Model with $\Lambda\propto H^{2}$}

We consider

\begin{equation}
\Lambda=\gamma_{2} H^{2}
\end{equation}

Proceeding as above we get,
\begin{equation}
\Lambda=8 \pi G
\frac{\gamma_{2}}{3-\gamma_{2}}\rho
\end{equation}

where $\gamma_{2}$ is a constant.\\

Solving equation (3.22), (3.23) and (3.29) we get,

\begin{equation}
G=\frac{C_{2}}{\rho^{\frac{\gamma_{2}}{3}}}
\end{equation}

where $C_{2}$ is a constant.\\

\begin{figure}
\includegraphics{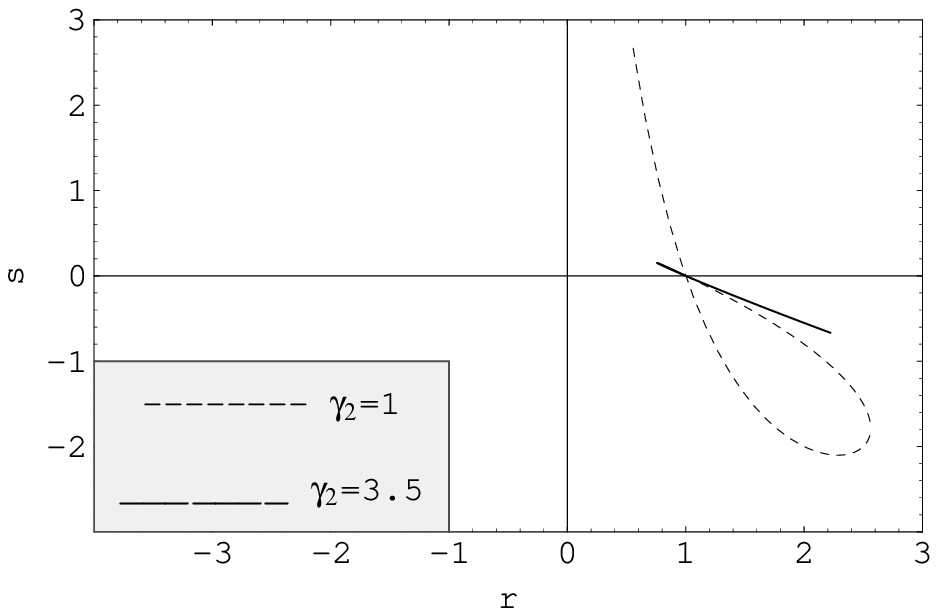}

\vspace{1mm}Fig 3.9: The variation of $s$ is plotted against $r$
for different values of $\gamma_{2}=1$ and $3.5$ and
$A=\frac{1}{3}, \alpha=1, B=1, C_{2}=1,~C~=~1$. \\
\vspace{4mm}

\end{figure}

Using equations (1.113), (3.1) and (3.23), we find the
state-finder parameters as,

\begin{equation}
r=1+\frac{(1+y)(3-\gamma_{2})[3\{A(1+\alpha)-y\alpha\}-(1+y)\gamma_{2}]}{2}
\end{equation}

and

\begin{equation}
s=\frac{(1+y)(3-\gamma_{2})[3\{A(1+\alpha)-y\alpha\}-(1+y)\gamma_{2}]}{(3-\gamma_{2})y-\gamma_{2}}
\end{equation}

where $y=\frac{p}{\rho}$.\\

Now $q=\frac{1}{2}[(3-\gamma_{2})y-(\gamma_{2}-1)]$. These
equations can further be resolved into a single relation of $r$
and $s$, which can be plotted diagrammatically in the ${r,s}$
plane. Here we get a discontinuity at $\gamma_{2}=3$. We have
plotted these values in the ${r,s}$ plane taking $\gamma_{2}=1$
and $3.5$ in figure 3.9 ($A=\frac{1}{3}, \alpha=1$). This case
explains the present acceleration of the Universe, starting from
radiation era to $\Lambda$CDM model.\\

\begin{figure}

\includegraphics{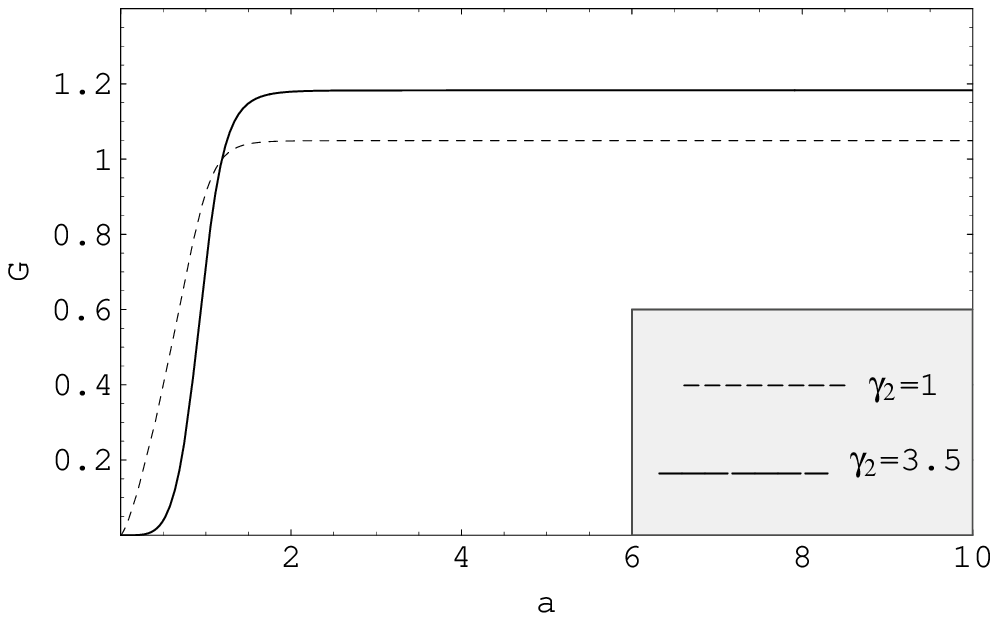}\\

\vspace{10mm} Fig 3.10: The variation of $G$ is plotted against
$a(t)$ for different values of $\gamma_{1}=1, 3.5$ and for
$\alpha$ = ~1, $A=1/3$, $C_{1}=1,~C~=1,~B~=~1$.\\

\vspace{1in}

\includegraphics{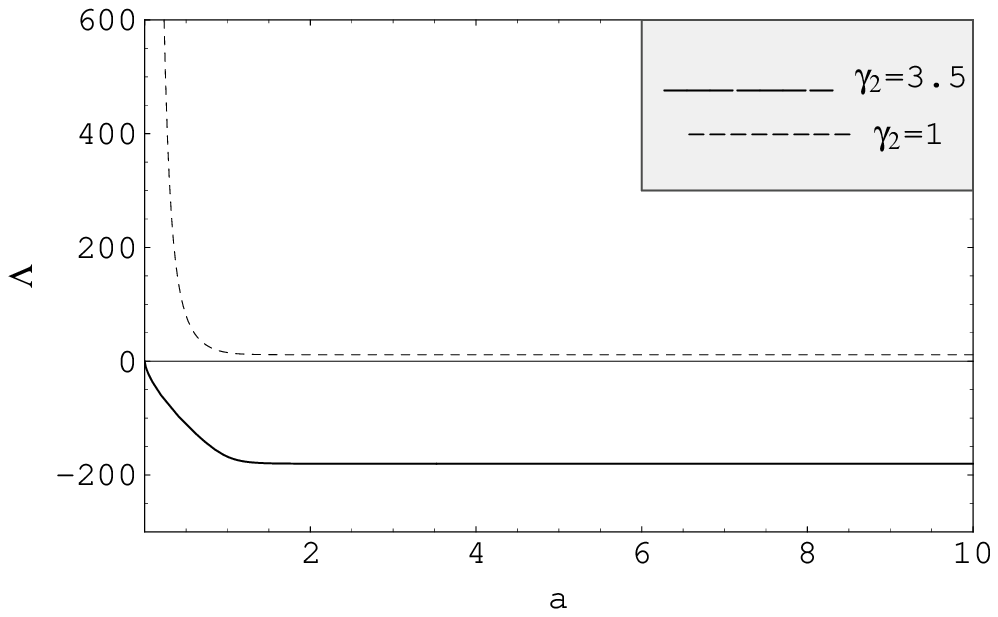}\\

\vspace{10mm}
 Fig 3.11: The variation of $\Lambda$ is plotted
against $a(t)$ for different values of $\gamma_{1}=1, 3, 3.5$ and
for $\alpha$ = ~1, $A~=~1/3$, $C_{2}~=~1, B~=~1,~C~=~1$.\\

\end{figure}

Also figures 3.10 and 3.11 show respectively the variation of $G$
and $\Lambda$ against the scale factor for the same values of the
constants. Here also like the previous case $G$ starting from a
very low initial value increases largely and then continues to be
constant near unity. On the other hand $\Lambda$ starting from a
large value decreases largely and continues to be constant after
a certain period of time.\\

\subsection{Model with $\Lambda\propto \frac{\ddot{a}}{a}$}

Here we consider

\begin{equation}
\Lambda=\gamma_{3} \frac{\ddot{a}}{a}
\end{equation}

where $\gamma_{3}$ is a constant.\\

Using equation (3.33) in equations (3.1) and (3.2), we get,

\begin{equation}
\Lambda=-\frac{4 \pi G \gamma_{3}}{3-\gamma_{3}}(\rho+3p)
\end{equation}

Also, $G$ can be solved to be,

\begin{equation}
G=C_{3}[\rho^{\frac{1+3A}{2-\gamma_{3}(1+A)}}\{2-\gamma_{3}(1+A)+\frac{B
\gamma_{3}}{\rho^{\alpha+1}}\}^{\{-\frac{3\alpha}{\gamma_{3}(1+\alpha)}+\frac{1+3A}{(1+\alpha)(2-\gamma_{3}(1+A))}\}}]^{\frac{\gamma_{3}}{3}}
\end{equation}

Using equations (1.113), (3.1), (3.22), (3.23), we find,

\begin{equation}
r=1+\frac{(1+y)(3-\gamma_{3})[6\{A(1+\alpha)-y\alpha\}+\gamma_{3}(1+y)]}{[2-\gamma_{3}(1+y)]^{2}}
\end{equation}

and

\begin{equation}
s=\frac{2(1+y)(3-\gamma_{3})[6\{A(1+\alpha)-y\alpha\}+\gamma_{3}(1+y)]}{3[2-\gamma_{3}(1+y)][\gamma_{3}+(\gamma_{3}+6)y}
\end{equation}

where $y=\frac{p}{\rho}$ and $C_{3}$ is a constant. Equations
(3.36) and (3.37) can further be resolved to get one single
relation between $r$ and $s$ and plotted diagrammatically taking
$\gamma_{3}=2$ and $3.5$ (figure 3.12). Since deceleration
parameter $q=-\frac{\ddot{a}}{a
H^{2}}=-\frac{\lambda}{\gamma_{3}H^{2}}=\frac{4\pi
G}{(3-\gamma_{3}) H^{2}}$, is negative in the present epoch, we
get $3-\gamma_{3}<0$, i.e., $\gamma_{3}>3$. Also for
$\gamma_{3}=3$ we get discontinuity. Both the models represent the
phases of the Universe starting from radiation era to $\Lambda$CDM
model. Again $G$ and $\Lambda$ can be plotted against $a$ (figures
3.13 and 3.14 respectively) . Unlike the previous cases this model
an opposite nature of $G$ and $\Lambda$, as $G$
decreases with time and $\Lambda$ increases with time.\\

\begin{figure}

\includegraphics{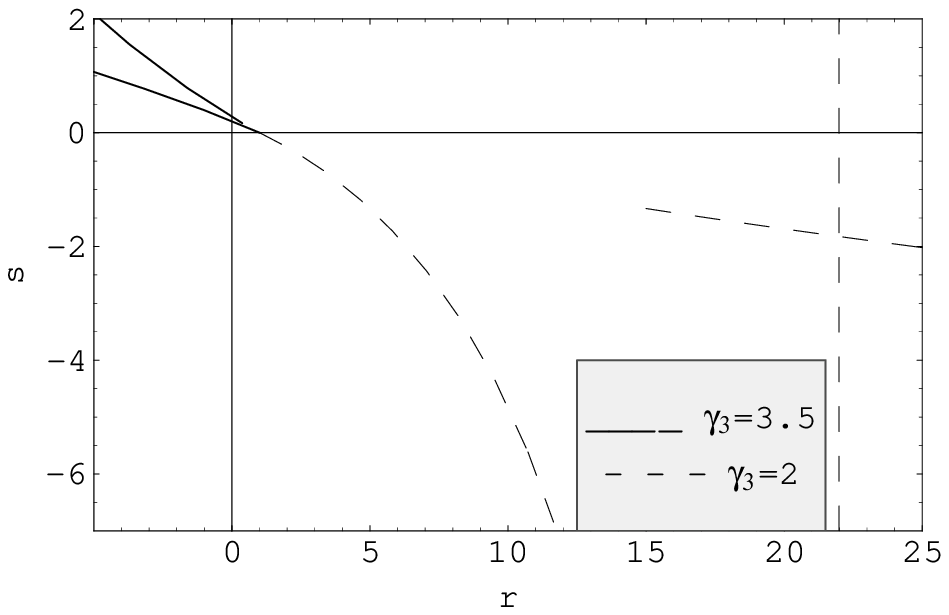}

\vspace{1mm}Fig 3.12: The variation of $s$ is plotted against $r$
for different values of $\gamma_{2}=2$ and $3.5$ and
$A=\frac{1}{3}, \alpha=1, B=1, C_{3}=1,~C~=~1$. \\

\end{figure}

\begin{figure}
\includegraphics{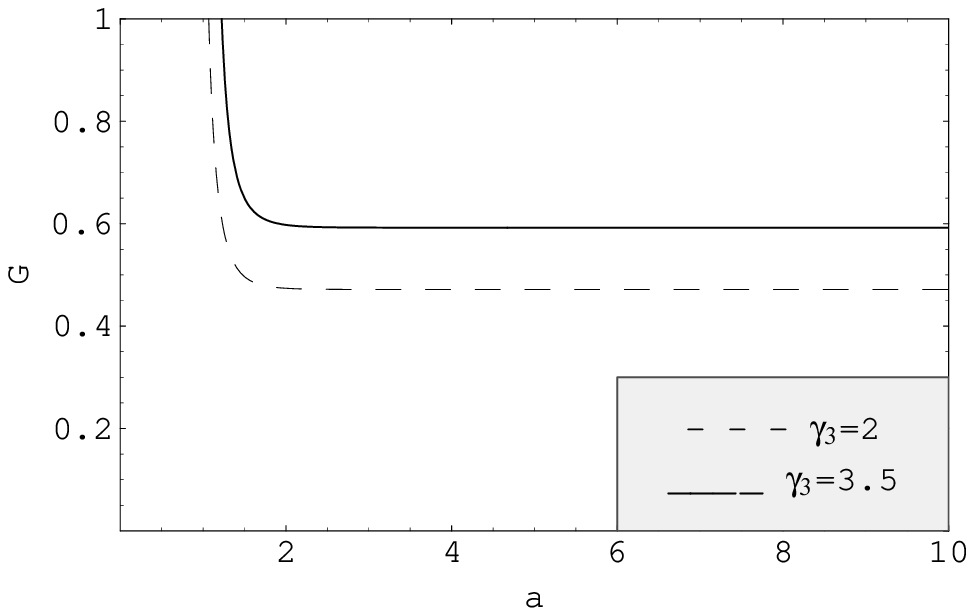}\\

\vspace{10mm} Fig 3.13: The variation of $G$ is plotted against
$a(t)$ for different values of $\gamma_{1}=2, 3.5$ and for
$\alpha$ = ~1, $A=1/3$, $C_{3}=1,~C~=1,~B~=~1$.\\

\vspace{1in}

\includegraphics{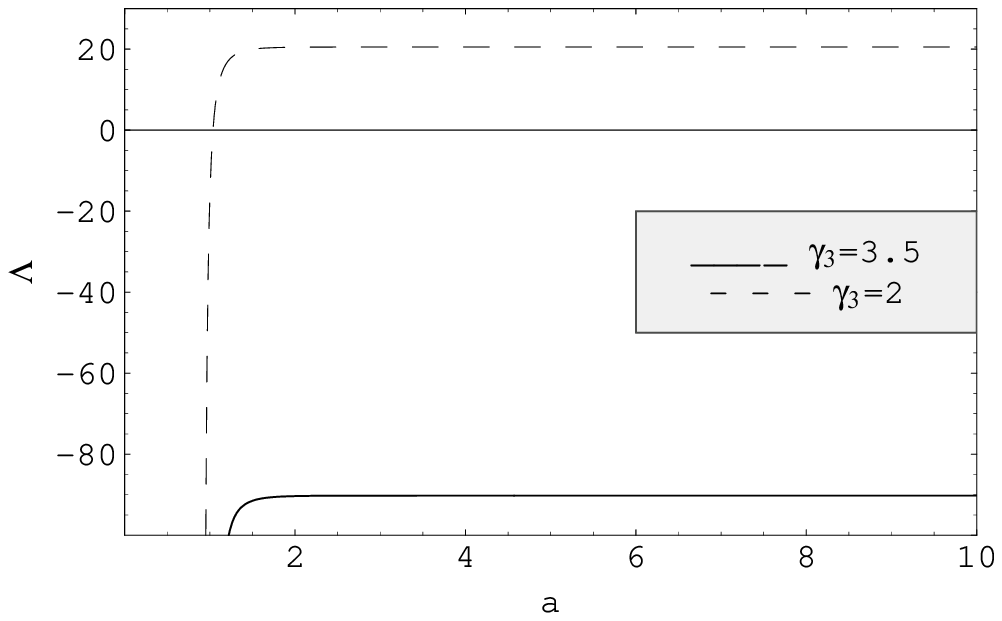}\\

\vspace{10mm} Fig 3.14: The variation of $\Lambda$ is plotted
against $a(t)$ for different values of $\gamma_{1}=2, 3, 3.5$ and
for $\alpha$ = ~1, $A~=~1/3$, $C_{3}~=~1, B~=~1,~C~=~1$.\\

\end{figure}

\section{Discussion}

Here we have considered three phenomenological models of
$\Lambda$, with or without keeping $G$ to be constant. Keeping
$G$ constant we always get accelerated expansion of the Universe.
For the first case, i.e., $\Lambda\propto \rho$ or more precisely,
$\Lambda=\beta_{1}\rho$, for particular choices of the constants
we get that the dark energy responsible for the present
acceleration is nothing but $\Lambda$. Also the density parameter
of the Universe for this case is given by,
$\Omega_{m}^{\beta_{1}}=\frac{8\pi G \rho}{3H^{2}}=\frac{8\pi
G}{8 \pi G +\beta_{1}}$ and the vacuum density parameter is
$\Omega_{\Lambda}^{\beta_{1}}=\frac{\Lambda}{3H^{2}}=\frac{\beta_{1}}{8\pi
G+\beta_{1}}$, so that
$\Omega_{total}=\Omega_{m}+\Omega_{\Lambda}=\Omega_{m}^{\beta_{1}}+\Omega_{\Lambda}^{\beta_{1}}=1$.
Also for $\Lambda\propto H^{2}$, i.e., $\Lambda=\beta_{2}~H^{2}$,
the density parameter and vacuum density parameter are given by,
$\Omega_{m}^{\beta_{2}}=\frac{3-\beta_{2}}{3}$ and
$\Omega_{\Lambda}^{\beta_{1}}=\frac {\beta_{2}}{3}$ respectively,
so that
$\Omega_{total}=\Omega_{m}^{\beta_{2}}+\Omega_{\Lambda}^{\beta_{1}}=1$.
Again for $\Lambda\propto \frac{\ddot{a}}{a}$ or
$\Lambda=\beta_{3}~\frac{\ddot{a}}{a}$, we have the corresponding
parameters as,
$\Omega_{m}^{\beta_{3}}=\frac{2(3-\beta_{3})}{3(2-\beta_{3}-\beta_{3}\frac{p}{\rho})}$,
$\Omega_{\Lambda}^{\beta_{3}}=\frac{-\beta_{3}(1+3\frac{p}{\rho})}{3(2-\beta_{3}-\beta_{3}\frac{p}{\rho})}$
and $\Omega_{total}=1$. Now
$\Omega_{total}=\Omega_{m}^{\beta_{3}}+\Omega_{\Lambda}^{\beta_{3}}=1$
for all the models. Also we can compare these models by taking,
$\Omega_{m}^{\beta_{1}}=\Omega_{m}^{\beta_{2}}$, so that
$\beta_{2}=\frac{3\beta_{1}}{(8\pi G +\beta_{1})}$. Now we would
like to take into account the present values of the density
parameter and vacuum parameter obtained by the recent
measurements. Considering $\Omega_{m 0}=0.33\pm.035$, we
calculate the present values of the proportional constants to be
$1.7397K \leq {\beta_{1}}^{0}\leq 2.3898 K,~1.905\leq
{\beta_{2}}^{0}\leq 2.115 $ and $3.7937 \leq {\beta_{3}}^{0}\leq
4.2099$, where $K= 8\pi G_{0}$ and $G_{0}$ is the present value of
the gravitational constant. Thus we get the value of
${\beta_{3}}^{0}$ to be lesser than the previous works. Again
considering $G$ to be time-dependent, we get the same values of
the parameters as that with $G$ constant, i.e., the ranges of
${\gamma_{1}}^{0}, {\gamma_{2}}^{0}, {\gamma_{3}}^{0}$ are same
as that of ${\beta_{1}}^{0}, {\beta_{2}}^{0}, {\beta_{3}}^{0}$
respectively. Here also we get cosmic acceleration and the nature
of variation $G$ and $\Lambda$ as well. We get two different
cases regarding the variation of $G$ and $\Lambda$. For the first
two cases we see that $G$ increases and $\Lambda$ decreases with
time, whereas for the third case $G$ decreases and $\Lambda$
increases with time. In all the cases the values become constant
after a certain period of time, i.e.,the present day values of
$G$ and $\Lambda$ are constants. Thus these models with the
phenomenological laws give us some interesting features of the
cosmic acceleration and some modified values of the parameters.
Also we get the natures of the Cosmological Constant and the
Gravitational Constant over the total age of the Universe. We can
also make use of the statefinder parameters to show the evolution
of the Universe starting from radiation era to
$\Lambda$CDM model.\\

%% file: chap4.tex
\large \baselineskip .85cm
\chapter{Generalized Cosmic Chaplygin Gas Model}
\label{chap4}\markright{\it CHAPTER~\ref{chap4}. Generalized
Cosmic Chaplygin Gas Model}

\section{Prelude}

Recently developed Generalized Cosmic Chaplygin gas (GCCG) is
studied as an unified model of dark matter and dark energy. To
explain the recent accelerating phase, the Universe is assumed to
have a mixture of radiation and GCCG. The mixture is considered
for without or with interaction. Solutions are obtained for
various choices of the parameters and trajectories in the plane
of the statefinder parameters and presented graphically.\\

In 2003, Gonz$\acute{a}$lez-Diaz have introduced the generalized
cosmic Chaplygin gas (GCCG) model in such a way that the resulting
models can be made stable and free from unphysical behaviours even
when the vacuum fluid satisfies the phantom energy condition. The
EOS of this model is

\begin{equation}
p=-{\rho}^{-\alpha}\left[ C + (\rho^{1+\alpha}-C)^{-\omega}\right]
\end{equation}

where $C=\frac{A}{1+\omega}-1$ with $A$ a constant which can take
on both positive and negative values and $-l<\omega<0$, $l$ being
a positive definite constant which can take on values larger than
unity.\\

The EOS reduces to that of current Chaplygin unified models for
dark matter and dark energy in the limit $\omega\rightarrow 0$
and satisfies the conditions: (i) it becomes a de Sitter fluid at
late time and when $\omega=-1$, (ii) it reduces to $p=w \rho$ in
the limit that the Chaplygin parameter $A\rightarrow 0$, (iii) it
also reduces to the EOS of current Chaplygin unified dark matter
models at high energy density and (iv) the evolution of density
perturbations derived from the chosen EOS becomes free from the
pathological behaviour of the matter power spectrum for
physically reasonable values of the involved parameters at late
time. This EOS shows dust era in the past and $\Lambda$CDM in the
future.\\

In this chapter, we consider the Universe is filled with the
mixture of radiation and GCCG. We also perform a statefinder
diagnostic to this model without and with interaction in
different cases.\\

\section{GCCG in presence of radiation}

The metric of a spatially flat isotropic and homogeneous Universe
in FRW model is equation (1.7). The Einstein field equations are
(choosing $8\pi G=c=1$)

\begin{equation}
3\frac{\dot{a}^{2}}{a^{2}}=\rho_{tot}
\end{equation}
and
\begin{equation}
6\frac{\ddot{a}}{a}=-(\rho_{tot}+3p_{tot})
\end{equation}

The energy conservation equation ($T_{\mu;\nu}^{\nu}=0$) is
\begin{equation}
\dot{\rho}_{tot}+3\frac{\dot{a}}{a}(\rho_{tot}+p_{tot})=0
\end{equation}

where, $\rho_{tot}$ and $p_{tot}$ are the total energy density and
the pressure of the Universe, given by,

\begin{equation}
\rho_{tot}=\rho+\rho_{r}
\end{equation}

and

\begin{equation}
p_{tot}=p+p_{r}
\end{equation}

with $\rho$ and $p$ are respectively the energy density and
pressure due to the GCCG satisfying the EOS (4.1) and $\rho_{r}$
and $p_{r}$ are the energy density and the pressure corresponding
to the radiation fluid with EOS,

\begin{equation}
p_{r}=\gamma \rho_{r}
\end{equation}

where $\gamma=\frac{1}{3}$.\\

Since GCCG can explain the evolution of the Universe starting from
dust era to $\Lambda$CDM, considering the mixture of GCCG with
radiation would make it possible to explain the evolution of the
Universe from radiation to $\Lambda$CDM.\\

\subsection{Non-interacting model}

In this case GCCG and the radiation fluid are conserved
separately. Conservation equation (4.4) yields,

\begin{equation}
\dot{\rho}+3\frac{\dot{a}}{a}(\rho+p)=0
\end{equation}

and

\begin{equation}
\dot{\rho_{r}}+3\frac{\dot{a}}{a}(\rho_{r}+p_{r})=0
\end{equation}

From equations (4.1), (4.7), (4.8), (4.9) we have

\begin{equation}
\rho=\left[C +
\left(1+\frac{B}{a^{3(1+\alpha)(1+\omega)}}\right)^{\frac{1}{1+\omega}}\right]^{\frac{1}{1+\alpha}}
\end{equation}

and

\begin{equation}
\rho_{r}=\rho_{0}~ a^{-3(1+\gamma)}
\end{equation}

For the two component fluids, statefinder parameters (1.113) takes
the following forms:

\begin{equation}
r=1+\frac{9}{2(\rho+\rho_{r})}\left[\frac{\partial
p}{\partial\rho}(\rho+p)+\frac{\partial
p_{r}}{\partial\rho_{r}}(\rho_{r}+p_{r})\right]
\end{equation}

and

\begin{equation}
s=\frac{1}{(p+p_{r})}\left[\frac{\partial
p}{\partial\rho}(\rho+p)+\frac{\partial
p_{r}}{\partial\rho_{r}}(\rho_{r}+p_{r})\right]
\end{equation}

Also the deceleration parameter $q$ has the form:

\begin{equation}
q=-\frac{\ddot{a}}{aH^{2}}=\frac{1}{2}+\frac{3}{2}\left(\frac{p+p_{r}}{\rho+\rho_{r}}\right)
\end{equation}

Now substituting $u=\rho^{1+\alpha}, ~ y=\frac{\rho_{r}}{\rho}$,
equation (4.12) and (4.13) can be written as,

\begin{eqnarray*}
r=1+\frac{9}{2(1+y)}\left[ \left(1 -
\frac{C}{u}-\frac{(u-C)^{-\omega}}{u} \right)\{ \frac{\alpha
C}{u}+\frac{\alpha}{u}(u-C)^{-\omega}\right.
\end{eqnarray*}
\begin{equation}
\left.+\omega(1+\alpha)(u-C)^{-\omega-1}\}+\gamma (1+\gamma) y
\right]
\end{equation}

and

\begin{equation}
s=\frac{2(r-1)(1+y)}{9\left[ \gamma y -
\frac{C}{u}-\frac{(u-C)^{-\omega}}{u}\right]}
\end{equation}

Normalizing the parameters we have shown the graphical
representation of the $\{r, s\}$ parameters in figure 4.1.

\begin{figure}

\includegraphics{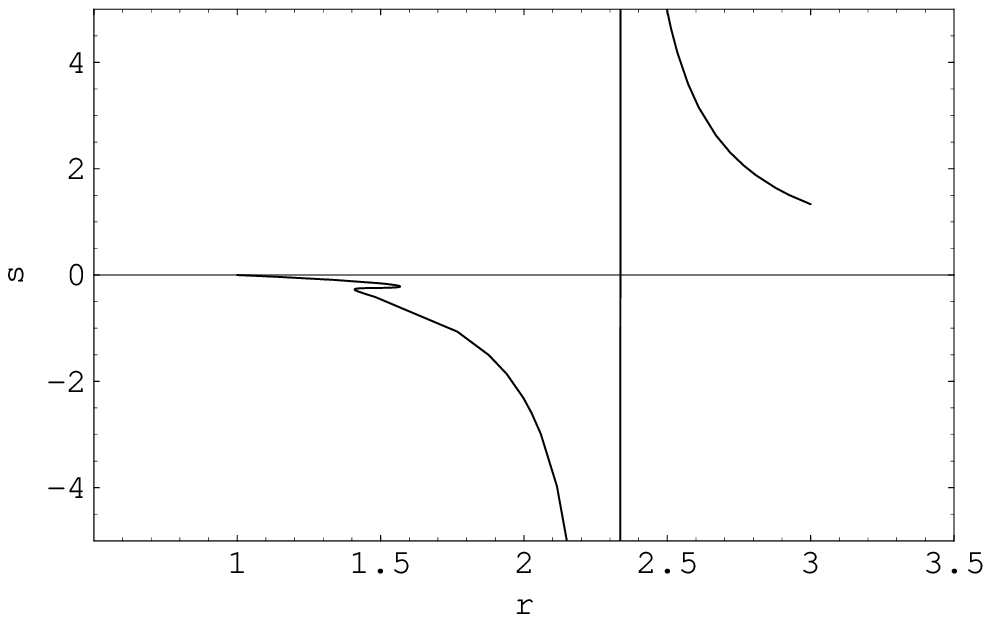}\\
\vspace{1mm}Fig 4.1: The variation of $s$ is plotted against $r$
for $ C=1,
B=1, \alpha=1, \omega=-2, \rho_{0}=1$.\\

\end{figure}

\subsection{Interacting Model}

We consider the GCCG interacting with radiation fluid through an
energy exchange between them. The equations of motion can be
written as,

\begin{equation}
\dot{\rho}+3\frac{\dot{a}}{a}(\rho+p)=-3H\delta
\end{equation}

and

\begin{equation}
\dot{\rho_{r}}+3\frac{\dot{a}}{a}(\rho_{r}+p_{r})=3H\delta
\end{equation}

where $\delta$ is a coupling function.\\

Let us choose,

\begin{equation}
\delta=\epsilon
\frac{(\rho^{1+\alpha}-C)^{-\omega}}{\rho^{\alpha}}
\end{equation}

Now equation (4.17) together with equation (4.1) gives,

\begin{equation}
\rho=\left[C + \left(1-\epsilon+ B
a^{3-(1+\alpha)(1+\omega)}\right)^{\frac{1}{1+\omega}}\right]^{\frac{1}{1+\alpha}}
\end{equation}

Also equations (4.7), (4.18) and (4.20) give

\begin{equation}
\rho_{r}=\rho_{0}~ a^{-3(1+\gamma)}+ 3~ \epsilon ~a
^{-3(1+\gamma)}I
\end{equation}

with

\begin{equation}
I=-\frac{1}{3 B (1+\alpha)}\int
\frac{dx}{(C+x)^{\frac{\alpha}{(1+\alpha)}}}\left\{\frac{x^{1+\omega}+\epsilon-1}{B}\right\}^{-\frac{1+\gamma}{(1+\omega)(1+\alpha)}-1}
\end{equation}

and

\begin{equation}
x=\left[ 1-\epsilon + B a^{-3(1+\omega)(1+\alpha)}
\right]^{\frac{1}{1+\omega}}
\end{equation}

From (4.20), we see that if $\epsilon=0$, i.e., $\delta=0$, then
the expression (4.20) reduces to the expression (4.10).\\

Now for the two component interacting fluids with equations of
motion (4.17) and (4.18), the $r, s$ parameters read:

\begin{equation}
r=1+\frac{9}{2(\rho+\rho_{r})}\left[\frac{\partial
p}{\partial\rho}(\rho+p+\delta)+\frac{\partial
p_{r}}{\partial\rho_{r}}(\rho_{r}+p_{r}-\delta)\right]
\end{equation}

and

\begin{equation}
s=\frac{2(r-1)(\rho+\rho_{r})}{9(p+p_{r})}
\end{equation}

Also the deceleration parameter $q$ has the form:

\begin{equation}
q=-\frac{1}{2}\left(1+3\frac{p+p_{r}}{\rho+\rho_{r}}\right)
\end{equation}

Now substituting $u=\rho^{1+\alpha}, ~ y=\frac{\rho_{r}}{\rho}$,
equation (4.12) and (4.13) can be written as,

\begin{equation}
r=1+\frac{9}{2(1+y)}\left[\frac{\partial p}{\partial
\rho}\left(1+\frac{p}{\rho}+\frac{\delta}{\rho}\right)+\gamma\left\{(1+\gamma)y-\frac{\delta}{\rho}\right\}
\right]
\end{equation}

and

\begin{equation}
s=\frac{2(r-1)(1+y)}{9\left( \frac{p}{\rho}+\gamma y \right)}
\end{equation}

where,
$$u=\left[C + \left(1-\epsilon+ B
a^{3-(1+\alpha)(1+\omega)}\right)^{\frac{1}{1+\omega}}\right] $$
$$ y=\frac{\rho_{0}}{\rho}~ a^{-3(1+\gamma)}+ 3~\frac{ \epsilon}{\rho} ~a
^{-3(1+\gamma)}I$$
$$\frac{p}{\rho}=-\frac{1}{u} \{C+(u-C)^{-\omega}\} $$
$$\frac{\delta}{\rho}=\epsilon \frac{(u-C)^{-\omega}}{u} $$
and $$ \frac{\partial p}{\partial \rho}=\frac{\alpha
C}{u}+\frac{\alpha}{u}(u-C)^{-\omega}+\omega(1+\alpha)(u-C)^{-\omega-1}$$

Now we find the exact solution for the ${r, s}$ parameters for
the following particular choices of $\omega$:

(i) If $-\frac{(1+\gamma)}{(1+\omega)(1+\alpha)}-1=0$, i.e.,
$\omega=\frac{-2-\gamma-\alpha}{1+\alpha}$, equation (4.21) can be
written as

\begin{equation}
\rho_{r}=\rho_{0}~a^{-3(1+\gamma)}-\frac{\epsilon}{B}a^{-3(1+\gamma)}
\rho
\end{equation}

as $I=-\frac{1}{3B}(c+x)^{\frac{1}{1+\alpha}}$\\
Normalizing the parameters, the corresponding statefinder
parameters are given in figure 4.2.

\begin{figure}

\includegraphics{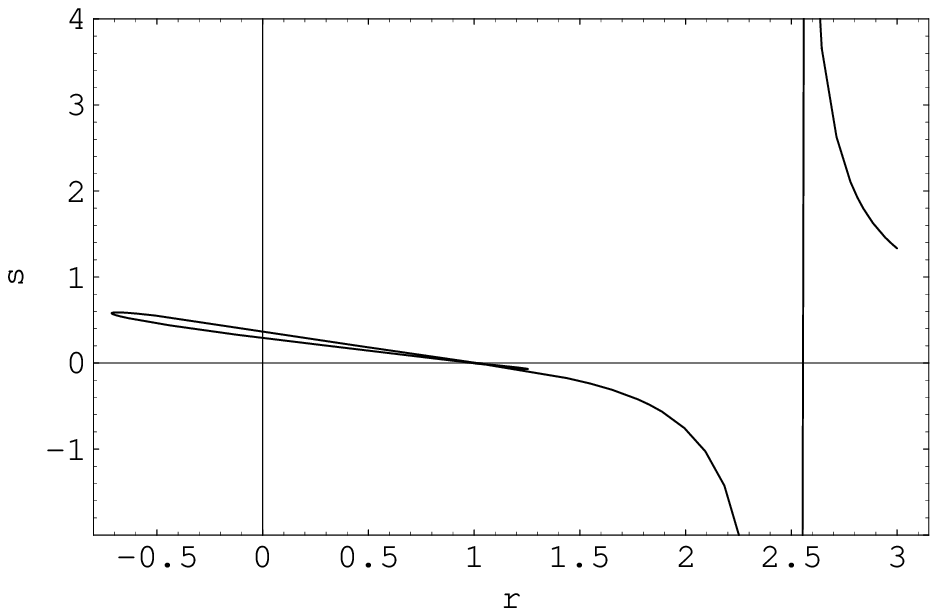}\\

\vspace{1mm} Fig 4.2: The variation of $s$ is plotted against $r$
for $ C=C_{0}=B=1,
\alpha=1, \omega=-2, \rho_{0}=1, \epsilon=\frac{1}{2}$.\\

\end{figure}

(ii) If $-\frac{(1+\gamma)}{(1+\omega)(1+\alpha)}-1=1$, i.e.,
$\omega=\frac{-3-\gamma-2\alpha}{2(1+\alpha)}$, equation (4.21)
can be written as

\begin{equation}
\rho_{r}=\rho_{0}~a^{-3(1+\gamma)}-\frac{\epsilon(\epsilon-1)}{B^{2}}
a^{-3(1+\gamma)}-\frac{\epsilon
a^{-3(1+\gamma)}}{B^{2}(1+\alpha)(2+\omega)
C^{\frac{\alpha}{1+\alpha}}} x^{2+\omega} ~_{2}F_{1}[2+\omega,
\frac{\alpha}{1+\alpha}, 3+\omega, -\frac{x}{C}]
\end{equation}

Normalizing the parameters, the corresponding statefinder
parameters are given in figure 4.3.

\begin{figure}

\includegraphics{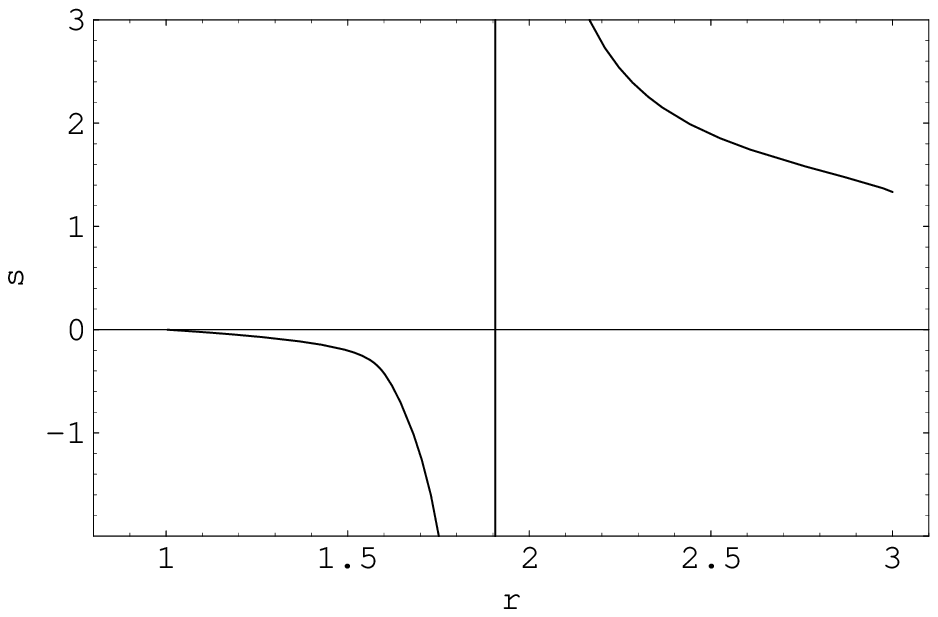}\\
\vspace{1mm} Fig 4.3: The variation of $s$ is plotted against $r$
for $ C=C_{0}=B=1, \alpha=1, \omega=-\frac{3}{2}, \rho_{0}=1,
\epsilon=\frac{1}{2}$. \\

\end{figure}

(iii) If $\omega=-2$, equation (4.21) can be written as

\begin{eqnarray*}
\rho_{r}=\rho_{0}~a^{-3(1+\gamma)}-\frac{
\epsilon}{(1+2\alpha-\gamma)} \frac{
a^{-3(1+\gamma)}}{x^{\frac{1+2\alpha-\gamma}{(1+\alpha)}}}
\frac{B^{-\frac{1+\gamma}{(1+\alpha)}}}{
C^{\frac{\alpha}{1+\alpha}}}
\end{eqnarray*}
\begin{equation}
Appell F_{1} \left[ \frac{1+2\alpha-\gamma}{(1+\alpha)},
\frac{\alpha}{1+\alpha}, \frac{\alpha-\gamma}{(1+\alpha)},
\frac{2+3\alpha-\gamma}{(1+\alpha)}, -\frac{x}{C}, x-x \epsilon
\right]
\end{equation}

Normalizing the parameters, the corresponding statefinder
parameters are given in figure 4.4.

\begin{figure}

\includegraphics{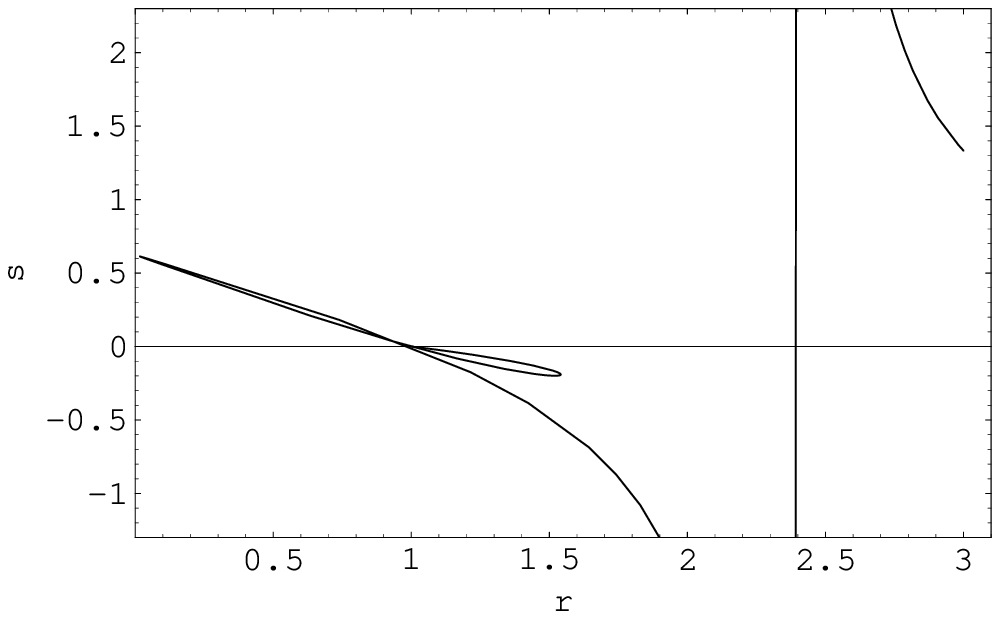}\\
\vspace{1mm} Fig 4.4: The variation of $s$ is plotted against $r$
for $ C=C_{0}=B=1, \alpha=1, \rho_{0}=1, \epsilon=\frac{1}{2}$.\\

\end{figure}

\section{Discussion}

Recently developed Generalized Cosmic Chaplygin gas (GCCG) is
studied as an unified model of dark matter and dark energy. In
this chapter, we have considered the matter in our Universe as a
mixture of the GCCG and radiation as GCCG can explain the
evolution of the Universe from dust era to $\Lambda$CDM. These
gases are taken both as non-interacting and interacting mixture.
In the first case we have considered a non-interacting model and
plotted the $r, s$ parameters. As expected this model represents
the evolution of the Universe from radiation era to $\Lambda$CDM
with a discontinuity at $r=2$ where it represents the dust era
(for $r=2$ implies the dust era, $r=1$ implies $\Lambda$CDM, $r<1$
phantom). In the second case the interaction term is chosen in a
very typical form to solve the corresponding conservation
equations analytically. Also the statefinder parameters are
evaluated for various choices of parameters and the trajectories
in the $\{r,s\}$ plane are plotted to characterize different
phases of the Universe. These trajectories show discontinuity at
same $r$ in the neighbourhood of $r=2$ and have peculiar behaviour
around $r=1$. The $\{r,s\}$ curves have two branches on two sides
of the asymptote. The branch on the right hand side of the
asymptote corresponds to decelerating phase before (or up to) dust
era, while the left hand side branch has a transition from
decelerating phase upto $\Lambda$CDM era. Some peculiarity has
been shown in figures 2 and 4 around $r=1$. In these two cases,
the model goes further from $\Lambda$CDM to phantom era and then
back to $\Lambda$CDM. Moreover, in figure 4.4, there is further
transition from $\Lambda$CDM to decelerating phase and then then
again back to $\Lambda$CDM. Thus we can conclude that the present
model describes a number of transitions from decelerating to
accelerating phase and vice-versa.

%% file: chap5.tex
\large \baselineskip .85cm
\chapter{Variable Modified Chaplygin Gas Model in presence of Scalar Field } \label{chap5}\markright{\it
CHAPTER~\ref{chap5}. Variable Modified Chaplygin Gas Model in
presence of Scalar Field}

\section{Prelude}

We have already studied the properties of GCG and MCG and their
roles in explaining the evolution of the Universe. Later Sthi
etal [2006] and Guo etal [2007] introduced inhomogeneity in the
EOS of MCG  given by equation (1.60) by considering $B$ to be a
function of the scale factor $a(t)$. This assumption is
reasonable since $B(a)$ is related to the scalar potential if we
take the Chaplygin gas as a Born-Infeld scalar field [Bento etal, 2003].\\

In this chapter we generalize the above model and present a new
form of the well known Chaplygin gas model by introducing
inhomogeneity in the EOS by considering both $A$ and $B$ in the
EOS (1.60) to be a function of the scale factor $a(t)$. We show
that this model can explain the evolution of the Universe
suitably by choosing different values of the parameters and also
can explain $\omega=-1$ crossing.\\

We have also seen that interaction models where the dark energy
weakly interacts with the dark matter have been studied to
explain the evolution of the Universe. These models describe an
energy flow between the components. To obtain a suitable
evolution of the Universe the decay rate should be proportional
to the present value of the Hubble parameter for good fit to the
expansion history of the Universe as determined by the Supernovae
and CMB data [Berger and Shojaei, 2006]. A variety of interacting
dark energy models have been proposed and studied for this
purpose [Zimdahl, 2005; Cai and Wang, 2005]. We therefore have
also considered a interaction of this model with the scalar field
by introducing a phenomenological coupling function which
describes the energy flow between them, thus showing the effect
of interaction in the evolution of the Universe. This kind of
interaction term has been studied in ref. [Cai and Wang, 2005].\\

\section{Field  Equations  and  Solutions}

The metric of a spatially flat homogeneous and isotropic universe
in FRW model is considered in eq. (1.7) The Einstein field
equations and energy conservation equation are given by in
equations (1.12) and (1.13) and (1.20).\\

Now, we extend MCG with equation of state (1.60) such that $A$ and
$B$ are positive function of the cosmological scale factor `$a$'
(i.e., $A=A(a), B=B(a)$). Then equation (1.60) reduces to,

\begin{equation}
p=A(a)\rho-\frac{B(a)}{\rho^{\alpha}} ~~~~\text{with}~~~~ 0\le
\alpha \le 1
\end{equation}

As we can see this is an inhomogeneous EOS [Brevik etal, 2007]
where the pressure is a function of the energy density $\rho$ and
the scale factor $a(t)$. Also if
$\rho=\left(\frac{B(a)}{A(a)}\right)^{\frac{1}{1+\alpha}}$,
this model reduces to dust model, pressure being zero.\\

Now, assume $A(a)$ and $B(a)$ to be of the form

\begin{equation}
A(a)=A_{0}a^{-n}
\end{equation}
and
\begin{equation}
B(a)=B_{0}a^{-m}
\end{equation}

where $A_{0}$, $B_{0}$, $n$ and $m$ are positive constants. If
$n=m=0$, we get back the modified Chaplygin gas [Debnath etal,
2004] and if $n=0$, we get back variable modified Chaplygin gas
(VMCG) model [Debnath, 2007]. Using equations (1.20), (5.1), (5.2)
and (5.3), we get the solution of $\rho$ as,

\begin{equation}
\rho=a^{-3} e^{\frac{3 A_{0}
a^{-n}}{n}}\left[C_{0}+\frac{B_{0}}{A_{0}} \left(\frac{3 A_{0}
(1+\alpha)}{n}\right)^{\frac{3(1+\alpha)+n-m}{n}}\Gamma(\frac{m-3(1+\alpha)}{n},\frac{3
A_{0}(1+\alpha)}{n}a^{-n})\right]^{\frac{1}{1+\alpha}}
\end{equation}

where $\Gamma(a,x)$ is the upper incomplete gamma function and $C_{0}$ is an integration constant .\\

Now, considering the equation of state
$$\omega_{eff}=\frac{p}{\rho}$$ for this fluid, we have,

\begin{equation}
\omega_{eff}=A_{0} a^{-n}-B_{0} a^{-\zeta}
e^{-\frac{3A_{0}(1+\alpha) a^{-n}}{n}}\left[C_{0}+
\left(\frac{3A_{0}(1+\alpha)}{n}\right)^{\frac{n-\zeta}{n}}
\frac{B_{0}}{A_{0}}\Gamma(\frac{\zeta}{n},\frac{3A_{0}(1+\alpha)a^{-n}}{n})\right]
\end{equation}

where $ \zeta=m-3(1+\alpha)$.\\

For small values of the scale factor $a(t)$, $\rho$ is very large
and
$$p=A\rho-\frac{B}{\rho^{\alpha}}\rightarrow A\rho$$ where
$A=A_{0}a^{-n}$ is a function of $a$, so that for small scale
factor we have very large pressure and energy densities.
Therefore initially
$$\frac{p}{\rho}=\omega_{eff}=A^{*} a^{-n} \le 1$$ where
$A^{*}$ is a constant, $$A^{*}=A_{0}.$$ If
$a={A_{0}}^{\frac{1}{n}}$, the Universe starts from stiff perfect
fluid, and if $a={3 A^{*}}^{\frac{1}{n}}$, the Universe starts
from radiation era. \\

\begin{figure}

\includegraphics{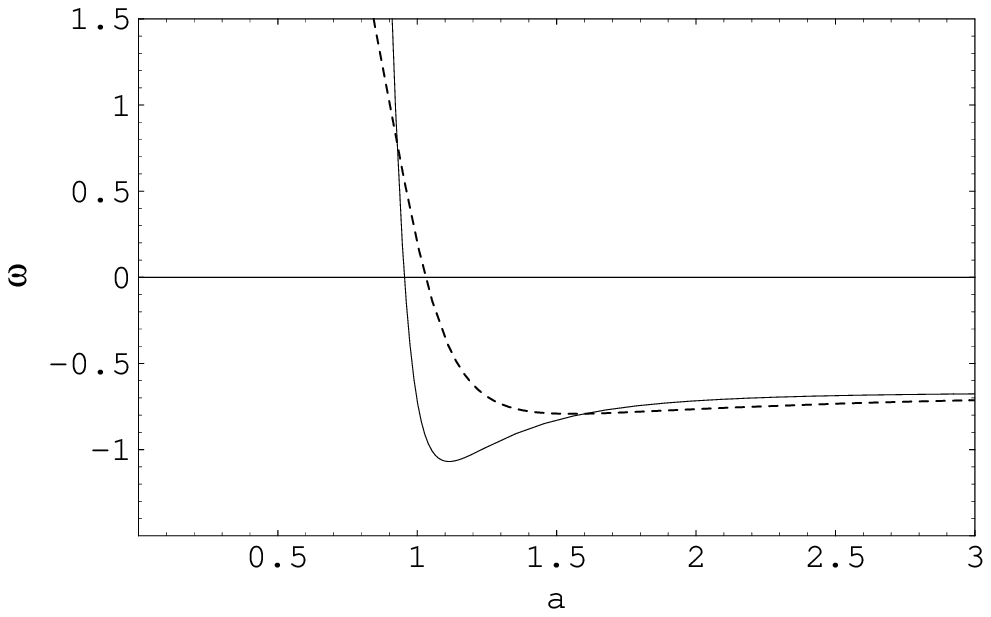}\\

\vspace{1mm}Fig 5.1: The variation of $\omega_{eff}$ is shown
against $a(t)$ for $ A_{0}=1, B_{0}=10, \alpha=1, m=2, C_{0}=1$
and $n=3$ (for dotted line), $n=10$ (for the dark line).\\

\end{figure}

Also for large values of the scale factor
$$p=A\rho-\frac{B}{\rho^{\alpha}}\rightarrow
-\frac{B}{\rho^{\alpha}}.$$ If $$\zeta=m-3(1+\alpha)<0$$ (as we
know that upper incomplete Gamma function $\Gamma(a,x)$ exists for
$a<0$), the second term dominates and hence
$\omega_{_{eff}}\rightarrow -B^{*}a^{-\zeta}$, where
$$B^{*}=B_{0} {\lim_{a\rightarrow\infty}}{
e^{-\frac{3A_{0}(1+\alpha) a^{-n}}{n}}\left[C_{0}+
\left(\frac{3A_{0}(1+\alpha)}{n}\right)^{\frac{n-\zeta}{n}}
\frac{B_{0}}{A_{0}}\Gamma(\frac{\zeta}{n},\frac{3A_{0}(1+\alpha)a^{-n}}{n})\right]^{-1}}$$
( ${\lim_{a\rightarrow\infty}}{e^{-\frac{3A_{0}(1+\alpha)
a^{-n}}{n}}}\rightarrow 1$ and
${\lim_{a\rightarrow\infty}}{\Gamma(\frac{\zeta}{n},\frac{3A_{0}(1+\alpha)a^{-n}}{n})}\rightarrow$large
value, for $\zeta<0$ ). This will represent dark energy if
$a>\left(\frac{1}{3B^{*}}\right)^{\frac{1}{3(1+\alpha)-m}}$,
$\Lambda$CDM if
$a=\left(\frac{1}{B^{*}}\right)^{\frac{1}{3(1+\alpha)-m}}$ and
phantom dark energy if
$a>\left(\frac{1}{B^{*}}\right)^{\frac{1}{3(1+\alpha)-m}}$.
Therefore we can explain the evolution of the Universe till the
phantom era depending on the various values of the parameters. We
have shown a graphical representation of $\omega_{eff}$ in fig 5.1
for different values of the parameters. We can see from fig 5.1
that $\omega_{eff}$ starting from a large values decreases with
$a$ crosses $\omega=-1$ for some choices of the parameters.\\

\section{Statefinder Diagnostics}

Now we analyse our model using statefinder parameters given by
equation (1.113) to investigate the validation of the model.\\

For this model

\begin{equation}
H^{2}=\frac{\dot{a}^{2}}{a^{2}}=\frac{1}{3}\rho
\end{equation}

and

\begin{equation}
q=-\frac{\ddot{a}}{aH^{2}}=\frac{1}{2}+\frac{3}{2}\frac{p}{\rho}
\end{equation}

So from equation (1.113) we get

\begin{equation}
r=1+\frac{9}{2}\left(1+\frac{p}{\rho}\right)\frac{\partial
p}{\partial\rho}-\frac{3}{2}\frac{a}{\rho}\frac{\partial
p}{\partial a}
~,~~~~~s=\frac{2(r-1)}{9\left(\frac{p}{\rho}\right)}
\end{equation}

so that, solving we get,

\begin{equation}
r=1+\frac{9}{2}(1+y)(A_{0} a^{-n}+\alpha B_{0} a^{-m}
x)+\frac{3}{2}(n A_{0} a^{-n}-m B_{0}a^{-m}x)
~,~~~~~s=\frac{2(r-1)}{9y}
\end{equation}

where, $ y=\frac{p}{\rho}=A_{0}a^{-n}-B_{0}a^{-m} x$ and
$x=\rho^{-(1+\alpha)}$, $\rho$ is given by equation (5.4).\\

\begin{figure}

\includegraphics{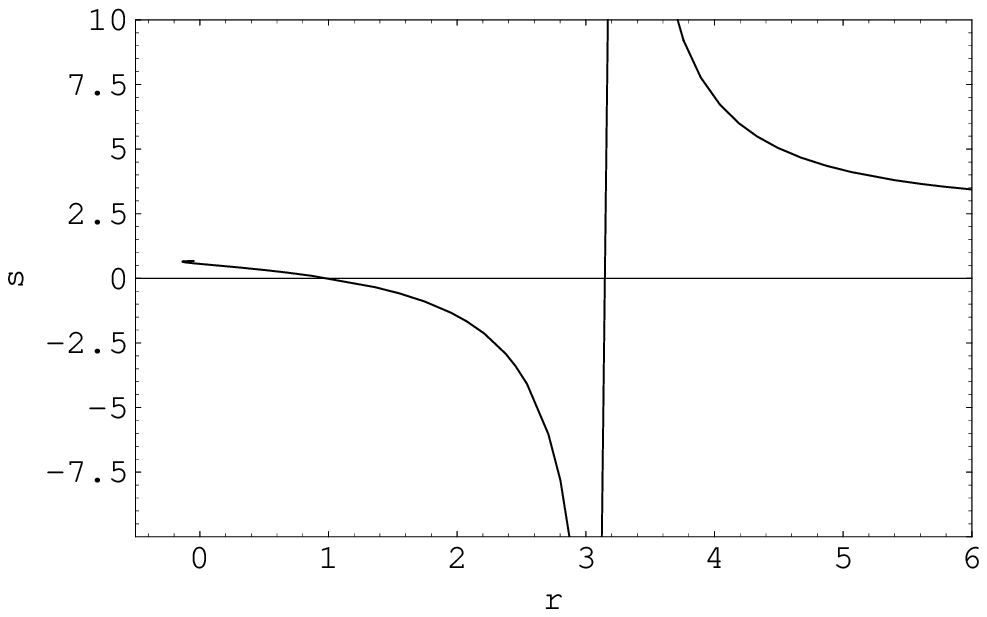}\\

\vspace{1mm}Fig 5.2: The variation of $s$ is plotted against $r$
for $A_{0}=1, B_{0}=1, \alpha=\frac{1}{2}, m=3, n=2, C_{0}=1$.\\

\end{figure}

We have plotted the $ \{r, s\}$ parameters in figure 5.2
normalizing the parameters and varying the scale factor $a(t)$.
We can see that the model starts from radiation era. Then we have
a discontinuity at the dust era (for radiation era: $s>0$ and
$r>1$; dust era: $r>1$ and $s\rightarrow \pm \infty$;
$\Lambda$CDM: $r=1$, $s=0$; phantom: $r<1$). The model reaches
$\Lambda$CDM at $r=1,~s=0$ and then crosses $\Lambda$CDM to
represent phantom dark energy. This model represents the phantom
dark energy, whereas, Modified Chaplygin Gas can explain the
evolution of the Universe from radiation to $\Lambda$CDM and
Variable Modified Chaplygin gas describes the
evolution of the Universe from radiation to quiessence model.\\

\section{New modified Chaplygin gas and interacting scalar field}

Now we consider model of interaction between scalar field and the
new modified Chaplygin Gas model, through a phenomenological
interaction term. Keeping into consideration the fact that the
Supernovae and CMB data determines that decay rate should be
proportional to the present value of the Hubble parameter [Berger
and Shojaei, 2006], we have chosen the interaction term likewise.
This interaction term describes the energy flow between the two
fluids. We have considered a scalar field to couple with the New
Modified Chaplygin gas given by EOS (5.1), (5.2) and (5.3). \\

Therefore now the conservation equation becomes equation (4.4).
For the interacting model, the equations of motion of the the new
fluid and scalar field read,

\begin{equation}
\dot{\rho}+3H(\rho+p)=-3H\rho\delta
\end{equation}

and

\begin{equation}
\dot{{\rho}}_{\phi}+3H({\rho}_{\phi}+p_{\phi})=3H\rho\delta
\end{equation}

($\delta$ is a constant).\\

Where the total energy density and pressure of the universe are
given by,

\begin{equation}
\rho_{tot}=\rho+\rho_{\phi}
\end{equation}

and

\begin{equation}
p_{tot}=p+p_{\phi}
\end{equation}

where, $\rho$ and $p$ are the energy density and pressure of the
extended modified Chaplygin gas model given by equations (5.1),
(5.2), (5.3), (5.4) and $\rho_{\phi}$ and $p_{\phi}$ are the
energy density and pressure due to the scalar field given by,

\begin{equation}
\rho_{\phi}=\frac{{\dot{\phi}}^{2}}{2}+V(\phi)
\end{equation}

and

\begin{equation}
p_{\phi}=\frac{{\dot{\phi}}^{2}}{2}-V(\phi)
\end{equation}

where, $V(\phi)$ is the relevant potential for the scalar field
$\phi$.\\

Thus from the field equations (4.2) and (4.3) and the conservation
equation (4.3), we get the solution for $\rho$ as

\begin{eqnarray*}
\rho=a^{-3(1+\delta)} e^{\frac{3 A_{0}
a^{-n}}{n}}\left[C_{0}+\frac{B_{0}}{A_{0}} \left(\frac{3 A_{0}
(1+\alpha)}{n}\right)^{\frac{3(1+\alpha)(1+\delta)+n-m}{n}}\right.
\end{eqnarray*}
\begin{equation}
\left. \Gamma(\frac{m-3(1+\alpha)(1+\delta)}{n},\frac{3
A_{0}(1+\alpha)}{n}a^{-n})\right]^{\frac{1}{1+\alpha}}
\end{equation}

where $C_{0}$ is an integration constant.\\

Further substitution in the above equations give,

\begin{equation}
V(\phi)=3H^{2}+\dot{H}+\frac{p-\rho}{2}
\end{equation}

To get an explicit form of the energy density and the potential
corresponding to the scalar field we consider a power law
expansion of the scale factor $a(t)$ as,

\begin{equation}
a=t^{\beta}
\end{equation}

so that, for $\beta>1$ we get accelerated expansion of the
Universe thus satisfying the observational constrains. If
$\beta=1$ or $\beta<1$ we get constant and decelerated expansion
respectively.\\

Using equations (5.6), (5.12) and (5.18), we get,

\begin{equation}
\rho_{\phi}=\frac{3 \beta^{2}}{t^{2}}-\rho
\end{equation}

where $\rho$ is given by equation (5.16).\\

Also the potential takes the form,

\begin{equation}
V=\frac{3n^{2}-n}{t^{2}}+\frac{p-\rho}{2}
\end{equation}

The graphical representation of $V$ against time is shown in
figure 5.3 normalizing the parameters. We see that the potential
decays with time.\\

\begin{figure}

\includegraphics{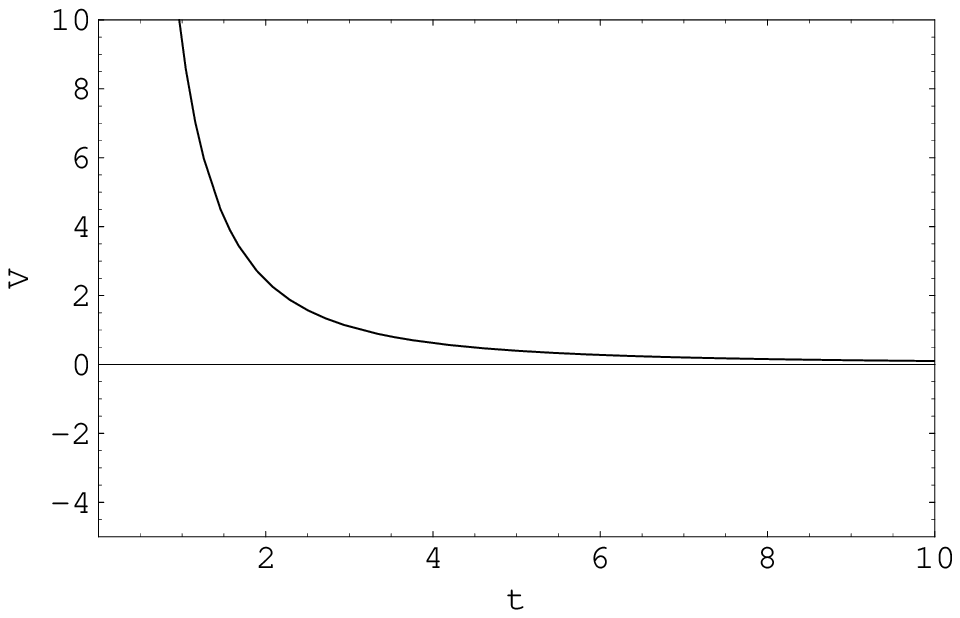}\\
\vspace{1mm} Fig 5.3: The variation of $V$ is plotted against $t$
for $ A_{0}=1, B_{0}=1, \alpha=\frac{1}{2}, m=3, n=2, C_{0}=0$.\\

\end{figure}

\section{Discussion}

Here we present a new variable modified Chaplygin gas model which
is an unified version of the dark matter and the dark energy of
the Universe. It behaves like dark matter at the initial stage and
later it explains the dark energy of the Universe. Unlike the
Generalized or Modified Chaplygin gas model, it can explain the
evolution of the Universe at phantom era depending on the
parameters. Also we have calculated the $\{r,s\}$ parameters
corresponding to this model. Normalizing the parameters such that
$m-3(1+\alpha)<0$, show the diagrammatical representation of
$\{r,s\}$ for our model (in Fig. 5.2), varying the scale factor.
We see that starting from the radiation era it crosses $\omega=-1$
and extends till phantom era. Also we can see that the
deceleration parameter starting from a positive point becomes
negative, indicating deceleration initially and acceleration at
later times. Again we have considered an interaction of this fluid
with that of scalar field by introducing a phenomenological
coupling term, so that there is a flow of energy between the field
and the fluid which decays with time, as in the initial stage the
fluid behaves more like dark matter and the field that of dark
energy, whereas in the later stage both explain the dark energy
present in the Universe. In Fig. 5.3, we have shown the nature of
the potential due to the scalar field considering a power law
expansion of the Universe to keep the recent observational support
of cosmic acceleration, and we see that the potential
decays with time.\\\\

%% file: chap6.tex
\large \baselineskip .85cm
\chapter{Dynamics of Tachyonic Field in Presence of Perfect Fluid}
\label{chap6}\markright{\it CHAPTER~\ref{chap6}. Dynamics of Tachyonic Field in Presence of Perfect Fluid }

\section{Prelude}

Recently tachyonic field with Lagrangian ${\cal
{L}}=-V(T)\sqrt{1+g^{\mu \nu}~\partial{_{\mu}}T
\partial{_{\nu}} T}$ [Sen, 2002] has gained a lot of importance as dark
energy model. The energy-momentum tensor of the tachyonic field
can be seen as a combination of two fluids, dust with pressure
zero and a cosmological constant with $p=-\rho$, thus generating
enough negative pressure such as to drive acceleration. Also the
tachyonic field has a potential which has an unstable maximum at
the origin and decays to almost zero as the field goes to
infinity. Depending on various forms of this potential following
this asymptotic behaviour a lot of works have been carried out on
tachyonic dark energy [Bagla etal, 2003; Copeland etal, 2005,
2006], tachyonic dark matter [Padmanabhan, 2002; Das etal, 2005]
and inflationary models [Sami, 2003]. Recently, interacting
tachyonic-dark matter model has also been studied [Herrera etal,
2004].\\

In this chapter, we consider a model which comprises of a two
component mixture. Firstly we consider a mixture of barotropic
fluid with tachyonic field without any interaction between them,
so that both of them retain their properties separately. Then we
consider an energy flow between them by introducing an interaction
term which is proportional to the product of the Hubble parameter
and the density of the barotropic fluid. We show that the energy
flow being considerably high at the beginning falls down
noticeably with the evolution of the Universe indicating a more
stable situation. Also in both the cases we find the exact
solutions for  the tachyonic field and the tachyonic potential and
show that the tachyonic potential follows the asymptotic behaviour
discussed above. Here the tachyonic field behaves as the dark
energy component whereas the dust acts as the cold dark matter.
Next we consider tachyonic dark matter, the Generalized Chaplygin
Gas (GCG) being the dark energy component. GCG, identified by the
equation of state (EOS) (1.53) has been considered as a suitable
dark energy model by several authors [Bento etal, 2002; Gorini
etal, 2003]. Here we consider the mixture of GCG with tachyonic
dark matter. Later we have also considered an interaction between
these two fluids by introducing a coupling term which is
proportional to the product of Hubble constant and the energy
density of the GCG. The coupling function decays with time
indicating a strong energy flow at the initial period and weak
interaction at later stage implying a stable situation. Here we
have found the exact solution of the tachyonic potential. To keep
the observational support of recent acceleration we have
considered a particular form of evolution of the Universe here as
\begin{equation}
a=t^{n}
\end{equation}
such that the deceleration parameter reads $q=-\frac{a
\ddot{a}}{{\dot{a}}^{2}}=-(1-\frac{1}{n})$, where $a$ is the scale
factor. Hence for $n>1$ we always get an accelerated expansion
and for $n=1$ we get a constant expansion of the Universe. This
kind of recipe has been studied in ref. [Padmanabhan, 2002].\\

\section{Field  Equations}

The action for the homogeneous tachyon condensate of string
theory in a gravitational background is given by,
\begin{equation}
S=\int {\sqrt{-g}~ d ^{4} x \left[\frac{\cal R}{16 \pi G}+{\cal
L_{tach}}\right]}
\end{equation}
where $\cal L$ is the Lagrangian density given by equation (1.66),
where $T$ is the tachyonic field, $V(T)$ is the tachyonic
potential and $\cal R$ is the Ricci Scalar. The energy-momentum
tensor for the tachyonic field is,

\begin{eqnarray}
\begin{array} {ccc}
T_{\mu \nu}=-\frac{2 \delta S}{\sqrt{-g}~ \delta g^{\mu
\nu}}=-V(T)\sqrt{1+g^{\mu \nu} \partial _{\mu}T \partial_{\nu}
T}g^{\mu \nu}+V(T) \frac{\partial _{\mu}T \partial_{\nu}
T}{\sqrt{1+g^{\mu \nu} \partial _{\mu}T \partial_{\nu} T}}\\\\\
=p_{T}~g_{\mu \nu}+(p_{T}+\rho_{T})u_{\mu} u_{\nu}\\\\
\end{array}
\end{eqnarray}

where the velocity $u_{\mu}$ is :

\begin{equation}
u_{\mu}=-\frac{\partial_{\mu}T}{\sqrt{-g^{\mu \nu} \partial
_{\mu}T
\partial_{\nu} T}}
\end{equation}

with $u^{\nu} u_{\nu}=-1$.\\

The energy density $\rho_{T}$ and the pressure $p_{T}$ of the
tachyonic field therefore are given by (1.71) and (1.72)
respectively. Hence the EOS parameter of the tachyonic field
becomes (1.73) and (1.74), which represents pure Chaplygin gas if $V(T)$ is constant.\\

Now the metric of a spatially flat isotropic and homogeneous
Universe in FRW model is presented by equation (1.7). The Einstein
field equations are (4.2) and (4.3), where, $\rho_{tot}$ and
$p_{tot}$ are the total energy density and the pressure of the
Universe. The energy conservation equation is
given by equation (4.4).\\

\section{Tachyonic Dark Energy in presence of Barotropic Fluid}

Now we consider a two fluid model consisting of tachyonic field
and barotropic fluid. The EOS of the barotropic fluid is given by,

\begin{equation}
p_{b}=\omega_{b} \rho_{b}
\end{equation}

where $p_{b}$ and $\rho_{b}$ are the pressure and energy density
of the barotropic fluid. Hence the total energy density and
pressure are respectively given by,

\begin{equation}
\rho_{tot}=\rho_{b}+\rho_{T}
\end {equation}

and

\begin{equation}
p_{tot}=p_{b}+p_{T}
\end {equation}

\subsection{Without Interaction}

First we consider that the two fluids do not interact with each
other so that they are conserved separately. Therefore, the
conservation equation (4.4) reduces to,

\begin{equation}
\dot{\rho}_{T}+3\frac{\dot{a}}{a}(\rho_{T}+p_{T})=0
\end{equation}

and

\begin{equation}
\dot{\rho}_{b}+3\frac{\dot{a}}{a}(\rho_{b}+p_{b})=0
\end{equation}

Equation (6.9) together with equation (6.5) gives,

\begin{equation}
\rho_{b}=\rho_{0}~a^{-3(1+\omega_{b})}
\end{equation}

Now, we consider a power law expansion of the scale factor $a(t)$
given by equation (6.1).\\

Using (6.1), equation (6.10) reduces to,

\begin{equation}
\rho_{b}=\rho_{0}~t^{-3n(1+\omega_{b})}
\end{equation}

Also the energy density corresponding to the tachyonic field
becomes,

\begin{equation}
\rho_{T}=\frac{1}{t^{2}}
\left[3n^{2}-\rho_{0}~t^{-3n(1+\omega_{b})+2}\right]
\end{equation}

Solving the equations the tachyonic field is obtained as,

\begin{eqnarray*}
T=\sqrt{1+\omega_{b}}t ~Appell~ F_{1}
\left[\frac{1}{3(1+\omega_{b})n-2}, \frac{1}{2}, -\frac{1}{2},
1+\frac{1}{3(1+\omega_{b})n-2}, \right.
\end{eqnarray*}

\begin{equation}
\left. \frac{3n^{2}}{\rho_{0}} t^{3(1+\omega_{b})n-2},
\frac{2n}{\rho_{0} (1+\omega_{b})} t^{3(1+\omega_{b})n-2} \right]
\end{equation}

where, $Appell ~F1[a,b_{1},b_{2},c,x,y]$ is the Appell
Hypergeometric function of two variables $x$ and $y$.\\

Also the potential will be of the form,

\begin{equation}
V(T)= \sqrt{\frac{3n^{2}}{t^{2}}-\rho_{0}~t^{-3n(1+\omega_{b})}}
\sqrt{\frac{3n^{2}}{t^{2}}-\frac{2n}{t^{2}}+\omega_{b}\rho_{0}~t^{-3n(1+\omega_{b})}}
\end{equation}

We can show the graphical representation of the potential against
time in figure 6.1. We can see that $V \rightarrow 0$ with time,
thus retaining the original property of the tachyon potential.\\

\begin{figure}

\includegraphics{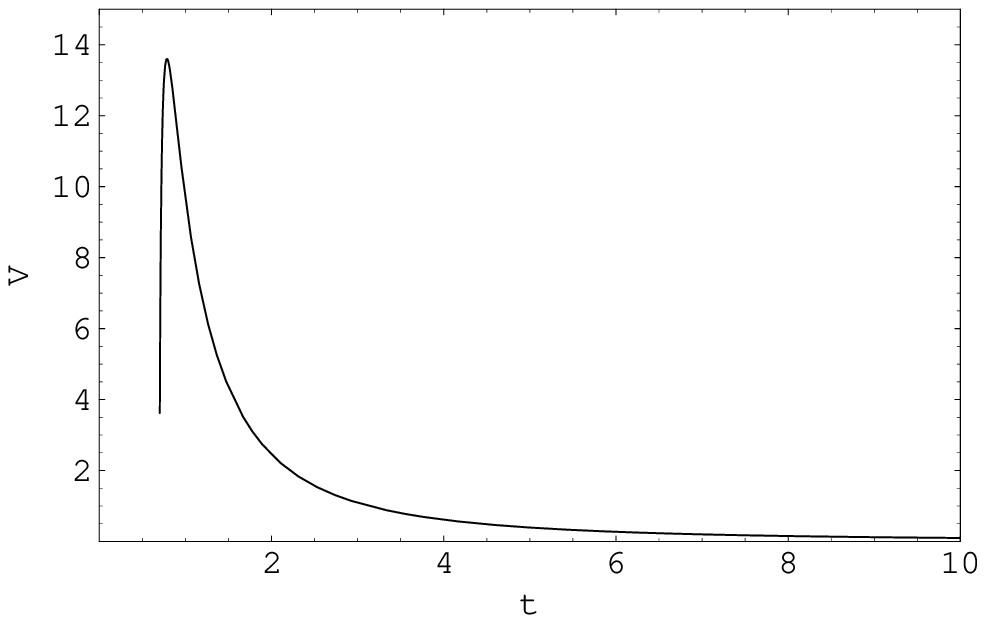}

\vspace{10mm} Fig 6.1: The variation of $V$ is plotted against $t$
for $ n=2, \rho_{0}=1, \omega_{b}=\frac{1}{2}$.\\

\end{figure}

\subsection{With Interaction}

Now we consider an interaction between the tachyonic field and
the barotropic fluid by introducing a phenomenological coupling
function which is a product of the Hubble parameter and the
energy density of the barotropic fluid. Thus there is an energy
flow between the two fluids. \\

Now the equations of motion corresponding to the tachyonic field
and the barotropic fluid are respectively,

\begin{equation}
\dot{\rho}_{T}+3\frac{\dot{a}}{a}(\rho_{T}+p_{T})=-3H\delta
\rho_{b}
\end{equation}

and

\begin{equation}
\dot{\rho_{b}}+3\frac{\dot{a}}{a}(\rho_{b}+p_{b})=3H\delta
\rho_{b}
\end{equation}

where $\delta$ is a coupling constant.\\

Solving equation (6.16) with the help of equation (6.5), we get,

\begin{equation}
\rho_{b}=\rho_{0}~a^{-3(1+\omega_{b}-\delta)}
\end{equation}

Considering the power law expansion (6.1), we get

\begin{equation}
\rho_{b}=\rho_{0}~t^{-3n(1+\omega_{b}-\delta)}
\end{equation}

Equation (6.2) and (6.18) give,

\begin{equation}
\rho_{T}=\frac{3n^{2}}{t^{2}}
-\rho_{0}~t^{-3n(1+\omega_{b}-\delta)}
\end{equation}

Solving the equations the tachyonic field is obtained as,

\begin{eqnarray*}
T=\sqrt{1+\omega_{b}}t ~Appell~ F_{1}
\left[\frac{1}{3(1+\omega_{b}-\delta)n-2}, \frac{1}{2},
-\frac{1}{2},\right.
\end{eqnarray*}
\begin{equation}
\left. 1+\frac{1}{3(1+\omega_{b}-\delta)n-2},
\frac{3n^{2}}{\rho_{0}} t^{3(1+\omega_{b}-\delta)n-2},
\frac{2n}{\rho_{0} (1+\omega_{b})} t^{3(1+\omega_{b}-\delta)n-2}
\right]
\end{equation}

Also the potential will be of the form,

\begin{equation}
V(T)=
\sqrt{\frac{3n^{2}}{t^{2}}-\rho_{0}~t^{-3n(1+\omega_{b}-\delta)}}
\sqrt{\frac{3n^{2}}{t^{2}}-\frac{2n}{t^{2}}+\omega_{b}\rho_{0}~t^{-3n(1+\omega_{b}-\delta)}}
\end{equation}

In this case also $V\rightarrow 0$ with time as shown in the
graphical representation of $V$ in figure 6.2.\\

\begin{figure}

\includegraphics{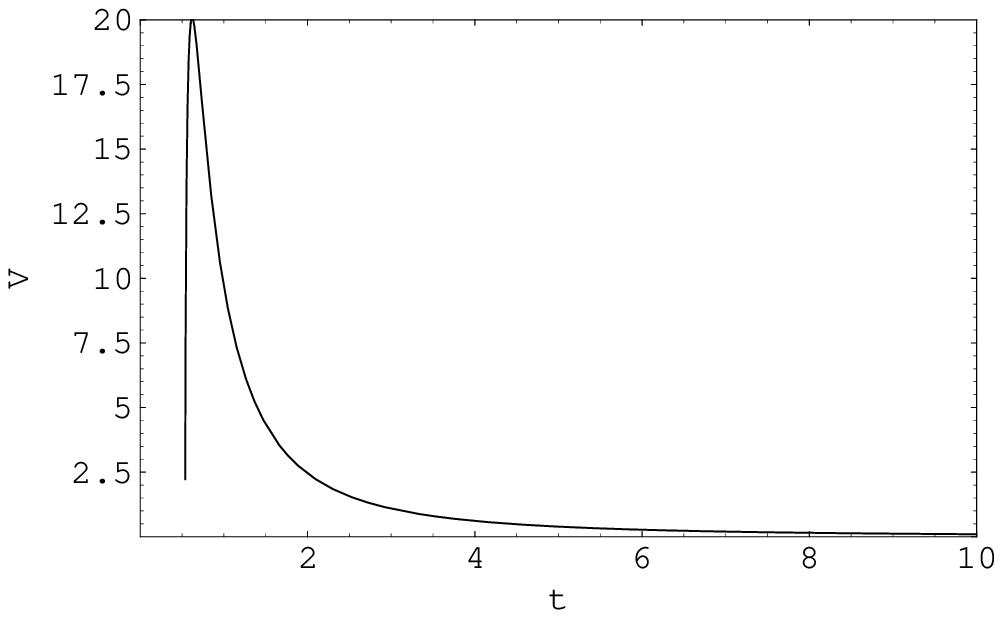}

\vspace{1mm} Fig 6.2: The variation of $V$ is plotted against $t$
for $ n=2, \rho_{0}=1, \omega_{b}=\frac{1}{2},
\delta=\frac{1}{2}$.\\

\end{figure}

\section{Tachyonic Dark Matter in presence of GCG}

Now we consider a two fluid model consisting of tachyonic field
and GCG. The EOS of GCG is given by,

\begin{equation}
p_{ch}=-B/{\rho}_{ch}^{\alpha}~~~~~~~~~~~\text~~~~~~~~~~~~ 0\le
\alpha \le 1, B>0.
\end{equation}

where $p_{ch}$ and $\rho_{ch}$ are the pressure and energy density
of GCG. Hence the total energy density and pressure are
respectively given by,

\begin{equation}
\rho_{tot}=\rho_{ch}+\rho_{T}
\end {equation}

and

\begin{equation}
p_{tot}=p_{ch}+p_{T}
\end {equation}

\subsection{Without Interaction}

First we consider that the two fluids do not interact with each
other so that they are conserved separately. Therefore, the
conservation equation (4.4) reduces to,

\begin{equation}
\dot{\rho}_{T}+3\frac{\dot{a}}{a}(\rho_{T}+p_{T})=0
\end{equation}

and

\begin{equation}
\dot{\rho}_{ch}+3\frac{\dot{a}}{a}(\rho_{ch}+p_{ch})=0
\end{equation}

Equation (6.26) together with equation (6.22) give,

\begin{equation}
\rho_{ch}=\left[B+\frac{\rho_{00}}{a^{3(1+\alpha)}}\right
]^{\frac{1}{(1+\alpha)}}
\end{equation}

Using (6.1), equation (6.27) reduces to

\begin{equation}
\rho_{ch}=\left[B+\rho_{00}t^{-3n(1+\alpha)}\right]^{\frac{1}{(1+\alpha)}}
\end{equation}

Hence the energy density of the tachyonic fluid is,

\begin{equation}
\rho_{T}=\frac{3n^{2}}{t^{2}}
-\left[B+\rho_{00}t^{-3n(1+\alpha)}\right]^{\frac{1}{(1+\alpha)}}
\end{equation}

Solving the equations the tachyonic field and the tachyonic
potential are obtained as,

\begin{equation}
T=\int\sqrt{\frac{\frac{2n}{t^{2}}-\rho_{00} t^{-3n(1+\alpha)}
[B+\rho_{00}t^{-3n(1+\alpha)}]^{-\frac{\alpha}{(1+\alpha)}}}{\frac{3n^{2}}{t^{2}}
-[B+\rho_{00}t^{-3n(1+\alpha)}]^{\frac{1}{(1+\alpha)}}}}dt
\end{equation}

Also the potential will be of the form,

\begin{equation}
V(T)=
\sqrt{\frac{3n^{2}}{t^{2}}-[B+\rho_{00}t^{-3n(1+\alpha)}]^{\frac{1}{(1+\alpha)}}}
\sqrt{\frac{3n^{2}}{t^{2}}-\frac{2n}{t^{2}}-B[B+\rho_{00}t^{-3n(1+\alpha)}]^{-\frac{\alpha}{(1+\alpha)}}}
\end{equation}

\begin{figure}

\includegraphics{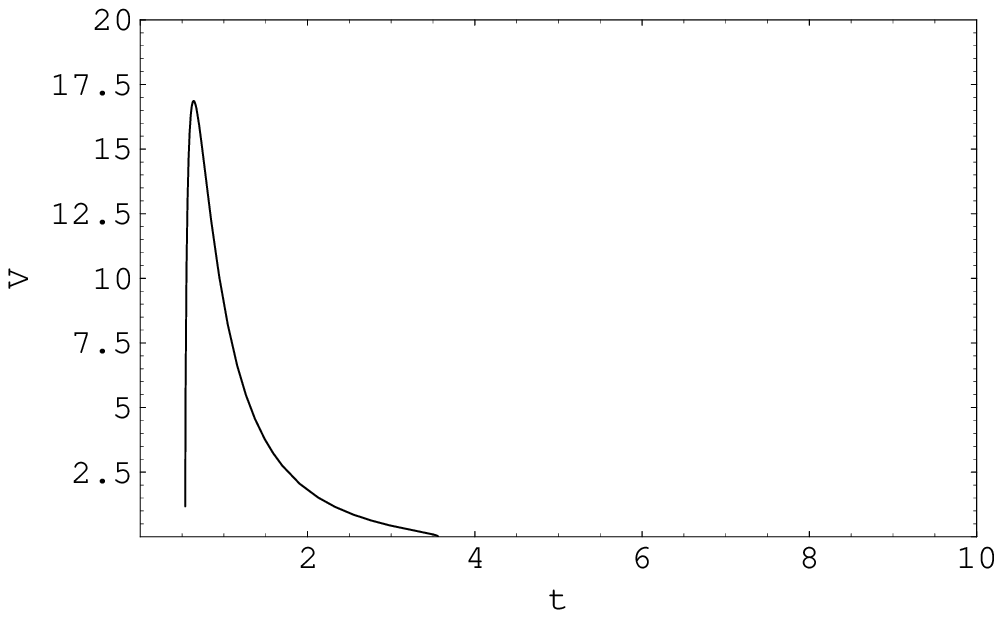}

\vspace{1mm} Fig 6.3: The variation of $V$ is plotted against $t$
for $ B=\frac{1}{2}, n=2, \rho_{00}=1, \alpha=\frac{1}{2}$.\\

\end{figure}

Like the mixture of tachyonic fluid with barotropic fluid in this
case also the potential $V$ starting from a low value increases
largely and then decreases to $0$ with time as shown in figure 6.3.\\

\subsection{With Interaction}

Now we consider an interaction between the tachyonic fluid and GCG
by phenomenologically introducing an interaction term as a product
of the Hubble parameter and the energy density of the Chaplygin
gas. Thus there is an energy flow between the two fluids. \\

\begin{figure}

\includegraphics{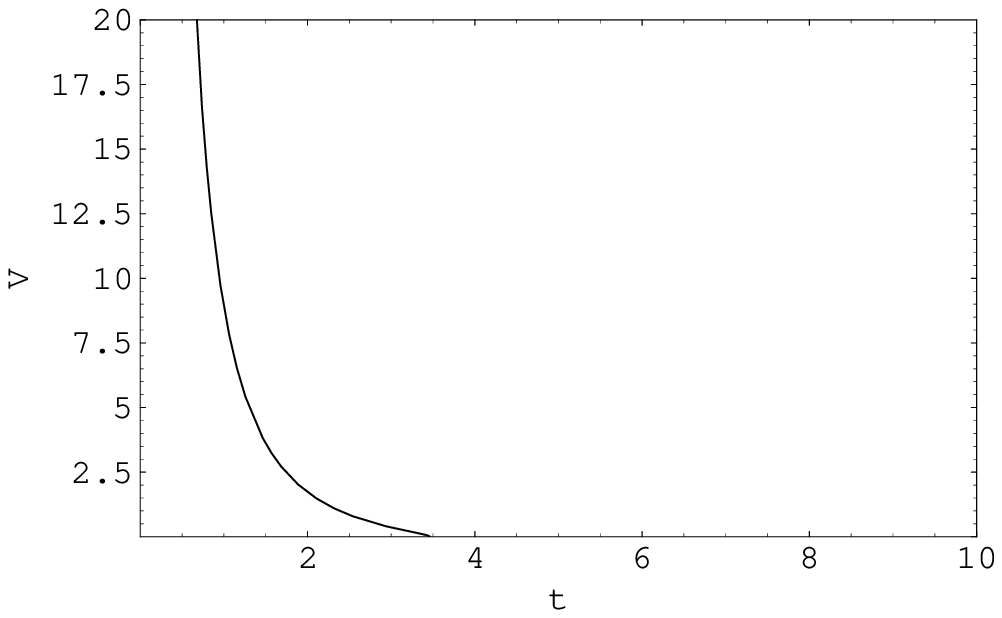}

\vspace{1mm} Fig 6.4: The variation of $V$ is plotted against $t$
for $ B=\frac{1}{2}, n=2, \rho_{00}=1, \alpha=\frac{1}{2},
\epsilon=\frac{1}{2}$. \\

\end{figure}

Now the equations of motion corresponding to the tachyonic field
and GCG are respectively,

\begin{equation}
\dot{\rho}_{T}+3\frac{\dot{a}}{a}(\rho_{T}+p_{T})=-3H\epsilon
\rho_{ch}
\end{equation}

and

\begin{equation}
\dot{\rho}_{ch}+3\frac{\dot{a}}{a}(\rho_{ch}+p_{ch})=3H\epsilon
\rho_{ch}
\end{equation}

where $\epsilon$ is a coupling constant.\\

Solving equation (6.33) with the help of equation (6.22) and
(6.1), we get,

\begin{equation}
\rho_{ch}=\left[\frac{B}{1-\epsilon}+\rho_{00}
t^{-3n(1+\alpha)(1-\epsilon)}\right]^{\frac{1}{(1+\alpha)}}
\end{equation}

Also the energy density of the tachyonic field will be read as,

\begin{equation}
\rho_{T}=\frac{3n^{2}}{t^{2}}
-\left[\frac{B}{1-\epsilon}+\rho_{00}
t^{-3n(1+\alpha)(1-\epsilon)}\right]^{\frac{1}{(1+\alpha)}}
\end{equation}

Solving the equations the tachyonic field is obtained as,

\begin{equation}
T=\int\sqrt{\frac{\frac{2n}{t^{2}}-\left[\frac{B \epsilon
}{1-\epsilon}+\rho_{00} t^{-3n(1+\alpha)(1-\epsilon)}\right]
\left[\frac{B}{1-\epsilon}+\rho_{00}
t^{-3n(1+\alpha)(1-\epsilon)}\right]^{-\frac{\alpha}{(1+\alpha)}}}{\frac{3n^{2}}{t^{2}}
-\left[\frac{B}{1-\epsilon}+\rho_{00}
t^{-3n(1+\alpha)(1-\epsilon)}\right]^{\frac{1}{(1+\alpha)}}}}dt
\end{equation}

Also the potential will be of the form,

\begin{eqnarray*}
V(T)=\sqrt{\frac{3n^{2}}{t^{2}}-\left[\frac{B}{1-\epsilon}+\rho_{00}
t^{-3n(1+\alpha)(1-\epsilon)}\right]^{\frac{1}{(1+\alpha)}}}
\end{eqnarray*}
\begin{equation}
\sqrt{\frac{3n^{2}}{t^{2}}-\frac{2n}{t^{2}}-B\left[\frac{B}{1-\epsilon}+\rho_{00}
t^{-3n(1+\alpha)(1-\epsilon)}\right]^{-\frac{\alpha}{(1+\alpha)}}}
\end{equation}

In this case the potential starting from a large value tends to
$0$ (figure 6.4).\\

\section{Discussion}

We have considered the flat FRW Universe driven by a mixture of
tachyonic field and a perfect fluid. We have considered barotropic
fluid and Chaplygin gas for this purpose. We have presented
accelerating expansion of our Universe due to interaction/without
interaction of the mixture of these fluids. We have found the
exact solution of the density and potential by considering a power
law expansion of the scale factor. We show that these potentials
represent the same decaying nature regardless the interaction
between the concerned fluids. Since we have considered a power law
expansion of the scale factor of the form $a=t^{n}$, we see that
for the present acceleration of the Universe to support the
observational data we need $n>1$. Now we consider the interaction
terms between these fluids. For the mixture of barotropic fluid
with tachyonic fluid, we see that the interaction term reduces the
potential. Also for the mixture of GCG with tachyonic fluid the
interaction parameter $\epsilon$ satisfying  $0<\epsilon<1$ so
that equation (6.34) exists for smaller values of $t$. In this
case also the interaction reduces the potential. Also if we
consider only tachyonic fluid with the power law expansion, we see
that the potential (which is obtained to be $V=\frac{3
n^{2}}{t^{2}} \sqrt{1-\frac{2}{3n}}$~) is greater than that we get
in mixtures. Also the potentials differ in the two cases we have
considered. For the mixture with GCG the potential decreases
faster than that in case of mixture with barotropic fluid.\\\\

%% file: chap7.tex
\large \baselineskip .85cm
\chapter{Interacting Model of Inhomogeneous EOS and Scalar Field}\label{chap7} \markright{\it
CHAPTER~\ref{chap7}.~Interacting Model of Inhomogeneous EOS and
Scalar Field}

\section{Prelude}

Presently we live in an epoch where the densities of the dark
energy and the dark matter are comparable. It becomes difficult
to solve this coincidence problem without a suitable interaction.
Generally interacting dark energy models are studied to explain
the cosmic coincidence problem [Cai and Wang, 2005]. Also the
transition from matter domination to dark energy domination can
be explained through an appropriate energy exchange rate.
Therefore, to obtain a suitable evolution of the Universe an
interaction is assumed and the decay rate should be proportional
to the present value of the Hubble parameter for good fit to the
expansion history of the Universe as determined by the Supernovae
and CMB data [Berger etal, 2006]. A variety of interacting dark
energy models have been proposed and studied for this purpose
[Zimdahl, 2005; Hu and Ling, 2006]. \\

Although a lot of models have been proposed to examine the nature
of the dark energy, it is not known what is the fundamental nature
of the dark energy. Usually models mentioned above are considered
for producing the present day acceleration. Also there is modified
gravity theories where the EOS depends on geometry, such as Hubble
parameter. It is therefore interesting to investigate models that
involve EOS different from the usual ones, and whether these EOS
is able to give rise to cosmological models meeting the present
day dark energy problem. In this chapter, we consider model of
interaction between scalar field and an ideal fluid with
inhomogeneous equation of state (EOS), through a phenomenological
interaction which describes the energy flow between them. Ideal
fluids with inhomogeneous EOS were introduced in [Nojiri etal,
2005, 2006; Elizalde etal, 2005]. Here we have considered two
exotic kind of equation of states which were studied in [Brevik
etal, 2004, 2007; Capozziello, 2006] with a linear inhomogeneous
EOS. Here we take the inhomogeneous EOS to be in polynomial form
to generalize the case. Also, the ideal fluid present here behaves
more like dark matter dominated by the scalar field so that the
total energy density and pressure of the Universe decreases with
time. Also the potential corresponding to the scalar field shows a
decaying nature. Here we have considered a power law expansion of
the scale factor, so that we always get a non-decelerated
expansion of the Universe for the power being greater than or
equal to unity. We have solved the energy densities of both the
scalar field and ideal fluid and the potential of the scalar
field. Also a decaying
nature of the interaction parameter is shown.\\

\section{Field Equations}

The metric of a spatially flat isotropic and homogeneous Universe
in FRW model is given by equation (1.7). The Einstein field
equations and energy conservation equation are in equations (4.2),
(4.3) and (4.4). Here, $\rho_{tot}$ and $p_{tot}$ are the total
energy density and the pressure of the Universe, given by,

\begin{equation}
\rho_{tot}=\rho_{\phi}+\rho_{d}
\end{equation}

and

\begin{equation}
p_{tot}=p_{\phi}+p_{d}
\end{equation}

with $\rho_{\phi}$ and $p_{\phi}$ are respectively the energy
density and pressure due to the scalar field given by equations
(5.14) and (5.15) respectively. Also, $\rho_{d}$ and $p_{d}$ are
the energy density and the pressure corresponding to the ideal
fluid with an inhomogeneous EOS,

\begin{equation}
p_{d}=\omega(t) {\rho}_{d}+{\omega}_{1}f(H,t)
\end{equation}

where, $\omega(t)$ is a function of $t$ and $f(H,t)$ is a function
of $H$ and $t$ ($H$ is the Hubble parameter $=\frac{\dot{a}}{a}$).\\

Now we consider the scalar field interacting with the ideal fluid
with inhomogeneous EOS through an energy exchange between them.
The equations of motion of the scalar field and the ideal fluid
can be written as,

\begin{equation}
\dot{{\rho}_{d}}+3H({\rho}_{d}+p_{d})=-3H{\rho}_{d}\delta
\end{equation}

and

\begin{equation}
\dot{{\rho}_{\phi}}+3H({\rho}_{\phi}+p_{\phi})=3H{\rho}_{d}\delta
\end{equation}

where $\delta$ is coupling constant.\\

\subsection{Case I: Model with EOS in power law form}

Taking into account the recent cosmological considerations of
variations of fundamental constants, one may start from the case
that the pressure depends on the time $t$ [Brevik etal, 2004].
Unlike the EOS studied in [Brevik etal, 2007] where the parameters
involved in EOS are linear in $t$, we consider rather a
polynomial form. First, we choose the EOS of the ideal fluid to
be,

\begin{equation}
p_{d}=a_{1} t^{-\alpha}{\rho}_{d}-c t^{-\beta}
\end{equation}

where, $a_{1}, c, \alpha, \beta$ are constants.\\

Here, we see that initially the pressure is very large and as
time increases pressure falls down, which is very much compatible
with the recent observational data.\\

We consider a Universe with power law expansion given by equation
(6.1), so as to get a non-decelerated expansion for $n \ge 1$, as
the deceleration parameter reduces to $q=-\frac{a\ddot{a}
}{\dot{a}^{2}}=\frac{1-n}{n}<0$.\\

Now equation (7.4) together with (7.6) and (6.1) gives the
solution for ${\rho}_{d}$ to be,

\begin{equation}
{\rho}_{d}=t^{-3n(1+\delta)}e^{\frac{3n a_{1}
t^{-\alpha}}{\alpha}} \left(\frac{3n
a_{1}}{\alpha}\right)^{\frac{3n(1+\delta)+\alpha-\beta}{\alpha}}\frac{c}{a_{1}}
~\Gamma(\frac{\beta-3n(1+\delta)}{\alpha},\frac{3n a_{1}
t^{-\alpha}}{\alpha})
\end{equation}

where, $\Gamma(a,x)$ is upper incomplete Gamma function.\\

Further substitution in the above equations give the solution for
${\rho}_{\phi}, {\dot{\phi}}^{2}$ and $V(\phi)$ to be,

\begin{equation}
{\rho}_{\phi}=3\frac{n^{2}}{t^{2}}-{\rho}_{d}
\end{equation}

\begin{equation}
{\dot{\phi}}^{2}=\frac{2n}{t^{2}}-\left[(1+a_{1}
t^{-\alpha}){\rho}_{d}-c t^{-\beta}\right]
\end{equation}

so that,

\begin{equation}
\phi=\phi_{0}+\int\sqrt{\frac{2n}{t^{2}}-\left[(1+a_{1}
t^{-\alpha}){\rho}_{d}-c t^{-\beta}\right]}~dt
\end{equation}

and

\begin{equation}
 V=\frac{3n^{2}-n}{t^{2}}+\frac{(-1+a_{1}
t^{-\alpha}){\rho}_{d}}{2}-\frac{c t^{-\beta}}{2}
\end{equation}

Since we have considered a power law expansion of the scale factor
so we can see from the above expressions that ${\rho}_{d}$ and
$\rho_{\phi}$ are decreasing functions of time so that the total
energy density as well as pressure decreases with time. The
evolution of the Universe therefore can be explained without any
singularity. Normalizing the parameters, we get the variation of
$V(\phi)$ against $\phi$ in figure 7.3. Equation (7.11) shows
that, for $\beta<2$, the potential being positive initially, may
not retain this as $t\rightarrow\infty$ (as the 3rd term dominates
over first term and the third and second term being negative for
large values); for $\beta=2$ the potential can be positive
depending on the value of $\left(3n^{2}-n-\frac{c}{2}\right)$ and
for, $\beta>2$ the potential can be either positive depending on
the choices of the constants, but always decreases with time.
Hence $\beta$ is completely arbitrary and depending on various
values of $\beta$ and the other constants, potential to be
positive, although it is always decreasing with time. Fig 7.3
shows the nature of the potential for arbitrarily chosen values of
the constants. Also if we consider $w_{d}=\frac{p_{d}}{\rho_{d}},
w_{\phi}=\frac{p_{\phi}}{\rho_{\phi}},
w_{tot}=\frac{p_{tot}}{\rho_{tot}}$, and plot them (figure 7.1)
against time, we see this represents an XCDM model and
therefore it makes a positive contribution to $\ddot{a}/a$.\\

\begin{figure}

\includegraphics{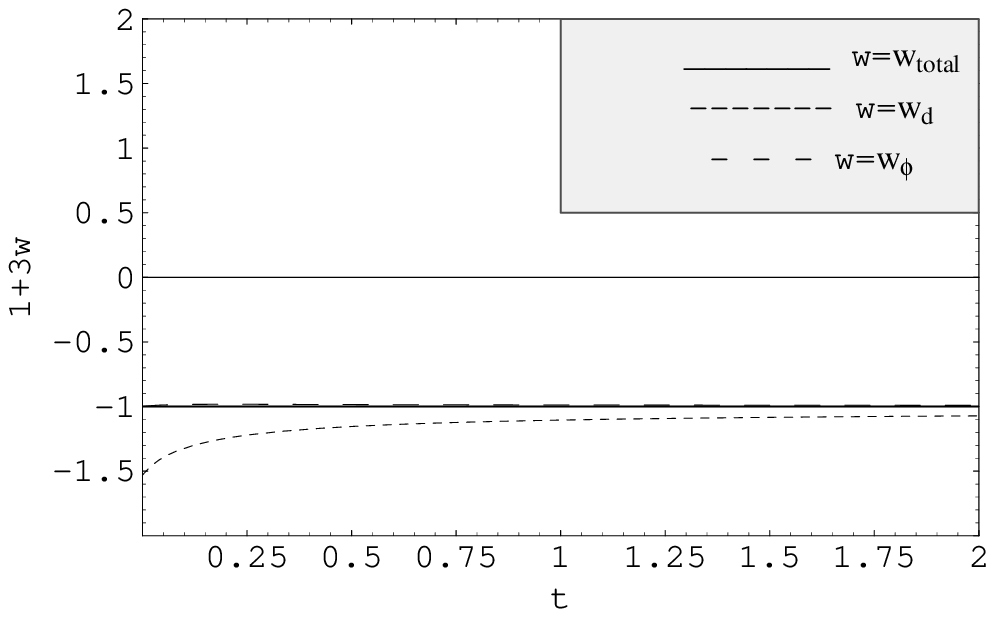}\\

\vspace{10mm} Fig 7.1: The variation of  $1+3w$ is plotted where
$w=w_{d}, w_{\phi}, w_{tot}$ against time, normalizing the
parameters as $n=2,  \alpha=1, \beta=2, a_{1}=.1, c=1,
\delta=.01$.\\

\vspace{1in}

\includegraphics{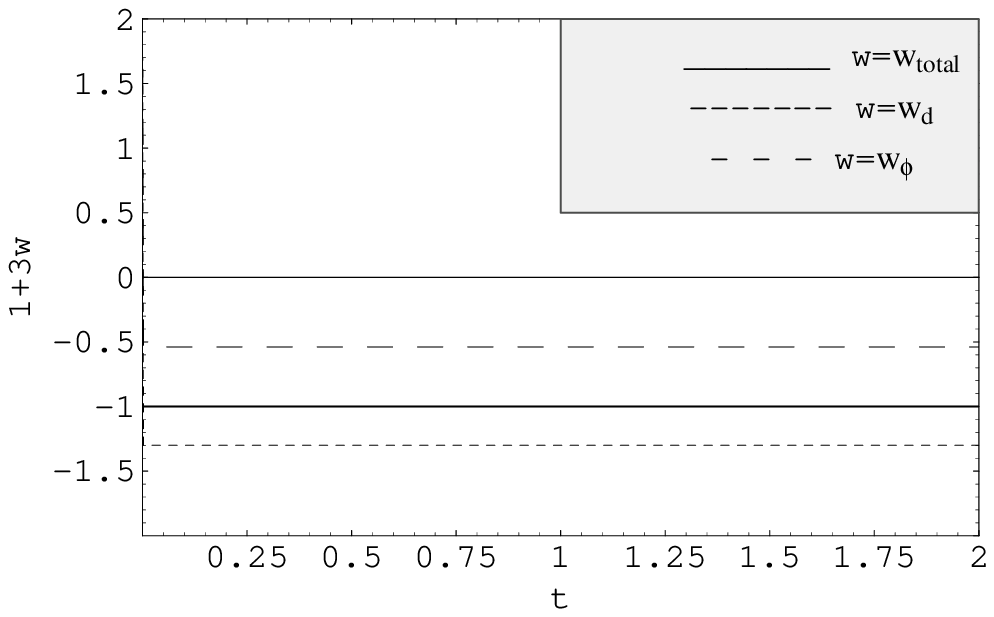}\\

 \vspace{10mm}
Fig 7.2: The variation of  $1+3w$ is plotted where $w=w_{d},
w_{\phi}, w_{tot}$ against time, normalizing the parameters as
$A=\frac{1}{3}, B=-2, \phi_{0}=1, \delta=.1, n = 2$.\\

\end{figure}

\subsection{Case II: Model with EOS depending on Hubble parameter}

Inhomogeneous dark energy EOS coming from geometry, for example,
$H$ can yield cosmological models which can avoid shortcomings
coming from coincidence problem and a fine-tuned sudden evolution
of the Universe from the early phase of deceleration driven by
dark matter to the present phase of acceleration driven by dark
energy. Furthermore, such models allow to recover also early
accelerated regimes with the meaning of inflationary behaviors
[Capozziello etal, 2006]. The following model is often referred to
as Increased Matter Model where the pressure depends on energy
density and $H$. A detailed discussion of this kind of
EOS can be found in ref. [Capozziello etal, 2006].\\

Now we choose the EOS of the ideal fluid to be,

\begin{equation}
p_{d}=A \rho_{d}+B H^{2}
\end{equation}

where, $A$ and $B$ are constants. \\

Considering the power law expansion (6.1) and using (7.4) and
(7.12), we get the solution for $\rho_d$ to be,

\begin{equation}
\rho_{d}=C_{0}
t^{-3n(1+A+\delta)}-\frac{3n^{3}B}{3n(1+A+\delta)-2}t^{-2}
\end{equation}

Further substitution in the related equations yields the solution
for $\rho_{\phi}, \phi, V(\phi)$ to be,

\begin{equation}
\rho_{\phi}=\frac{3n^{2}}{t^{2}}-\rho_{d}
\end{equation}

\begin{equation}
\phi=\phi_{0}+\frac{2}{2-K_{3}}\left[ \sqrt{K_{1}+K_{2}
t^{2-K_{3}}}-\sqrt{K_{1}} \sinh^{-1}
\left(\sqrt{\frac{K_{1}}{K_{2}}} x\right)\right]
\end{equation}

where, $x=t^{\frac{K_{3}}{2}-1}, K_{2}=-C_{0}(1+A),
K_{1}=\frac{6n^{2}(1+A+\delta)-4n-3Bn^{3}\delta+2B n^{2}
}{K_{3}-2}, K_{3}=3n(1+A+\delta)$\\

and

\begin{equation}
V=\frac{3n^{2}-n}{t^{2}}+\frac{A-1}{2} \rho_{d}+\frac{B n^{2}}{2
t^{2}}
\end{equation}

\begin{figure}

\includegraphics{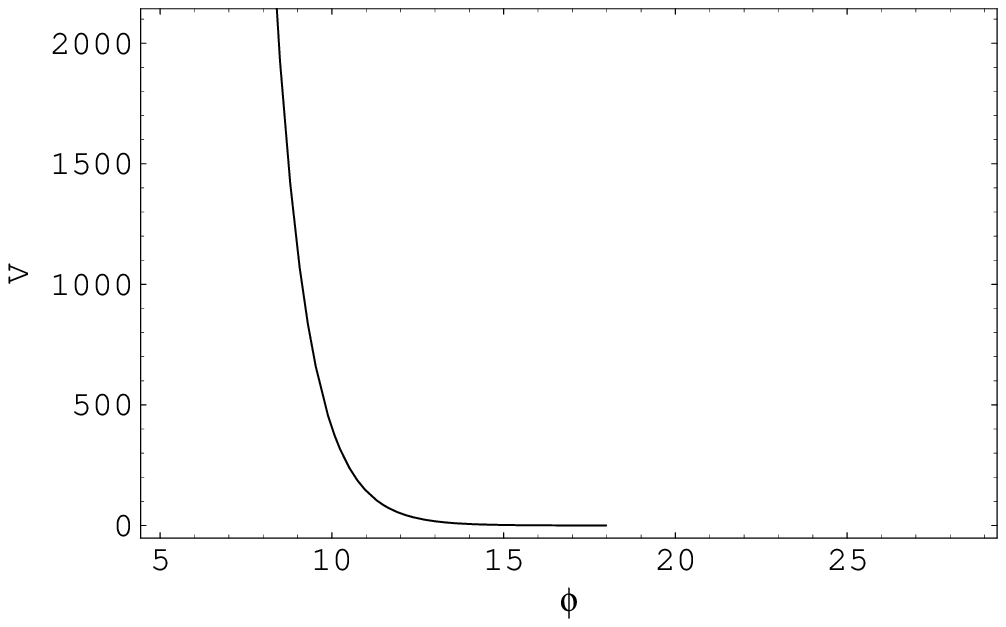}\\

\vspace{10mm} Fig 7.3: The variation of $V$ has been plotted
against $\phi$ normalizing the parameters as $n=2,  \alpha=1,
\beta=2, a_{1}=.1, c=1, \delta=.01$.\\

\vspace{1in}

\includegraphics{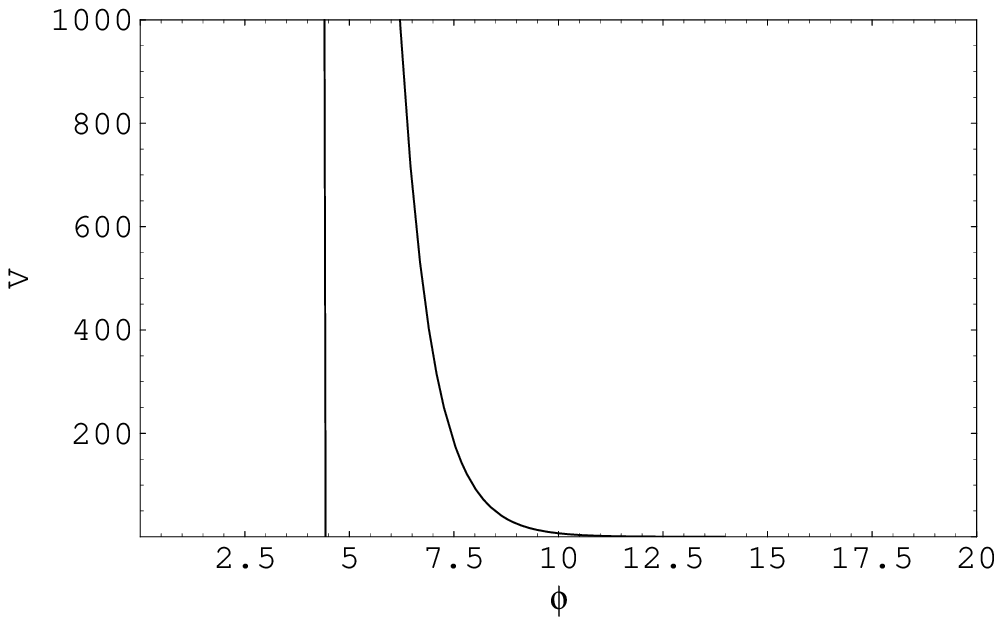}\\

\vspace{10mm} Fig 7.4: The variation of $V$ has been plotted
against $\phi$ normalizing the parameters as $A=\frac{1}{3},
B=-2, \phi_{0}=1, \delta=.1, n = 2$.\\

\end{figure}

Equation (7.15) shows that $K_{1}$ must be positive and hence
$K_{2}$ also must be positive for a valid expression. Also
equation (7.13) says that $C_{0}$ must be positive, otherwise
$\rho_{d}$ becomes negative initially. Therefore expression of
$K_{2}$ says that $A$ must be negative, in fact, $A<-1$, such that
depending on the value of $B$ pressure can be positive or
negative. Normalizing the parameters, we get the variation of $V$
against $\phi$ in figure 7.4. The figure shows a decaying nature
of the potential. Also if we consider
$w_{d}=\frac{p_{d}}{\rho_{d}},
w_{\phi}=\frac{p_{\phi}}{\rho_{\phi}},
w_{tot}=\frac{p_{tot}}{\rho_{tot}}$, and plot them (figure 7.2)
against time, like the previous case, we see this represents an
XCDM model and therefore it makes a positive
contribution to $\ddot{a}/a$.\\

\section{Discussion}

In this chapter, we study a cosmological model of the Universe in
which the scalar field has an interaction with an ideal fluid with
inhomogeneous EOS. The interaction is introduced
phenomenologically by considering  term parameterized by the
product of the Hubble parameter, the energy density of the ideal
fluid and a coupling constant in the equations of motion of the
fluid and the scalar field. This type of phenomenological
interaction term has been investigated in [Cai and Wang, 2005].
This describes an energy flow between the scalar field and the
ideal fluid. Also we consider a power law form of the scale factor
$a(t)$ to keep the recent observational support of cosmic
acceleration. For the first model putting $c=0, \alpha=0$ we get
the results for barotropic fluid. Here for $\alpha$ and $\beta$ to
be positive, the ideal fluid and the scalar field behave as dark
energy. Also we see that the interaction term decreases with time
showing strong interaction at the earlier stage and weak
interaction later. Also the potential corresponding to the scalar
field is positive and shows a decaying nature. In the second model
where $p_{d}$ is a function of $\rho_{d}$ and the Hubble parameter
$H$, we see that the energy density and the pressure of the ideal
fluid and that of the scalar field always decreases with time.
From figures 7.3 and 7.4, we see that, the potential function $V$
decreases for both decelerating ($n<1$) and accelerating phase
($n>1$). Also from the values of density and pressure terms, it
can be shown that the individual fluids and their mixtures satisfy
strong energy condition for $n<1$ and violate for $n>1$. A
detailed discussion of the potential of a scalar field can be
found in ref. [Cardenas etal, 2004]. We see that the coupling
parameter shows a decaying nature in both the cases implying
strong interaction at the early times and weak interaction later.
Thus following the recipe provided in ref. [Padmanabhan, 2002] we
can establish a model which can be a suitable alternative to dark
energy explaining the decaying energy flow between the scalar
field and the fluid and giving rise to a decaying potential. As a
scalar field with potential to drive acceleration is a common
practice in cosmology [Padmanabhan, 2002], the potential presented
here can reproduce enough acceleration together with the ideal
fluid, thus explaining the evolution of the Universe. Also we have
considered inhomogeneous EOS interacting with the scalar field
which can represent an alternative to the usual dark energy model.
However, stability analysis and spatial inhomogeneity analysis
[Peebles etal, 1988; Rara etal, 1988] are more complicated for our
investigation, since we are considering the ideal fluid with two
types of equation of states and are analysing whether they can be
considered as an alternative to dark energy. Also we have seen
that the conservation equation (7.4) together with the given form
of the pressure (7.6) and (7.12) are difficult to solve unless we
consider the power law form (6.1). Once the power law form is
considered, we can easily find exact solution of $\rho_{d}$ [from
equation (7.4))] and hence $\rho_{\phi}$ [from equation(4.2)],
which lead to the given expression for the potential $V(\phi)$
[from eqs. (5.14), (5.15)] analytically. Though this is the
backward approach, but otherwise if we start from $V(\phi)$ i.e.,
say $V(\phi)=V_{0}Exp(-k \phi)$, we cannot find any exact solution
of $\rho_{d}, \rho_{\phi}, p_{d},p_{\phi}, \phi, a $. So we can
only draw conclusions graphically, not analytically. For example,
Ellis et al [1991] have discussed for the model with radiation
and scalar field and found exact solutions in the backward approach. \\

%% file: chap8.tex
\large \baselineskip .85cm
\chapter{Perfect Fluid Dynamics in Brans-Dicke Theory}\label{chap8} \markright{\it
CHAPTER~\ref{chap8}.~Perfect Fluid Dynamics in Brans-Dicke Theory}

\section{Prelude}

Brans-Dicke (BD) theory has been proved to be very effective
regarding the recent study of cosmic acceleration [Banerjee and
Pavon, 2001]. As we have already discussed BD theory is explained
by a scalar function $\phi$ and a constant coupling constant
$\omega$, often known as the BD parameter. This can be obtained
from general theory of relativity (GR) by letting $\omega
\rightarrow \infty$ and $\phi$ = constant [Sahoo and Singh, 2003].
This theory has very effectively solved the problems of inflation
and the early and the late time behaviour of the Universe.
Banerjee and Pavon [2001] have shown that BD scalar tensor theory
can potentially solve the quintessence problem. The generalized BD
theory [Bergmann, 1968; Nordtvedt, 1970; Wagoner, 1970] is an
extension of the original BD theory with a time dependent coupling
function $\omega$. In Generalized BD theory, the BD parameter
$\omega$ is a function of the scalar field $\phi$. Banerjee and
Pavon have shown that the generalized BD theory can give rise to a
decelerating radiation model  where the big-bang nucleosynthesis
scenario is not adversely affected. Modified BD theory with a
self-interacting potential have also been introduced in this
regard. Bertolami and Martins [2000] have used this theory to
present an accelerated Universe for spatially flat model. All
these theories conclude that $\omega$ should have a low negative
value in order to solve the cosmic acceleration problem. This
contradicts the solar system experimental bound $\omega\geq500$.
However Bertolami and Martins [2000] have obtained the solution
for accelerated expansion with a potential ${\phi}^{2}$ and large
$|\omega|$, although they have not considered the positive energy
conditions for
the matter and scalar field.\\

In this chapter, we investigate the possibilities of obtaining
accelerated expansion of the Universe in BD theory where we have
considered a self-interacting potential $V$ which is a function of
the BD scalar field $\phi$ itself and a variable BD parameter
which is also a function of $\phi$. We show all the cases of
$\omega$ = constant, $\omega=\omega(\phi)$, $V=0$ and $V=V(\phi)$
to consider all the possible solutions. We examine these solutions
for both barotropic fluid and the GCG, to get a generalized view
of the results in the later case. We analyze the conditions under
which we get a negative $q$ (deceleration parameter, $-\frac{a
\ddot{a}}{{\dot{a}^{2}}}$) in all the models of the Universe. For
this purpose we have shown the graphical representations of these
scenario for further discussion.\\

\section{Field Equations}

The self-interacting BD theory is described by the action
(choosing $8\pi G_{0}=c=1$), given by equation (1.107), where
$V(\phi)$ is the self-interacting potential for the BD scalar
field $\phi$ and $\omega(\phi)$ is modified version of the BD
parameter which is a function of $\phi$ [Sahoo and Singh, 2003].
The matter content of the Universe is composed of perfect fluid
given by equation (1.42). From the action (1.107), we obtain the
field equations (1.108), where,

\begin{equation}
^{\fbox{}}~\phi=\frac{1}{3+2\omega(\phi)}T-\frac{1}{3+2\omega(\phi)}\left[2V(\phi)-\phi
\frac{dV(\phi)}{d\phi}\right]-\frac{\frac{d\omega(\phi)}{d\phi}}{3+2\omega(\phi)}{\phi,}_{\mu}
{\phi}^{,\mu}
\end{equation}

and $T=T_{\mu \nu}g^{\mu \nu}$.\\

The line element for Friedman-Robertson-Walker space-time is given
by equation (1.7). The Einstein field equations for the metric
(1.7) and the wave equation for the BD scalar field $\phi$ are
respectively given by equation (1.111), (1.112) and

\begin{equation}
\ddot{\phi}+3\frac{\dot{a}}{a}
\dot{\phi}=\frac{\rho-3p}{3+2\omega(\phi)}+\frac{1}{3+2\omega(\phi)}\left[2V(\phi)-\phi
\frac{dV(\phi)}{d\phi}\right]-\dot{\phi}\frac{\frac{d\omega(\phi)}{dt}}{3+2\omega(\phi)}
\end{equation}

The energy conservation equation is (1.20).\\

Now we consider two types of fluids, first one being the
barotropic perfect fluid and the second one is GCG.\\

\section{Model with Barotropic Fluid in the Background}

Here we consider the Universe to be filled with barotropic fluid
with EOS

\begin{equation}
p=\gamma \rho~~~~~~~~~~~~~~~~(-1\le\gamma\le 1)
\end{equation}

The conservation equation (1.20) yields the solution for $\rho$
as,

\begin{equation}
\rho=\rho_{0} a^{-3(\gamma+1)}
 \end{equation}

where $\rho_{0}(>0)$ is an integration constant.\\

\subsection{Solution Without Potential}

\begin{figure}

\includegraphics{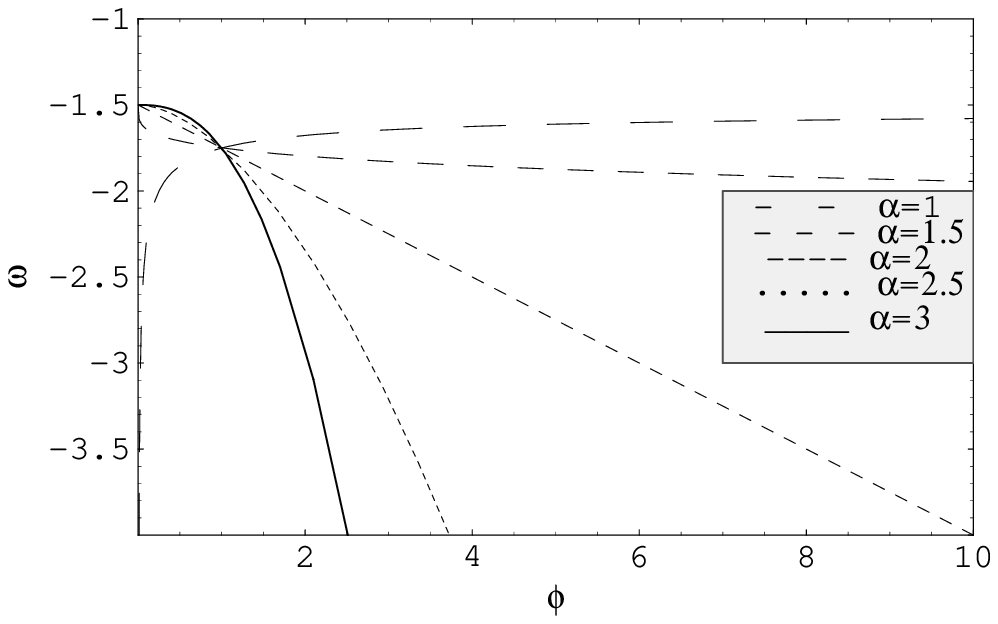}\\

\vspace{10mm} Fig 8.1: The variation of $\omega$ (for barotropic
fluid) has been plotted against $\phi$ for different values of
$\alpha=1, 1.5, 2, 2.5,3$ in a flat ($k=0$) dust filled
($\gamma=0$ and $\beta=-2({\text dust})$) epoch, normalizing the
parameters as
$a_{0}=\rho_{0}=\phi_{0}=1$.\\

\vspace{1in}

\includegraphics{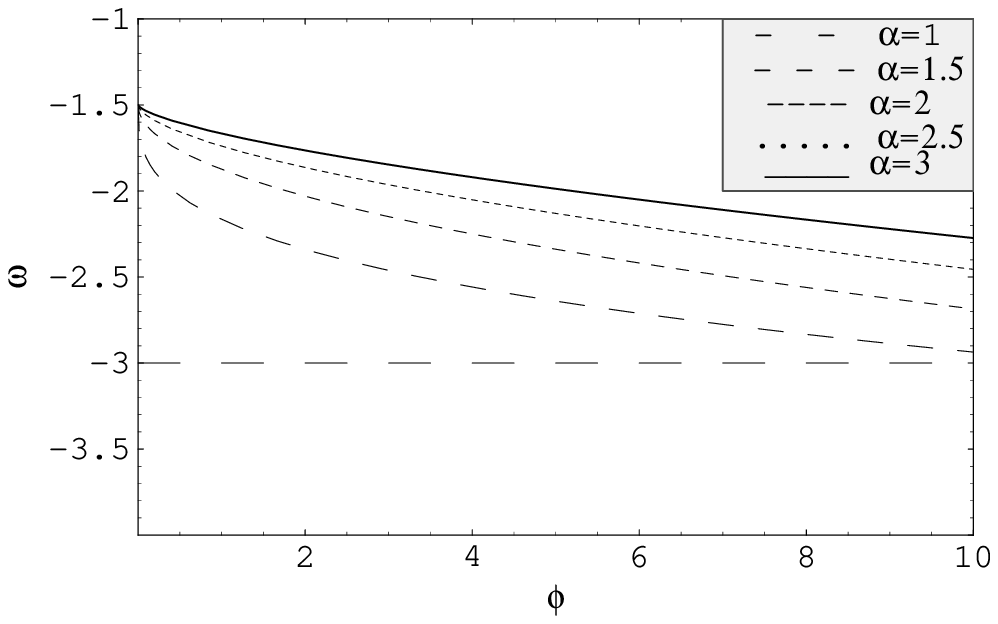}\\

\vspace{10mm} Fig 8.2: The variation of $\omega$ (for barotropic
fluid) has been plotted against $\phi$ for different values of
$\alpha=1, 1.5, 2, 2.5,3$ in a closed model ($k=1$) in radiation
($\gamma=\frac{1}{3}$ and $\beta=-2\alpha$) era, normalizing the
parameters as $a_{0}=\phi_{0}=1,
~\rho_{0}=6$.\\

\end{figure}

{\bf{Case I}:} First we choose $\omega(\phi)=\omega$ = constant.\\

Now we consider power law form of the scale factor

\begin{equation}
a(t)=a_{0} t^{\alpha}     ~~~~~(\alpha \ge 1)
\end{equation}

In view of equations (8.3) and (8.4), the wave equation (8.2)
leads to the solution for $\phi$ to be

\begin{equation}
\phi=\frac{\rho_{0}
{a_{0}}^{-3(1+\gamma)}t^{2-3\alpha(1+\gamma)}}{(2\omega+3)(1-3\alpha
\gamma) \left[2-3\alpha(1+\gamma)\right] }
\end{equation}

For $k\neq 0$ we get from the field equations (1.111) and (1.112),
the value of $\alpha=1$ and

\begin{equation}
(3\gamma+1)\left[\frac{\omega}{2}(\gamma-1)(3\gamma+1)-1-\frac{k}{{a_{0}}
^{2} }\right]=0
\end{equation}

We have seen that $\gamma \ne -\frac{1}{3}$ and we have

\begin{equation}
\omega=\frac{2(1+\frac{k}{{a_{0}} ^{2}})}{(\gamma-1)(3\gamma+1)}
\end{equation}

Since $\omega$ must be negative for $-\frac{1}{3}<\gamma<1$, we
have seen that for this case the deceleration parameter $q=0$,
i.e., the universe is in a state of uniform expansion. For $k=0$,
the field equations yield

\begin{equation}
\left[2-3\alpha(\gamma+1)\right]\left[2(2\alpha-1)+\omega(\gamma-1)\{2-3\alpha(\gamma+1)\}\right]=0
\end{equation}

From equation (8.9) we have two possible solutions for $\alpha$:\\
$\alpha=\frac{2}{3(\gamma+1)}$ ~~  for $-1<\gamma < -\frac{1}{3}$\\
 and\\
$\alpha=\frac{2\left[1+\omega(1-\gamma)\right]}{\left[4+3\omega(1-\gamma^{2})\right]}$
~~ for $-\frac{1}{3}<\gamma<1$\\

For these values of $\alpha$, we have seen that $\omega<0$ and
the deceleration parameter $q<0$. Thus for $k=0$ with the power
law form of the scale factor $a=a_{0} t ^{\alpha}$ it is possible
to get the accelerated expansion of the Universe.\\

\begin{figure}

\includegraphics{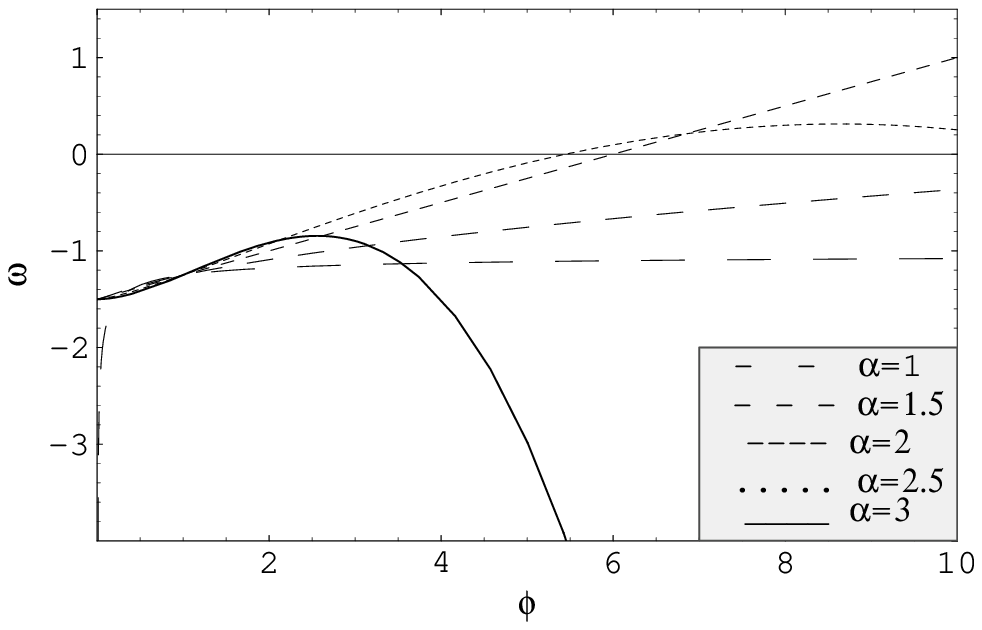}\\

\vspace{10mm} Fig 8.3: The variation of $\omega$ (for dust) has
been plotted against $\phi$ for respectively closed model of the
Universe in the present dust filled epoch, i.e., $\gamma=0$ and
$\beta=-2$. We take different values of $\alpha=1, 1.5, 2, 2.5,3$
and
normalize the parameters as $a_{0}=\rho_{0}=\phi_{0}=1$.\\

\vspace{1in}

\includegraphics{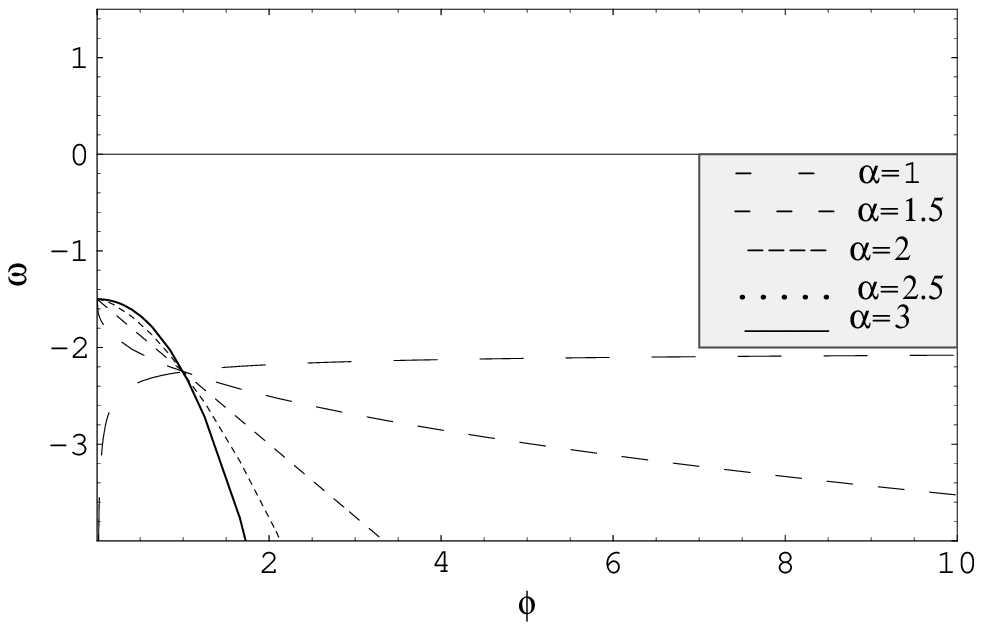}\\

\vspace{10mm} Fig 8.4: The variation of $\omega$ (for dust) has
been plotted against $\phi$ for respectively open model of the
Universe in the present dust filled epoch, i.e., $\gamma=0$ and
$\beta=-2$. We take different values of $\alpha=1, 1.5, 2, 2.5,3$
and
normalize the parameters as $a_{0}=\rho_{0}=\phi_{0}=1$.\\

\end{figure}

{\bf{Case II}:}  Now we choose $\omega=\omega(\phi)$ to be
variable. Here we consider the power law form of $\phi$ as

\begin{equation}
\phi(t)=\phi_{0} t^{\beta}
\end{equation}

with  the power law from of $a(t)$ given by equation (8.5).\\

Proceeding as above we get

\begin{equation}
\omega=\frac{\alpha \beta+2\alpha+\beta-\beta ^{2}}{\beta
^{2}}-\frac{1+\gamma}{\beta ^{2}} \rho_{0} {a_{0}}^{-3(1+\gamma)}
{\phi_{0}}^{\frac{3\alpha(\gamma+1)-2}{\beta}}
\phi^{-\frac{3\alpha(\gamma+1)+\beta-2}{\beta}}+\frac{2k}{{a_{0}}^{2}
\beta^{2} {\phi_{0}} ^{\frac{2(1-\alpha)}{\beta}}}
\phi^{\frac{2(1-\alpha)}{\beta}}
\end{equation}

Now for acceleration $q<0$ implies that $\alpha>1$.
Using the other equations we arrive at two different situations:\\

$(i)$ First considering the flat Universe model, i.e., $k=0$, we
get, $\beta=1-3\alpha$, i.e., $\beta<-2$ for
$\gamma>\frac{1}{3}$~,  $\beta=-2\alpha$, i.e., $\beta<-2$ (as
$\alpha>1$) for $\gamma=\frac{1}{3}$ and  $\beta=-2$ for
$\gamma<\frac{1}{3}$. That is cosmic acceleration
can be explained at all the phases of the Universe with different values of $\beta$ where $\phi=\phi_{0} t^{\beta}$\\

$(ii)$ If we consider the non-flat model of the Universe, i.e.,
$k\ne 0$, we are left with two options. For closed model of the
Universe, i.e., for $k=1$ we can explain cosmic acceleration for
the radiation phase only and for that $\beta=-2\alpha$ giving
$\beta<-2$ and $6\phi_{0} {a_{0}}^{2}=\rho_{0}$, whereas we do
not get any such possibility
for the open model of the Universe.\\

Now preferably taking into account the recent measurements
confirming the flat model of the Universe, if $\beta=-2$ we see
that we have an accelerated expansion of the Universe after the
radiation period preceded by a decelerated expansion before the
radiation era and a phase of uniform expansion at the radiation
era itself. Also if $\beta<-2$ cosmic acceleration is followed by
a deceleration phase as $\alpha< 1$ for $\gamma<\frac{1}{3}$.\\

\subsection{Solution With Potential}

\begin{figure}

\includegraphics{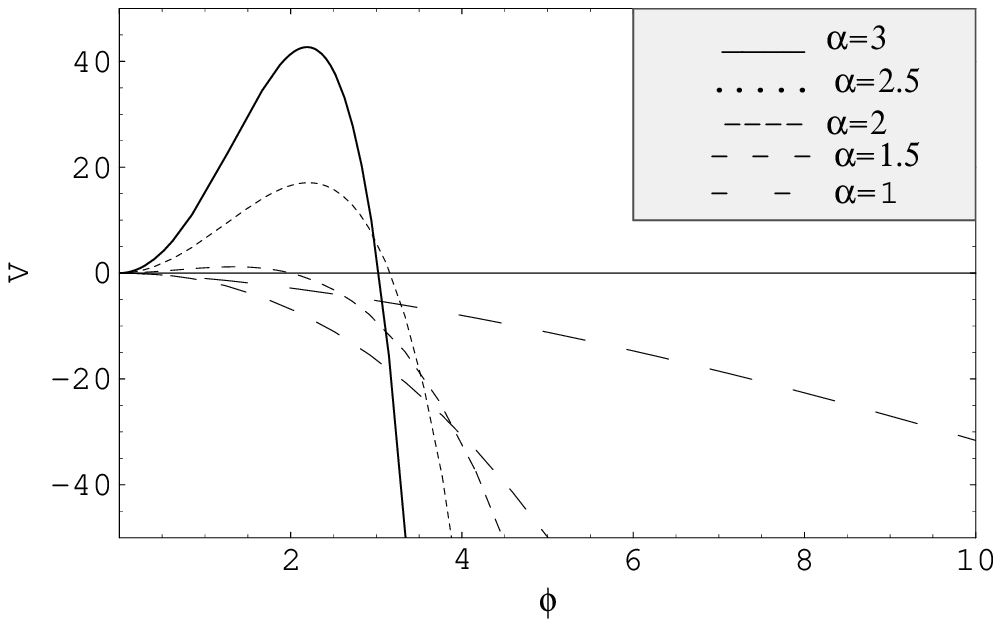}\\
\vspace{.5mm} Fig 8.5: Variation of $V$ (for dust) has been
plotted against $\phi$ for flat ($k=0$) model of the Universe in
dust filled epoch ($\gamma=0$). We have considered different
values of $\alpha=1, 1.5, 2, 2.5,3$ and $\beta=-2$ and normalize
the parameters as $a_{0}=\rho_{0}=\phi_{0}=1$.\\

\vspace{2mm}

\includegraphics{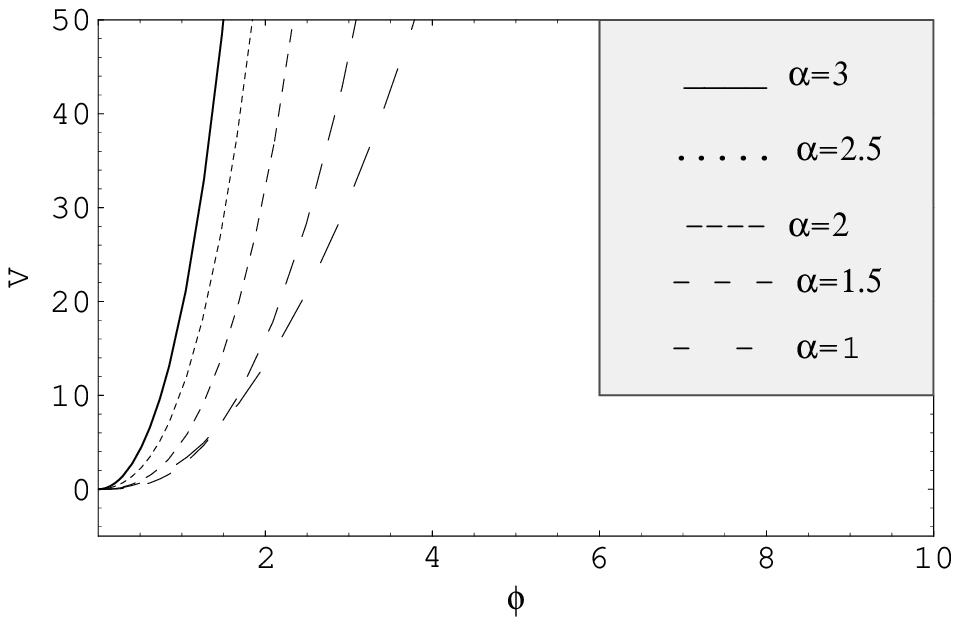}\\
\vspace{.5mm} Fig 8.6: Variation of $V$ (for dust) has been
plotted against $\phi$ for closed ($k=1$) model of the Universe
in dust filled epoch ($\gamma=0$). We have considered different
values of $\alpha=1, 1.5, 2, 2.5,3$ and $\beta=-2$ and normalize
the parameters as $a_{0}=\rho_{0}=\phi_{0}=1$.\\

\vspace{2mm}

\includegraphics{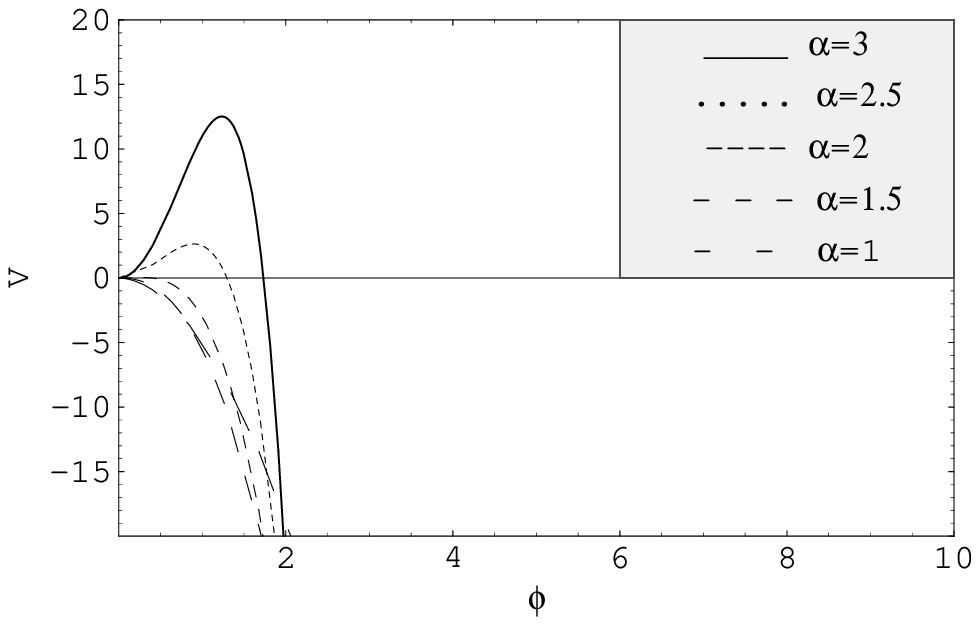}\\
\vspace{.5mm} Fig 8.7: Variation of $V$ (for dust) has been
plotted against $\phi$ for open ($k=-1$) model of the Universe in
dust filled epoch ($\gamma=0$). We have considered different
values of $\alpha=1, 1.5, 2, 2.5,3$ and $\beta=-2$ and normalize
the parameters as $a_{0}=\rho_{0}=\phi_{0}=1$.\\

\end{figure}

{\bf{Case I}:} Let us choose $\omega(\phi)=\omega=$ constant.\\

In this case instead of considering equations (8.5) and (8.10) we
consider only one power law form

\begin{equation}
\phi=\phi_{0} a^{\alpha}
\end{equation}

Using equation $(8.12)$ in equations $(1.111)$ and $(1.112)$ we
get

$$
\dot{a}={\left[2k+2(1+\gamma)\frac{\rho_{0}}{\phi_{0}}\frac{a^{-3\gamma-\alpha-1}}{\{3\gamma
\alpha+6\gamma-\alpha^{2}+7\alpha+6-2\omega \alpha^{2}\}
}\right]}^{\frac{1}{2}} $$

Putting $k=0$, we get

\begin{equation}
a=A t^{\frac{2}{3+\alpha+3\gamma}}
\end{equation}

where $A={\left[\frac{\rho_{0}
(1+\gamma)(3+\alpha+3\gamma)^{2}}{2\rho_{0}\{6(1+\gamma)+
\alpha(7+3\gamma)-\alpha^{2}(1+2\omega)\}}\right]}^{\frac{1}{3+\alpha+3\gamma}}$.\\

Therefore, $\phi=B t^{\frac{2\alpha}{3+\alpha+3\gamma}}$ where,
$B=\phi_{0} A^{\alpha}$.\\

Now, if $\frac{2}{3+\alpha+3\gamma}\ge 1$, we get

\begin{equation}
\alpha\le -(1+3\gamma)
\end{equation}

Substituting these values in $(1.111), ~(1.112), ~(8.2)$, the
solution for the potential $V$ is obtained as,
$V=\frac{B'}{\phi^{\frac{3+3\gamma}{\alpha}}}$ where,
$B'=-\frac{2B\{6-18\alpha+6\omega \alpha+6\omega \alpha
\gamma-18\gamma}{3(3+3\gamma+\alpha)^{2}(1+\gamma)}$.\\

Also, the deceleration parameter reduces to, $q=-\frac{a
\ddot{a}}{{\dot{a}}^{2}}=\frac{3\gamma+\alpha+1}{2}\le 0$ (using equation (8.14))\\

Hence, the present Universe is in a state of expansion with
acceleration.\\

Also, we get
$\omega=-\frac{6\gamma(1+\gamma)}{\alpha}-\frac{3+\alpha}{2\alpha}$
and, $\alpha=-\frac{3(1+2\gamma)^{2}}{1+2\omega}$.\\

Also, $\gamma\ge -1 \Rightarrow \alpha \le 2$ and $\omega \ge
-\frac{5}{4}$. For the present Universe (i.e., taking $\gamma=0$)
and the $\Lambda$CDM model,
$\omega=-\frac{3+\alpha}{2\alpha}$.\\\\

\begin{figure}

\includegraphics{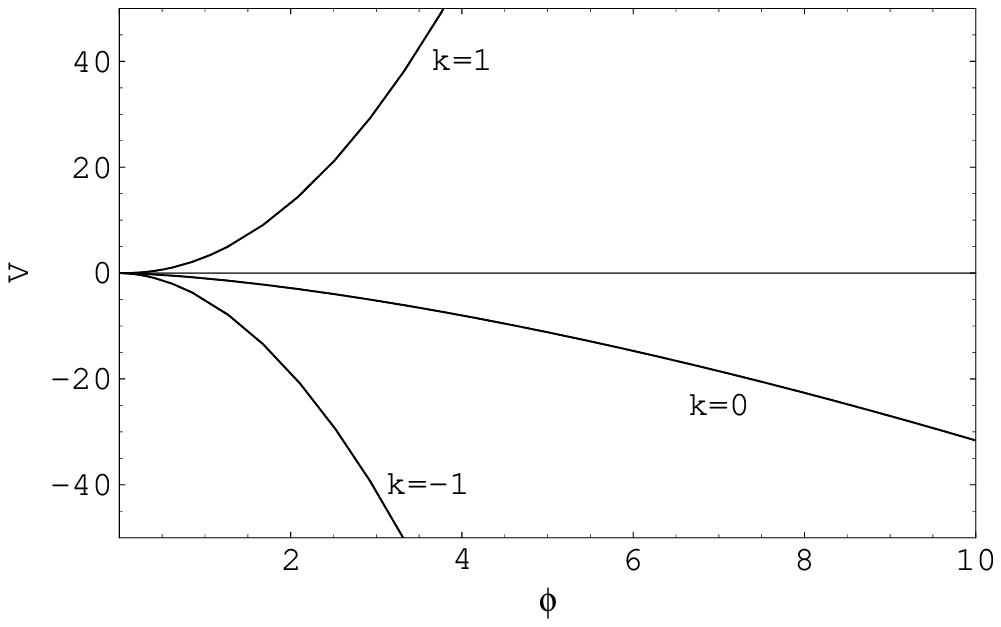}\\

\vspace{10mm}Fig 8.8: Variation of $V$ (for dust) is plotted
against the variation of $\phi$ for all the models of the
Universe. We have considered different values of $\alpha=1, 1.5,
2, 2.5,3$ and the present dust filled epoch, i.e., $\gamma=0$,
normalizing the parameters as $a_{0}=\rho_{0}=\phi_{0}=1$ and
$\beta=-2\alpha$. The results for different values of $\alpha$
coincides with each
other in each model of the Universe.\\

\vspace{1in}

\includegraphics{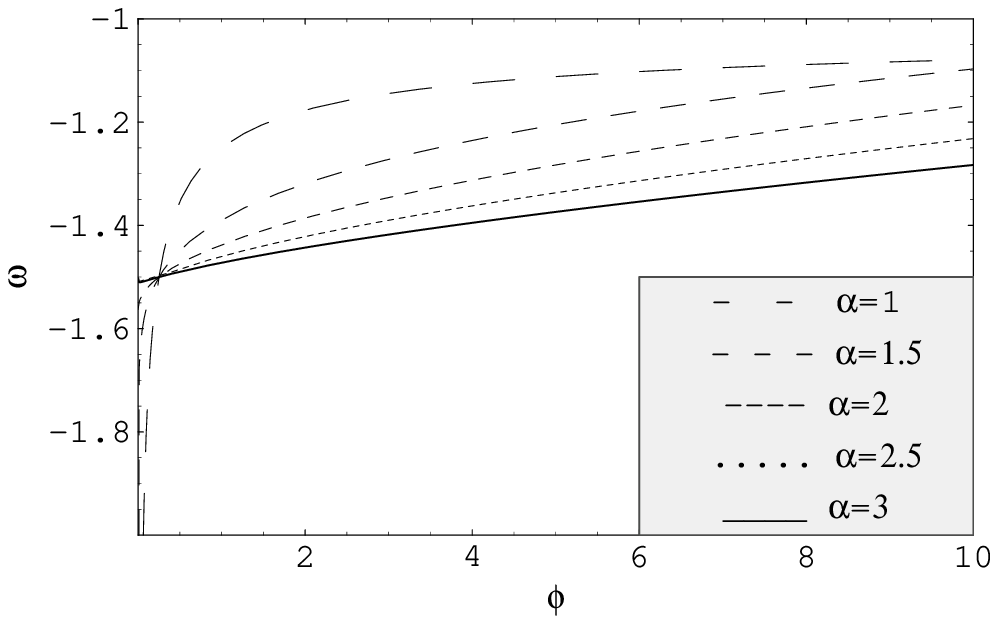}\\

\vspace{10mm}Fig 8.9: Variation of $\omega$ (for dust) is plotted
for closed ($k=1$) model of the Universe. We have considered
different values of $\alpha=1, 1.5, 2, 2.5,3$ and the present
dust filled epoch, i.e., $\gamma=0$, normalizing the parameters as
$a_{0}=\rho_{0}=\phi_{0}=1$ and $\beta=-2\alpha$.\\

\end{figure}

{\bf{Case II}:}  Now we choose $\omega(\phi)$ to be dependent on
$\phi$. Again we consider the power law forms (8.5) and (8.10).
Solving the equations in a similar manner, we get

\begin{equation}
\omega=\frac{\alpha \beta+2\alpha+\beta-\beta ^{2}}{\beta
^{2}}-\frac{1+\gamma}{\beta ^{2}} \rho_{0} {a_{0}}^{-3(1+\gamma)}
{\phi_{0}}^{\frac{3\alpha(\gamma+1)-2}{\beta}}
\phi^{-\frac{3\alpha(\gamma+1)+\beta-2}{\beta}}+\frac{2k}{{a_{0}}^{2}
\beta^{2} {\phi_{0}} ^{\frac{2(1-\alpha)}{\beta}}}
\phi^{\frac{2(1-\alpha)}{\beta}}
\end{equation}

and

\begin{equation}
V(\phi)=(2\alpha+\beta)(3\alpha+2\beta-1)
{\phi_{0}}^{\frac{2}{\beta}}
\phi^{\frac{\beta-2}{\beta}}-(1-\gamma) \rho_{0}
{a_{0}}^{-3(1+\gamma)}
{\phi_{0}}^{\frac{3\alpha(\gamma+1)}{\beta}}
\phi^{-\frac{3\alpha(\gamma+1)}{\beta}}+\frac{4k}{{a_{0}}^{2}}\phi^{\frac{\beta-2\alpha}{\beta}}{\phi_{0}}^{\frac{2\alpha}{\beta}}
\end{equation}

Substituting these values in equation (8.2), we get

\begin{equation}
\text {either}~~~~ \beta=-2   ~~~~~\text {or} ~~~~\beta=-2\alpha
\end{equation}

Therefore for cosmic acceleration $q<0\Rightarrow \alpha>1$ and
$\beta\le -2$.\\

Therefore for the present era,

\begin{eqnarray*}
\omega=-\frac{3}{2}-\frac{\rho_{0}{a_{0}}^{-3}}{4\phi_{0}}
t^{\frac{3\alpha-4}{2}}~~~~~\text {and}~~~V=2(\alpha-1)(3\alpha-5)
\phi_{0} t -t^{\frac{3\alpha}{2}}\rho_{0}{a_{0}}^{-3} ~~~\text
{if}~~~ \beta=-2
\end{eqnarray*}
\begin{equation}
\omega=-\frac{3}{2}-\frac{\rho_{0}{a_{0}}^{-3}}{\phi_{0}}
t^{\frac{\alpha-2}{2\alpha}}~~~~\text{and} ~~~
V=-\phi_{0}{a_{0}}^{-3}t^{\frac{3}{2}}~~~\text{if} ~~~\beta<-2
\end{equation}

Also for vacuum dominated era,

\begin{eqnarray*}
\omega=-\frac{3}{2} ~~~~\text{and}~~~~ V=
2(\alpha-1)(3\alpha-5)t-2\rho_{0}~~~~\text {for} ~~~\beta=-2
\end{eqnarray*}
\begin{equation}
\omega=-\frac{3}{2}    ~~~\text {and} ~~~~V=-2\rho_{0}~~~\text
{for} ~~~\beta<-2
\end{equation}

\section{Model with GCG in the Background}

Here we consider the Universe to be filed with Generalized
Chaplygin Gas with EOS

\begin{equation}
p=-\frac{B}{\rho^{n}}
\end{equation}

Here the conservation equation (1.20) yields the solution for
$\rho$ as,

\begin{equation}
\rho=\left[B+\frac{C}{a^{3(1+n)}}\right]^{\frac{1}{(1+n)}}
\end{equation}

where $C$ is an integration constant.

\subsection{Solution Without Potential}

{\bf{Case I}:} First we choose $\omega(\phi)=\omega=$ constant.\\

We consider the power law form

\begin{equation}
\phi=\phi_{0} a^{\alpha}
\end{equation}

Equations (1.111), (1.112) and (8.2) give,

\begin{equation}
(2\omega \alpha-6)\ddot{a}+(\omega \alpha^{2}+4\omega
\alpha-6)\frac{{\dot{a}}^{2}}{a}=\frac{6k}{a}
\end{equation}

which yields the solution,

\begin{equation}
\dot{a}=\sqrt{\frac{6k}{P(\omega \alpha-3)}+K_{0}a^{-P}}
\end{equation}

where $P=\frac{\omega \alpha^{2}+4\omega \alpha-6}{\omega
\alpha-3}$ and $K_{0}$ is an integration constant.\\

First we consider $P>0$. Multiplying both sides of equation (8.24)
by $a^{P}$ after squaring it, we get $K_{0}=0$, therefore giving,
$a=\sqrt{\frac{6k}{(\omega \alpha-3)P}}t$.\\

Hence for flat Universe, we get, $a=$ constant.\\

For open model, we must have $\omega \alpha<3$ and
$a=\sqrt{\frac{6}{(3-\omega \alpha)P}}t$, whereas, for closed
model, $\omega \alpha>3$ and $a=\sqrt{\frac{6}{(\omega
\alpha-3)P}}t$. In all cases $q=0$, i.e., we get uniform
expansion.\\

If $P=0$, $a \ddot{a}=\frac{3}{\omega \alpha-3}k$, i.e.,
$\dot{a}^{2}=\frac{6k}{\omega \alpha-3}\ln{a}+K_{0}$.\\

If $k=0$, $a=\sqrt{K_{0}}t+C_{0}$, ($C_{0}$ is an integration
constant) causing $q=0$, i.e., uniform expansion again.\\

\begin{figure}

\includegraphics{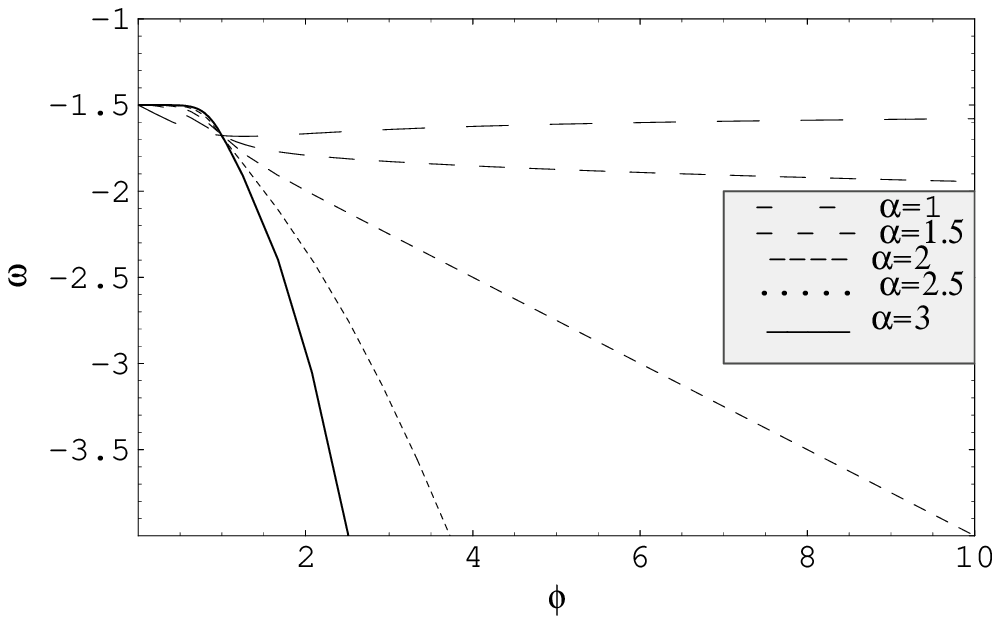}\\
\vspace{.5mm} Fig 8.10: Variation of $\omega$ (for GCG) is shown
against $\phi$ for flat ($k=0$) model of the Universe. We have
considered different values of $\alpha=1, 1.5, 2, 2.5,3$ and
$\beta=-2$ and normalize the parameters as
$a_{0}=\rho_{0}=\phi_{0}=B=C=1$ and
$n=1$.\\

\vspace{2mm}

\includegraphics{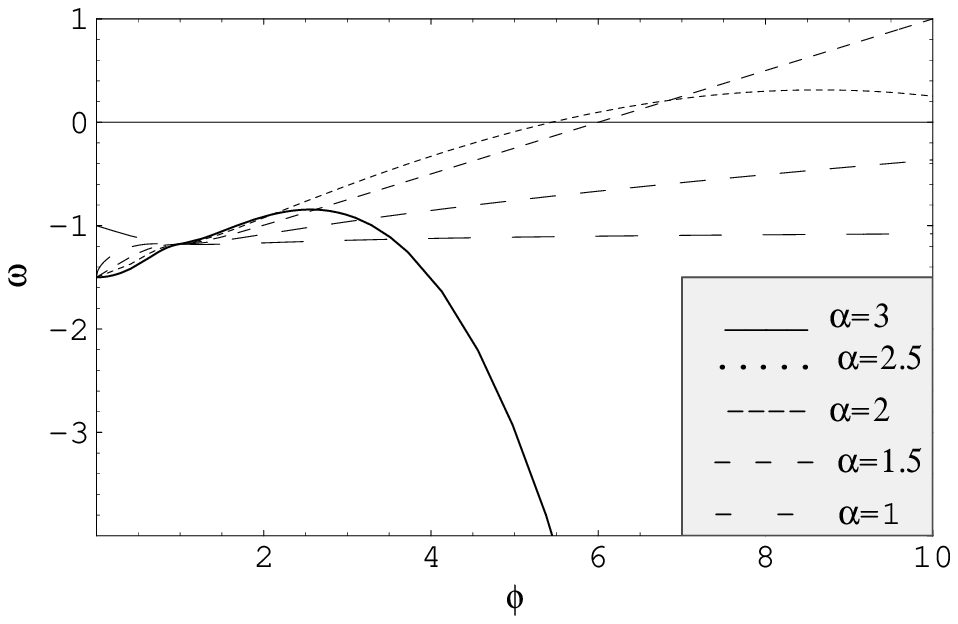}\\
\vspace{.5mm} Fig 8.11: Variation of $\omega$ (for GCG) is shown
against $\phi$ for closed ($k=1$) model of the Universe. We have
considered different values of $\alpha=1, 1.5, 2, 2.5,3$ and
$\beta=-2$ and normalize the parameters as
$a_{0}=\rho_{0}=\phi_{0}=B=C=1$ and $n=1$.\\

\vspace{2mm}

\includegraphics{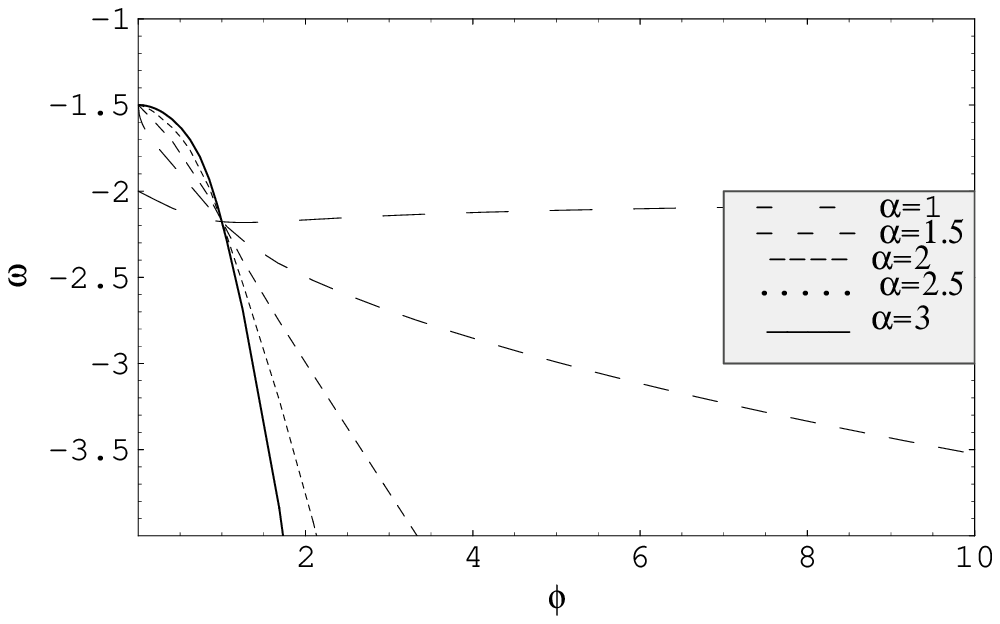}\\
\vspace{.5mm} Fig 8.12: Variation of $\omega$ (for GCG) is shown
against $\phi$ for open ($k=-1$) model of the Universe. We have
considered different values of $\alpha=1, 1.5, 2, 2.5,3$ and
$\beta=-2$ and normalize the parameters as
$a_{0}=\rho_{0}=\phi_{0}=B=C=1$ and $n=1$.\\

\end{figure}

{\bf{Case II}:} Now we consider $\omega=\omega(\phi)$, i.e.,
$\omega$ dependent on $\phi$.\\

Also the power law forms considered will be (8.5) and (8.10).
Solving the equations we get,

\begin{equation}
\omega(\phi)=\frac{\alpha \beta+2\alpha+\beta-\beta ^{2}}{\beta
^{2}}-\frac{C{a_{0}}^{-3(1+n)}{\phi_{0}}^{\frac{3\alpha(1+n)-2}{\beta}}
\phi^{\frac{-3\alpha(1+n)-\beta+2}{\beta}}} {\beta^{2}\left[B +
C{a_{0}}^{-3(1+n)}{\phi_{0}}^{\frac{3\alpha(1+n)}{\beta}}
\phi^{\frac{-3\alpha(1+n)}{\beta}}\right]^{\frac{n}{1+n}}}
+\frac{2k \phi^{\frac{2(1-\alpha)}{\beta}}}{{a_{0}}^{2}
\beta^{2}{\phi_{0}}^{\frac{2(1-\alpha)}{\beta}}}
\end{equation}

Also substituting these values  in the given equations, we get,
either $n=-1$ or $B=0$ and also $k=0$. If $n=-1$, we get back
barotropic fluid, and if $B=0$, we get dust filled Universe. In
both the cases the Generalized Chaplygin gas does not seem to
have any additional effect on the cosmic
acceleration.\\

\subsection{Solution With Potential}

{\bf{Case I}:} Let us choose $\omega(\phi)=\omega=$ constant.\\

We again consider the power law forms (8.5) and (8.10). We get the
solution for $V(\phi)$ to be

\begin{equation}
V(\phi)=(2\alpha+\beta)(3\alpha+2\beta-1)
{\phi_{0}}^{\frac{2}{\beta}}
\phi^{\frac{\beta-2}{\beta}}+\frac{-2B-C{a_{0}}^{-3(1+n)}
{\phi_{0}}^{\frac{3\alpha(1+n)}{\beta}}\phi^{\frac{-3\alpha(1+n)}{\beta}}}{\left[B
+ C{a_{0}}^{-3(1+n)}{\phi_{0}}^{\frac{3\alpha(1+n)}{\beta}}
\phi^{\frac{-3\alpha(1+n)}{\beta}}\right]^{\frac{n}
{(1+n)}}}+\frac{4k}{{a_{0}}^{2}}\phi^{\frac{\beta-2\alpha}{\beta}}{\phi_{0}}^{\frac{2\alpha}{\beta}}
\end{equation}

Substituting these values in the other equations we get that
$n=-1$, i.e., the equation of state of Generalized Chaplygin Gas
takes the form of that of barotropic fluid. Also we get,
$\alpha=1$, which implies $q=0$, i.e., uniform expansion of the
Universe.\\

\begin{figure}

\includegraphics{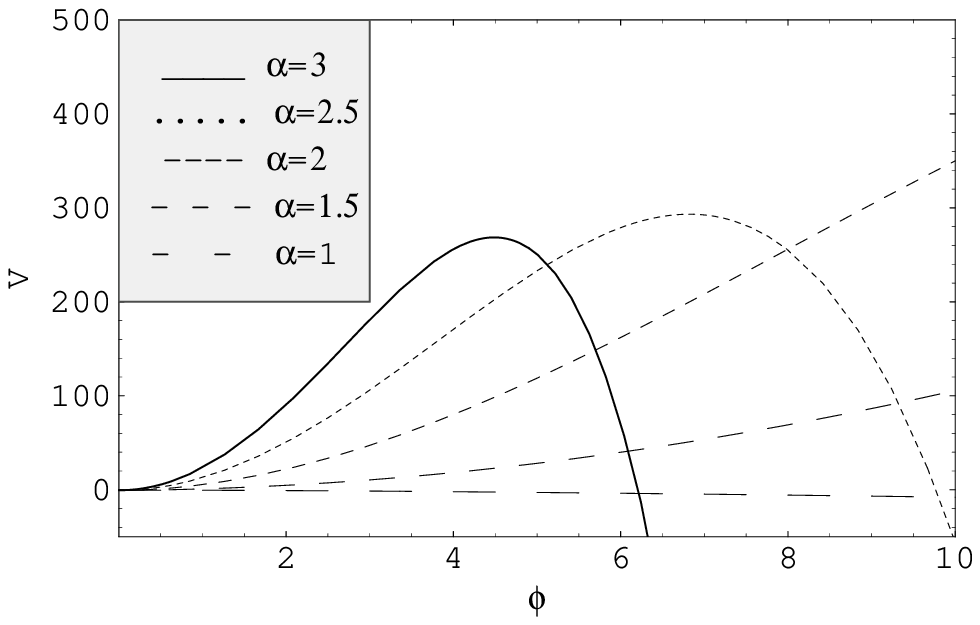}\\
\vspace{.5mm} Fig 8.13: Variation of $V$ (for GCG) has been
plotted against $\phi$ for flat ($k=0$) model of the Universe. We
have considered different values of $\alpha=1, 1.5, 2, 2.5,3$ and
$\beta=-2$ and normalize the parameters as
$a_{0}=\rho_{0}=\phi_{0}=B=C=1$ and
$n=1$.\\

\vspace{2mm}

\includegraphics{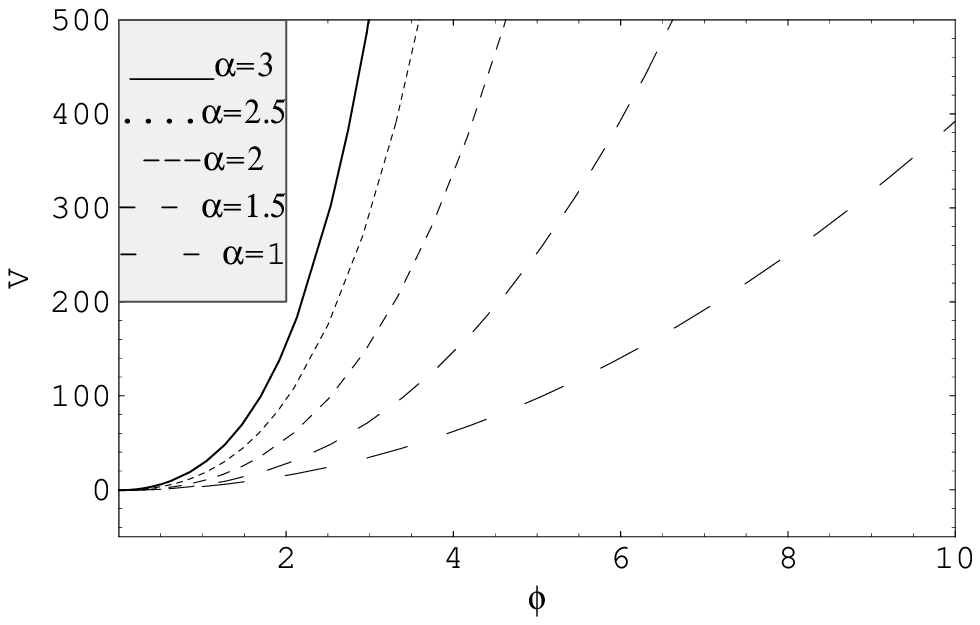}\\
\vspace{.5mm} Fig 8.14: Variation of $V$ (for GCG) has been
plotted against $\phi$ for closed ($k=1$) model of the Universe.
We have considered different values of $\alpha=1, 1.5, 2, 2.5,3$
and $\beta=-2$ and normalize the parameters as
$a_{0}=\rho_{0}=\phi_{0}=B=C=1$ and
$n=1$.\\

\vspace{2mm}

\includegraphics{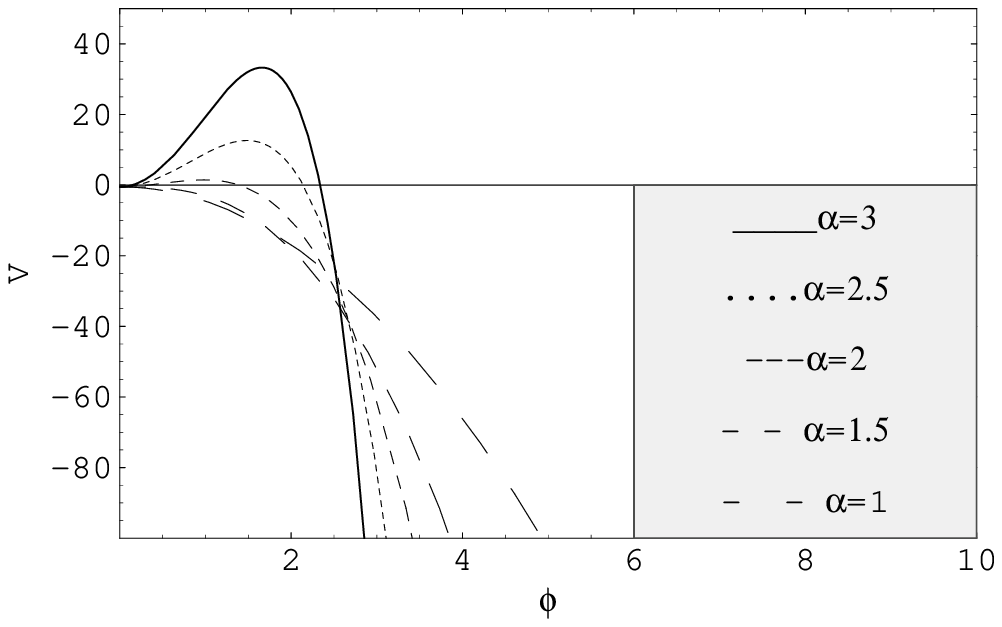}\\
\vspace{.5mm} Fig 8.15: Variation of $V$ (for GCG) has been
plotted against $\phi$ for open ($k=-1$) model of the Universe.
We have considered different values of $\alpha=1, 1.5, 2, 2.5,3$
and $\beta=-2$ and normalize the parameters as
$a_{0}=\rho_{0}=\phi_{0}=B=C=1$ and
$n=1$.\\

\end{figure}

\begin{figure}

\includegraphics{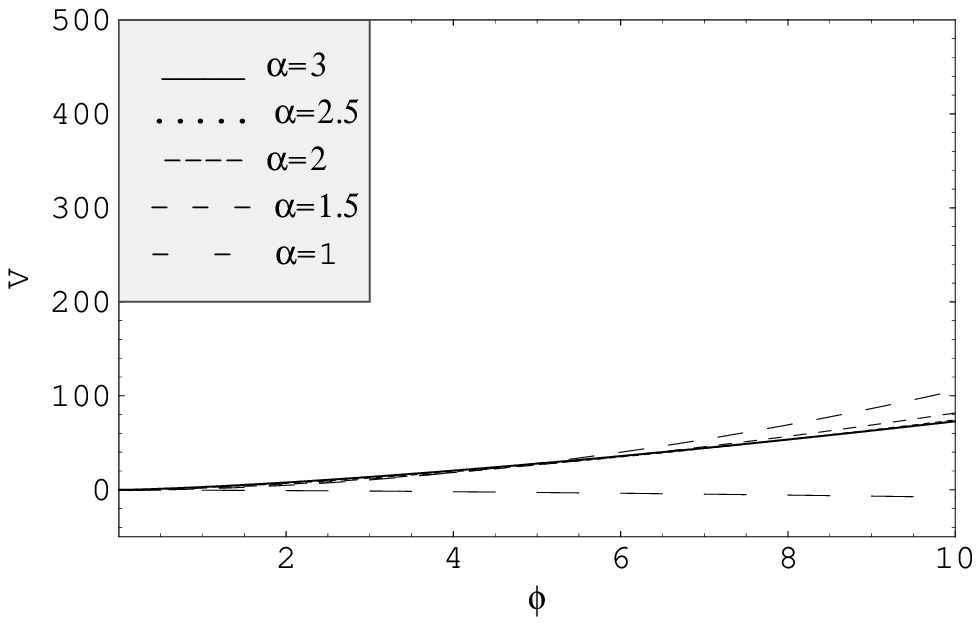}\\
\vspace{.3mm} Fig 8.16: The variation of $V$ (for GCG) is shown
against $\phi$ for flat ($k=0$) model of the Universe. We have
considered different values of $\alpha=1, 1.5, 2, 2.5,3$, $n=1$
and $\beta=-2\alpha$ and normalize the parameters as
$a_{0}=\rho_{0}=\phi_{0}=B=C=1$.\\

\vspace{1.5mm}

\includegraphics{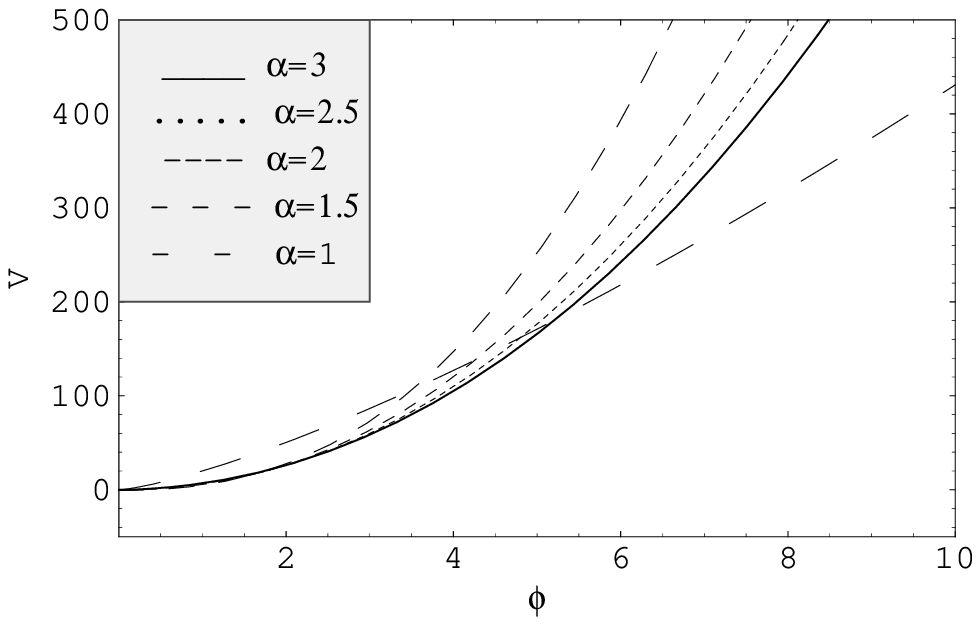}\\
\vspace{.3mm} Fig 8.17: The variation of $V$ (for GCG) is shown
against $\phi$ for closed ($k=1$) model of the Universe. We have
considered different values of $\alpha=1, 1.5, 2, 2.5,3$, $n=1$
and $\beta=-2\alpha$ and normalize the parameters as
$a_{0}=\rho_{0}=\phi_{0}=B=C=1$.\\

\vspace{1.5mm}

\includegraphics{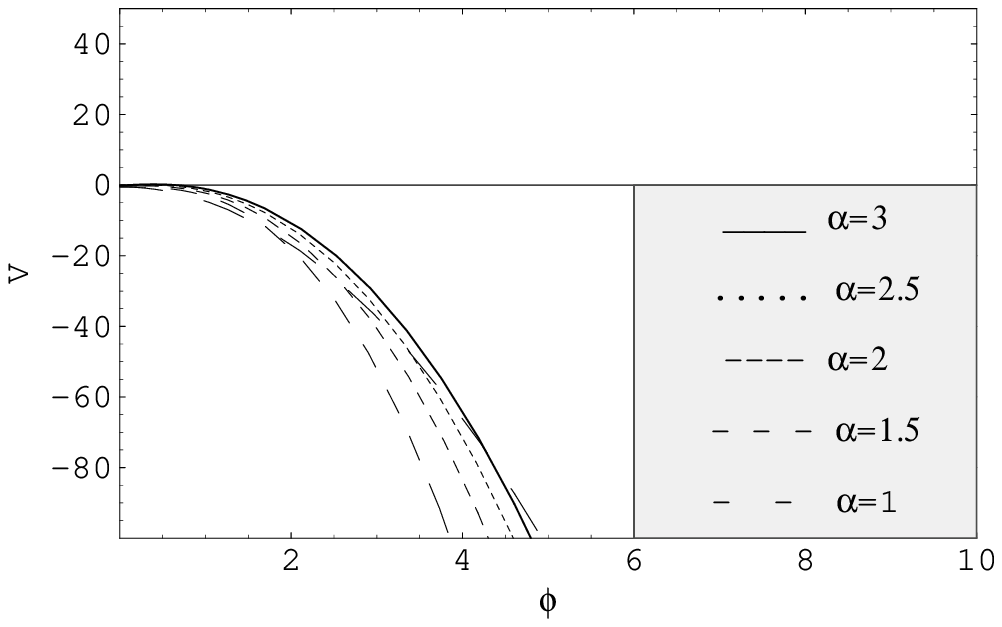}\\
\vspace{.3mm} Fig 8.18: The variation of $V$ (for GCG) is shown
against $\phi$ for open ($k=-1$) model of the Universe. We have
considered different values of $\alpha=1, 1.5, 2, 2.5,3$, $n=1$
and $\beta=-2\alpha$ and normalize the parameters as
$a_{0}=\rho_{0}=\phi_{0}=B=C=1$.\\

\end{figure}

{\bf{Case II}:} Now we choose $\omega(\phi)$ to be dependent on $\phi$.\\

Again we consider the power law forms, (8.5) and (8.10). Solving
the equations we get the solutions for Brans-Dicke parameter and
self-interacting potential as same as equations (8.25) and (8.26)
respectively.\\

Substituting these values in equation (8.2), we get

\begin{equation}
\text {either}~~~~ \beta=-2   ~~~~~\text {or} ~~~~\beta=-2\alpha
\end{equation}

Therefore for cosmic acceleration $q<0\Rightarrow \alpha>1$ and
$\beta\le -2$.\\

Therefore for the dust dominated era,

\begin{eqnarray*}
\omega=-\frac{3}{2}-\frac{\rho_{0}
{a_{0}}^{-3}\phi^{\frac{3\alpha-4}{2}}}{4{\phi_{0}}^{\frac{3\alpha-2}{2}}}~~~~~\text
{and}~~~V=\frac{2(\alpha-1)(3\alpha-5)}{\phi_{0}}\phi^{2}-\rho_{0}{a_{0}}^{-3}
\frac{\phi^{\frac{3\alpha}{2}}}{{\phi_{0}}^{\frac{3\alpha}{2}}}
~~~\text {if}~~~ \beta=-2
\end{eqnarray*}
\begin{equation}
\omega=-\frac{3}{2}-\frac{\rho_{0}
{a_{0}}^{-3}\phi^{\frac{\alpha-2}{2\alpha}}}{4
\alpha^{2}{\phi_{0}}^{\frac{3\alpha-2}{2\alpha}}}~~~~\text{and}
~~~
V=-\rho_{0}{a_{0}}^{-3}\frac{\phi^{\frac{3}{2}}}{{\phi_{0}}^{\frac{3}{2}}}~~~\text{if}
~~~\beta<-2
\end{equation}

Also for vacuum dominated era,

\begin{eqnarray*}
\omega=-\frac{3}{2} ~~~~\text{and}~~~~ V=
2(\alpha-1)(3\alpha-5)\frac{\phi^{2}}{\phi_{0}}-2\left[\rho_{vac}\right]_{\beta=-2}~~~~\text
{for} ~~~\beta=-2
\end{eqnarray*}
\begin{equation}
\omega=-\frac{3}{2}    ~~~\text {and}
~~~~V=-2\left[\rho_{vac}\right]_{2\alpha+\beta=0}~~~\text {for}
~~~\beta<-2
\end{equation}

\begin{figure}

\includegraphics{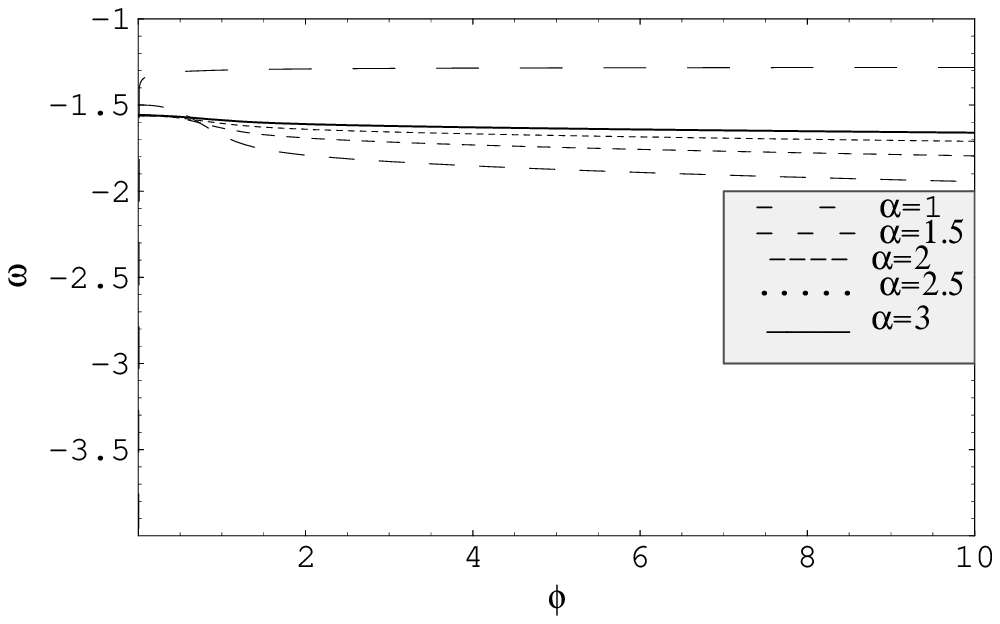}\\

\vspace{10mm} Fig 8.19: Variation of $\omega$ (for GCG) is shown
against $\phi$ for different values of $\alpha=1, 1.5, 2, 2.5,3$
in a flat ($k=0$) model of the Universe. Here we have considered
$\beta=-2\alpha$ and $n=0$ and normalize the parameters as
$a_{0}=\rho_{0}=\phi_{0}=b=C=1$.\\

\vspace{1in}

\includegraphics{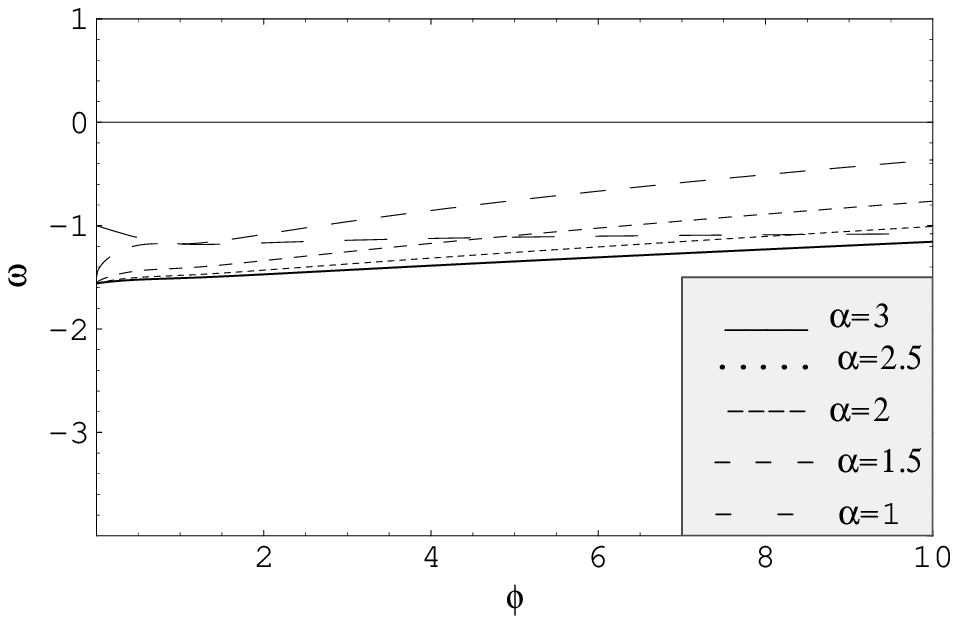}\\

\vspace{10mm} Fig 8.20: Variation of $\omega$ (for GCG) is shown
against $\phi$ for different values of $\alpha=1, 1.5, 2, 2.5,3$
in a closed ($k=1$) model of the Universe. Here we have considered
$\beta=-2\alpha$ and $n=0$ and normalize the parameters as
$a_{0}=\rho_{0}=\phi_{0}=b=C=1$.\\

\end{figure}

\section{Discussion}

We are considering Friedman-Robertson-Walker model in Brans-Dicke
Theory with and without potential ($V$). Also we have considered
the Brans-Dicke parameter ($\omega$) to be constant and variable.
We take barotropic fluid and Generalized Chaplygin Gas as the concerned fluid.\\

Using barotropic equation of state, we get,\\
 $(i)$ for $V=0$ and
$\omega=$constant, $\omega<0$ and $q<0$ for some values of
$\alpha$, giving rise to cosmic acceleration, \\
$(ii)$ for $V=0$ and $\omega=\omega(\phi)$, we obtain cosmic
acceleration depending on some values of $\alpha$ and $\beta$. In
this case we get acceleration for closed model also at the
radiation phase. We can show the variation of $\omega(\phi)$
against the variation of $\phi$ here [figure 8.1, 8.2]. Figure
8.1 shows that as the value of $\alpha$ increases $\omega$
decreases steadily against the variation of $\phi$. For
$\alpha>1$, we have accelerated expansion. The figure shows that
the greatest value of $\omega$ can be $-\frac{3}{2}$ and it
decreases further as $\phi$ increases, \\
$(iii)$ for $V=V(\phi)$ and $\omega=$constant, we get acceleration
in the flat model irrespective of the values of $\alpha$, \\
$(iv)$ for $V=V(\phi)$ and $\omega=\omega(\phi)$, cosmic
acceleration is obtained for $\beta\le 2$. Here we can represent
the variation of $\omega$ and $V$ against the variation of $\phi$
for $\beta=-2$ and $\beta=-2\alpha$. For $\beta=-2$, the variation
of $\omega$ against $\phi$ is same as figure 8.1 and that for
closed and open models are given in figure 8.3 and 8.4. Here we
can see that for open model $\omega$ starting at $-\frac{3}{2}$
decreases further, whereas for closed model $\omega$ starting at
$-\frac{3}{2}$ increases to be positive for $\alpha=2$. Figures
8.5, 8.6 and 8.7 show that variation of $V$ against the variation
of $\phi$ for $\beta=-2$ in respectively flat, closed and open
models of the Universe. Here we can see that only for the closed
model the potential increases positively, in the other two cases
the potential becomes negative after a certain point. Figure 8.8
shows the variation of $V$ against $\phi$ for $\beta=-2\alpha$.
Again positive potential energy is obtained for only the closed
model. The variation of $\omega$ is shown in figure 8.9 for $k=1$
and we can see that $\omega$ increases starting
at $-\frac{3}{2}$.\\

Using Generalized Chaplygin Gas , we get, \\
$(i)$ for $V=0$ and $\omega=$constant, uniform expansion is
obtained, \\
$(ii)$ for $V=0$ and $\omega=\omega(\phi)$, Generalized Chaplygin
Gas does not seem to have any effect of itself, \\
$(iii)$ for $V=V(\phi)$ and $\omega=$constant, we get $q=0$ giving
uniform expansion, \\
$(iv)$ for $V=V(\phi)$ and
$\omega=\omega(\phi)$ cosmic acceleration is obtained for
$\beta\le 2$ as previously obtained for barotropic fluid. Figures
8.10, 8.11, and 8.12 show the variation of $\omega$ for $\beta=-2$
in flat closed and open models and the natures of the graphs do
not vary much from that for barotropic fluid. Figures 8.13, 8.14
and 8.15 show the variation of $V$ for flat, closed and open
models respectively. Here for open model we get a negative
potential after a certain point, whereas for closed model we get a
positive potential always. For spatially flat model a positive $V$
is obtained for $\alpha=1.5, 2$. Figures 8.16, 8.17 and 8.18 show
the variation of $V$ for the models of the Universe for
$\beta=-2\alpha$. Positive potential is obtained for closed model
and flat model shows positive potential for $\alpha>1$. For open
model we get negative $V$ again. Figures 8.19 and 8.20 show the
variation of $\omega$ for flat and closed models respectively
($\beta=-2\alpha$). For flat model $\omega$ starting at
$-\frac{3}{2}$ decreases further and for closed model it
increases slowly from $-\frac{3}{2}$.\\

We have used BD theory to solve the problem of cosmic
acceleration. Here we use barotropic fluid and Generalized
Chaplygin Gas. Although the problem of fitting the value of
$\omega$ to the limits imposed by the solar system experiments
could not be solved fully, for closed Universe and $\beta=-2$ and
$\alpha>1$, $\omega$ starting from $-\frac{3}{2}$ increases and
for large $\phi$, we get $\omega>500$, for both barotropic fluid
and Generalized Chaplygin Gas. Also for flat Universe filled with
barotropic fluid taking $\omega=$constant and $V=V(\phi)$, we get
the Bertolami-Martins [2000] solution, i.e, $V=V(\phi^{2})$ and
$q_{0}=-\frac{1}{4}$ for $a=At^\frac{4}{3}$. But taking
Generalized Chaplygin Gas, we get accelerated expansion only when
both $\omega$ and $V$ are functions of the scalar field $\phi$.
For $\beta=-2$ we get cosmic acceleration in the closed model,
whereas, $\beta=-2\alpha$ gives acceleration in both closed and
flat models of the Universe, although for flat Universe $\omega$
varies from $-\frac{3}{2}$ to $-2$ and for closed Universe
$\omega$ takes large values for large $\phi$. In the end we see
that for all the cases accelerated expansion can be achieved for
closed model of the Universe for large values of $\omega$. Also
the present day acceleration of the Universe can also be explained
successfully, although in this
case $\omega$ cannot meet the solar system limits.\\\\

%% file: chap04.tex
\addcontentsline{toc}{part}{Short Discussions and Concluding
Remarks}
 \baselineskip
.81cm

\markright{}

\begin{center}
 { \huge {\bf Short Discussions and Concluding
Remarks} }
\end{center}
\vspace{.7in}

This thesis concentrates on the accelerated expansion of the
Universe recently explored by measurements of redshift and
luminosity-distance relations of type Ia Supernovae. This work
also deals with the dark energy problem, which is recently one of
the most widely investigated problems in Cosmology. A few dark
energy models have been considered for this purpose. These models
have been discussed, the equations have been solved for exact
solutions to show the significance of these models to solve the
dark energy problem. Statefinder diagnostics play very important
role here, as the statefinder parameters have been solved and
plotted to show the evolution of the Universe.\\

In chapter 1, standard cosmology and the FRW model of the Universe
have been described. It also discusses dark matter and dark energy
and various candidates of dark energy. A brief introduction to
Brans-Dicke cosmology has also been given. This
chapter ends with a short note on the statefinder parameters.\\

Chapter 2 deals with a model of the universe filled with modified
Chaplygin gas and barotropic fluid. The field equations have been
solved to show its role in acceleration of the universe.
Statefinder parameters have been solved and plotted to show the
different phases of evolution of the Universe. This model has
also been discussed from field theoretical point of view.\\

In chapter 3 the role of dynamical cosmological constant has been
explored with Modified Chaplygin Gas as the background fluid.
Various phenomenological models for $\Lambda$, viz.,
$\Lambda\propto\rho, \Lambda\propto\frac{\dot{a}^{2}}{a^{2}}$  and
$\Lambda\propto\frac{\ddot{a}}{a}$ have been considered for this
purpose. These models have been studied in presence of the
gravitational constant $G$ to be constant or time dependent.
Natures of $G$ and $\Lambda$ have been shown over the total age
of the Universe. Statefinder analysis has been done to show the
evolution of the Universe.\\

Recently developed Generalized Cosmic Chaplygin gas (GCCG) is
studied in chapter 4 as an unified model of dark matter and dark
energy. To explain the recent accelerating phase, the Universe is
assumed to have a mixture of radiation and GCCG. The mixture is
considered for without or with interaction. Solutions are
obtained for various choices of the parameters and trajectories
in the plane of the statefinder parameters and presented
graphically. For particular choice of interaction parameter, the
role of statefinder parameters have been shown in various cases
for the evolution of the Universe.\\

In chapter 5 a new form of the well known Chaplygin gas model has
been presented by introducing inhomogeneity in the EOS. This
model explains $\omega=-1$ crossing. Also a graphical
representation of the model using $\{r,s\}$ parameters have been
given to show the evolution of the Universe. An interaction of
this model with the scalar field has also been investigated
through a phenomenological coupling function. A decaying nature
of the potential has been shown for this model.\\

In chapter 6 tachyonic field has been depicted as dark energy
model to represent the present acceleration of the Universe. For
this purpose a mixture of tachyonic fluid with barotropic fluid
has been assumed. Also a mixture of the tachyonic fluid has been
considered with Generalized Chaplygin Gas to show the role of the
later as a dark energy candidate in presence of tachyonic matter.
A particular form of the scale factor has been assumed to solve
the equations of motion and get the exact solutions of the
density, tachyonic potential and the tachyonic field. A coupling
term has also been introduced in both the models to represent the
energy transfer between the two fluids. The interaction term and
the tachyonic potential has been analysed and plotted to show
their nature in the evolution of the Universe.\\

Chapter 7 deals with inhomogeneous EOS. A model of interaction
has been studied with scalar field and the inhomogeneous ideal
fluid. Two forms of the ideal fluid have been analysed. A power
law expansion for the scale factor has been assumed to solve the
equations for the energy densities. This model shows a decaying
nature of the scalar field potential and the interaction parameter.\\

In chapter 8, Brans-Dicke theory has been used to investigate the
possibility of obtaining cosmic acceleration. For this purpose a
constant and a variable $\omega$ (Brans-Dicke parameter) have been
considered. A self-interacting potential has been introduced to
show its role in the evolution of the Universe. This model has
been studied in presence of barotropic fluid and Generalized
Chaplygin Gas. Power law forms of the scale factor and the scalar
field have been assumed to solve the field equations. It has been
shown that accelerated expansion can also be achieved for high
values of $\omega$ for closed Universe.\\

%% file: chap05.tex
\addcontentsline{toc}{part}{\bf References of the Papers Presented
in the Thesis}
 \baselineskip
.81cm

\markright{}

\begin{center}
 { \huge {\bf References of the Papers Presented
in the Thesis} }
\end{center}
\vspace{.8in}

\begin{enumerate}

\item  {\bf IS MODIFIED CHAPLYGIN GAS ALONG WITH
       BAROTROPIC FLUID RESPONSIBLE FOR ACCELERATION OF THE UNIVERSE?:}
       -\\
       {\bf\it Writambhara Chakraborty and Ujjal Debnath}\\
       {\it Modern Physics Letters A}, {\bf 22} 1805-1812
       (2007).\\
       Pre-print: {\it gr-qc}/ 0611094.\\

\item  {\bf GENERALIZED COSMIC CHAPLYGIN GAS MODEL WITH R WITHOUT
INTERACTION:} -\\ {\bf\it
       Writambhara Chakraborty, Ujjal Debnath and Subenoy Chakraborty}\\
       {\it Gravitation and Cosmology}, {\bf 13} 293-297
       (2007).\\
       Pre-print: {\it gr-qc}/ 0711.0079.\\

\item  {\bf Effect of dynamical cosmological constant in presence
       of modified Chaplygin gas for accelerating universe:} -\\
       {\bf\it Writambhara Chakraborty and Ujjal Debnath}\\
       {\it Astrophysics and Space Science}, {\bf 313} 409-417
       (2008).\\
       Pre-print: {\it gr-qc}/ 0705.4147.\\

\item  {\bf Interaction between scalar field and ideal fluid with inhomogeneous
       equation of state:} - \\{\bf\it Writambhara Chakraborty and Ujjal Debnath}\\
       {\it Physics Letters B}, {\bf 661} 1-4
       (2008).\\
       Pre-print: {\it gr-qc}/ 0802.3751.\\

\item  {\bf Role of a tachyonic field in accelerating the Universe
       in the presence of a perfect fluid:} - \\{\bf\it Writambhara Chakraborty and Ujjal Debnath}\\
       {\it Astrophysics and Space Science}, {\bf 315} 73-78
       (2008).\\
       Pre-print: {\it gr-qc}/ 0804.4801.\\

\item  {\bf Role of Brans-Dicke Theory
       with or without Self-Interacting Potential in Cosmic
       Acceleration:} -\\ {\bf\it Writambhara Chakraborty and Ujjal Debnath}\\
       {\it International Journal of Theoretical Physics}, {\bf 48} 232-247
       (2009).\\
       Pre-print: {\it gr-qc}/ 0807.1776.\\

\item  {\bf A New Variable Modified Chaplygin Gas Model
Interacting with Scalar Field:} -\\ {\bf\it
       Writambhara  Chakraborty and Ujjal Debnath}\\
       (Communicated in {\it Gravitation and Cosmology})\\

\end{enumerate}

%% file: chap06.tex
\addcontentsline{toc}{part}{\bf  Bibliography}
 \baselineskip
.81cm \markright{ Bibliography}